\documentclass{pasj01}
\Received{$\langle$2021 June 8$\rangle$}
\Accepted{$\langle$2021 November 2$\rangle$}
\Published{$\langle$publication date$\rangle$}

\begin{document}

\title{Galaxy Morphologies Revealed with Subaru HSC and Super-Resolution Techniques I: \\ Major Merger Fractions of\\ $L_{\rm UV}\sim3-15\, L_{\rm UV}^*$ Dropout Galaxies at $z\sim4-7$\thanks{Based on data collected at the Subaru Telescope and retrieved from the HSC data archive system, which is operated by Subaru Telescope and Astronomy Data Center at National Astronomical Observatory of Japan.}}
\author{Takatoshi \textsc{Shibuya}\altaffilmark{1}}
\author{Noriaki \textsc{Miura}\altaffilmark{1}}
\author{Kenji \textsc{Iwadate}\altaffilmark{1}}
\author{Seiji \textsc{Fujimoto}\altaffilmark{2, 3}} 
\author{Yuichi \textsc{Harikane}\altaffilmark{4, 5}} 
\author{Yoshiki \textsc{Toba}\altaffilmark{6, 7, 8}} 
\author{Takuya \textsc{Umayahara}\altaffilmark{1}}
\author{Yohito \textsc{Ito}\altaffilmark{1}}

\email{tshibuya@mail.kitami-it.ac.jp}

\altaffiltext{1}{Kitami Institute of Technology, 165, Koen-cho, Kitami, Hokkaido 090-8507, Japan}
\altaffiltext{2}{Cosmic Dawn Center (DAWN), Jagtvej 128, DK2200 Copenhagen N, Denmark}
\altaffiltext{3}{Niels Bohr Institute, University of Copenhagen, Lyngbyvej 2, DK2100 Copenhagen \O, Denmark}
\altaffiltext{4}{Institute for Cosmic Ray Research, The University of Tokyo, 5-1-5 Kashiwanoha, Kashiwa, Chiba 277-8582, Japan}
\altaffiltext{5}{Department of Physics and Astronomy, University College London, Gower Street, London WC1E 6BT, UK}
\altaffiltext{6}{Department of Astronomy, Kyoto University, Kitashirakawa-Oiwake-cho, Sakyo-ku, Kyoto 606-8502, Japan}
\altaffiltext{7}{Academia Sinica Institute of Astronomy and Astrophysics, 11F of Astronomy-Mathematics Building, AS/NTU, No.1, Section 4, Roosevelt Road, Taipei 10617, Taiwan}
\altaffiltext{8}{Research Center for Space and Cosmic Evolution, Ehime University, 2-5 Bunkyo-cho, Matsuyama, Ehime 790-8577, Japan}

\KeyWords{galaxies: structure --- galaxies: formation --- galaxies: evolution
 --- galaxies: high-redshift}

\maketitle

\begin{abstract}

We perform a super-resolution analysis of the Subaru Hyper Suprime-Cam (HSC) images to estimate the major merger fractions of $z\sim4-7$ dropout galaxies at the bright end of galaxy UV luminosity functions (LFs). Our super-resolution technique improves the spatial resolution of the ground-based HSC images, from $\sim1^{\prime\prime}$ to $\lesssim0.\!\!^{\prime\prime}1$, which is comparable to that of the {\it Hubble Space Telescope}, allowing us to identify $z\sim4-7$ bright major mergers at a high completeness value of $\gtrsim90$\%. We apply the super-resolution technique to $6412$, $16$, $94$, and $13$ very bright dropout galaxies at $z\sim4$, $5$, $6$, and $7$, respectively, in a UV luminosity range of $L_{\rm UV}\sim3-15\, L_{\rm UV}^*$ corresponding to $-24\lesssim M_{\rm UV}\lesssim-22$. The major merger fractions are estimated to be $f_{\rm merger}\sim10-20$\% at $z\sim4$ and $\sim50-70$\% at $z\sim5-7$, which shows no $f_{\rm merger}$ difference compared to those of a control faint galaxy sample. Based on the $f_{\rm merger}$ estimates, we verify contributions of source blending effects and major mergers to the bright-end of double power-law (DPL) shape of $z\sim4-7$ galaxy UV LFs. While these two effects partly explain the DPL shape at $L_{\rm UV}\sim3-10\, L_{\rm UV}^*$, the DPL shape cannot be explained at the very bright end of $L_{\rm UV}\gtrsim10\, L_{\rm UV}^*$, even after the AGN contribution is subtracted. The results support scenarios in which other additional mechanisms such as insignificant mass quenching and low dust obscuration contribute to the DPL shape of galaxy UV LFs. 

\end{abstract}


\section{Introduction}\label{sec_intro}

The shape of the rest-frame ultraviolet (UV) luminosity functions (LFs) provides important clues for understanding physical mechanisms of the galaxy formation and evolution. Extensive searches with the {\it Hubble Space Telescope} (hereafter, {\it Hubble}) have identified $>10,000$ galaxies fainter than an absolute UV magnitude of $M_{\rm UV}\sim-22$ (e.g., \cite{2010ApJ...725L.150O, 2013ApJ...773...75O, 2012ApJ...760..108B, 2013ApJ...763L...7E, 2013ApJ...768..196S, 2013MNRAS.432.2696M, 2015ApJ...810...71F, 2016MNRAS.459.3812M, 2016MNRAS.456.3194P, 2015ApJ...803...34B, 2021arXiv210207775B} and reference therein), suggesting that galaxy UV LFs at redshifts $z\sim2-10$ basically follow the Schechter function with the bright-end exponential cut-off in the number density \citep{1976ApJ...203..297S}. Complementary to these {\it Hubble} observations, wide-area imaging surveys with ground-based telescopes have uncovered the shape of UV LFs at $z\gtrsim4$ in a high UV luminosity regime of $-24\lesssim M_{\rm UV}\lesssim-23$ where the number density of galaxies and low-luminosity AGNs is comparable (e.g., \cite{2018PASJ...70S..10O, 2018ApJ...863...63S, 2018PASJ...70S..34A, 2019ApJ...883..183M, 2017ApJ...851...43S, 2019ApJ...883...99S, 2020MNRAS.494.1771A, 2014MNRAS.440.2810B,2015MNRAS.452.1817B, 2020MNRAS.493.2059B, 2021MNRAS.502..662B, 2021arXiv210613813F, 2021arXiv210801090H}). The most intriguing feature in the bright-end shape of galaxy UV LFs is that the number density at $-24\lesssim M_{\rm UV}\lesssim-23$ could deviate from the Schechter functions even if the AGN contribution is subtracted based on spectroscopic data and/or the source extendedness. The bright-end number density excess of galaxy UV LFs has been often characterized by the double power-law (DPL) function with $M_{\rm UV}\lesssim-22$ bright and $M_{\rm UV}\gtrsim-21$ faint power-law slopes (e.g., \cite{2012MNRAS.426.2772B, 2014MNRAS.440.2810B}). Several scenarios have been proposed to explain the bright-end DPL shape of galaxy UV LFs at high redshifts: 1) insignificant mass quenching (e.g., \cite{1977ApJ...215..483B, 2006MNRAS.365...11C, 2010ApJ...721..193P, 2019ApJ...878..114R}), 2) low dust obscuration (e.g., \cite{2015MNRAS.452.1817B, 2020MNRAS.493.2059B, 2021arXiv210800830V}), 3) gravitational lensing magnification (e.g., \cite{2011Natur.469..181W, 2011ApJ...742...15T, 2015ApJ...805...79M, 2015MNRAS.450.1224B}), and 4) hidden AGN activity (e.g., \cite{2021ApJ...910L..11K}).

\begin{longtable}{*{6}{c}}
\caption{Sample of Bright Dropout Galaxies for Our Super-Resolution Study\footnotemark[$*$]}\label{tab_sample}
 \hline
 Redshift & Field & $m_{\rm UV}$ & $M_{\rm UV}$ & $L_{\rm UV}/L_{\rm UV}^*$ & $N_{\rm gal}$ \\
  & & (mag) & (mag) & & \\
 (1) & (2) & (3)$^d$ & (4)$^d$ & (5)$^d$ & (6) \\
 \endhead
 \hline
  $4$ & UltraDeep & $22.00-23.69$ & $-23.94-(-22.25)$ & $15.00-3.16$ & $60$ \\ 
         & Deep          & $21.94-23.69$ & $-24.00-(-22.25)$ & $15.85-3.16$ & $1038$ \\ 
          & Wide          & $22.90-23.69$ & $-23.04-(-22.25)$ & $6.55-3.16$ & $5314$ \\ 
  $5$ & UltraDeep & $21.87-23.59$ & $-24.50-(-22.77)$ & $25.12-5.11$ & $3$ \\ 
          & Deep          & $23.37-23.59$ & $-22.99-(-22.77)$ & $6.25-5.11$ & $13$ \\ 
          & Wide$^b$          & ---             & ---        & ---       & --- \\ 
  $6$ & UltraDeep & $22.67-23.94$ & $-24.00-(-22.73)$ & $15.85-4.92$ & $0$ \\  
          & Deep          & $22.67-23.94$ & $-24.00-(-22.73)$ & $15.85-4.92$ & $14$ \\ 
          & Wide          & $22.67-23.94$ & $-24.00-(-22.73)$ & $15.85-4.92$ & $80$ \\ 
  $7^a$ & UltraDeep$^c$ & ---            & ---       & ---       & ---          \\ 
          & Deep          & $<24.50$ & $<-22.41$ & $>3.66$ & $1$ \\ 
          & Wide          & $<24.50$ & $<-22.41$ & $>3.66$ & $12$ \\ 
  Total & ---              & ---             & ---       & ---        & $6535$ \\ 
 \hline
\multicolumn{6}{l}{\footnotemark[$*$](1) Redshift of dropout galaxies. (2) Layers of the HSC-SSP fields. } \\
\multicolumn{6}{l}{(3) Range of apparent UV magnitude. (4) Range of absolute UV magnitude. } \\
\multicolumn{6}{l}{(5) Range of UV luminosity in units of $L_{\rm UV}^*$. (6) Number of dropout galaxies. } \\
\multicolumn{6}{l}{$^a$ The bright magnitude limit is not placed due to no estimate of the AGN fraction. See \citet{2018PASJ...70S..10O}. } \\
\multicolumn{6}{l}{$^b$ The $m_{\rm UV}$ range cannot be defined due to a relatively-high contamination fraction of low-$z$ interlopers} \\
\multicolumn{6}{l}{(i.e., $>50$\%) at $M_{\rm UV}<M_{\rm UV, th}$. See Section \ref{sec_data} in this paper and Figure 5 in \citet{2018PASJ...70S..10O}. } \\
\multicolumn{6}{l}{$^c$ No dropout galaxies are identified in  \citet{2018PASJ...70S..10O}. } \\
\multicolumn{6}{l}{$^d$ The bright and faint thresholds of the $M_{\rm UV}$ ranges are mainly determined by} \\
\multicolumn{6}{l}{criteria of the AGN fraction and the contamination fraction, respectively (see Section \ref{sec_data}).} \\
\end{longtable}

In addition to these scenarios, galaxy mergers could play significant roles in shaping the bright end of galaxy UV LFs by two effects: 1) the source blending effect and 2) the star-formation rate (SFR) enhancement. Individual galaxy components of a merger system tend to be blended at the low spatial resolution, in which case the blended sources are considered as bright single galaxies (e.g., \cite{2017MNRAS.466.3612B}). On the other hand, the UV luminosity could be enhanced/brighten as a consequence of the SFR enhancement caused by galaxy mergers and interactions (e.g., \cite{2008AJ....135.1877E, 2011ApJ...728..119W, 2013MNRAS.435.3627E, 2015MNRAS.450.1546B, 2019MNRAS.485.5631C}). The predominance of major mergers at high-$z$ might influence the fraction of bright galaxies at the bright-end of UV LFs (e.g., \cite{1975ApJ...202L.113O, 2020MNRAS.494.1366S}). These effects might reproduce the DPL functional form by flattening the bright-end exponential decline of the Schechter function. However, it is difficult to investigate morphologies of $z\gtrsim4$ bright galaxies at sub-structure levels of merger systems with low spatial resolution images of ground-based telescopes. Follow-up observations with {\it Hubble} are a promising approach (e.g., \cite{2013ApJ...773..153J, 2013AJ....145....4W, 2017MNRAS.466.3612B}), but unable to be efficiently performed for bright galaxies that are patchy distributed over $\gtrsim100$-deg$^2$-scale wide-area survey fields. For the reason, image processing techniques (e.g., super-resolution and machine learning) are critical to address such questions about morphological properties with a statistical sample of high-$z$ bright galaxies. 

In this paper, we study the relation between the bright-end shape of UV LFs and galaxy mergers with a super-resolution technique and the Subaru/Hyper Suprime-Cam (HSC)-Strategic Survey Program (SSP) data \citep{2018PASJ...70S...4A,2018PASJ...70S...8A, 2012SPIE.8446E..0ZM, 2018PASJ...70S...1M, 2018PASJ...70S...2K, 2018PASJ...70...66K, 2018PASJ...70S...3F}. This is the first paper in a series investigating the galaxy morphology with techniques of the super-resolution. The structure of this paper is as follows. We describe details of the HSC-SSP data and dropout galaxy samples in Section \ref{sec_data}. Section \ref{sec_analysis} explains the super-resolution technique, a method to identify galaxy mergers, the completeness and the contamination rate for the galaxy merger identification, and how to estimate galaxy merger fractions. In Section \ref{sec_results}, we present the galaxy merger fractions for bright dropout galaxies. In Section \ref{sec_discuss}, we discuss the possibility that the number density excess is reproduced by galaxy mergers. Section \ref{sec_summary} summarizes our findings. 

Throughout this paper, we assume a flat universe with the cosmological parameters of $(\Omega_{\rm m}, \Omega_\lambda, h)=(0.3, 0.7, 0.7)$. All magnitudes are given in the AB system \citep{1974ApJS...27...21O, 1983ApJ...266..713O}.

\section{Data}\label{sec_data}

We use a sample of dropout galaxies at $z\sim4-7$ \citep{2018PASJ...70S..10O, 2018PASJ...70S..11H, 2018PASJ...70S..12T}. The dropout galaxies are identified with the $\sim100$ deg$^2$-area HSC-SSP S16A data \citep{2018PASJ...70S...4A,2018PASJ...70S...8A}. 
The HSC-SSP is a large imaging survey project with a wide-field camera HSC on the Subaru telescope \citep{2018PASJ...70S...4A, 2018PASJ...70S...8A}. The HSC-SSP survey comprises three layers: UltraDeep, Deep and Wide. The effective survey areas of the HSC-SSP S16A  data used for the dropout galaxy selection are $2.4$, $17.7$, $82.6$ deg$^2$ for the UltraDeep, Deep and Wide fields, respectively. These fields are observed with five broadband filters of $g$, $r$, $i$, $z$, and $y$. The typical $5\sigma$ limiting magnitudes are $i\sim26.5$, $\sim26$, and $\sim25.5-26$ for the UltraDeep, Deep and Wide fields, respectively. The dropout galaxies are selected in the color selection criteria of $g-r>1.0$, $r-i<1.0$, $g-r>1.5(r-i)+0.8$ for $g$-dropouts, $r-i>1.2$, $i-z<0.7$, $r-i>1.5(i-z)+1.0$ for $r$-dropouts, $i-z>1.5$, $z-y<0.5$, $i-z>2.0(z-y)+1.1$ for $i$-dropouts, and $z-y>1.6$ for $z$-dropouts. The total numbers of dropout galaxy candidates are $540011$, $38944$, $537$, and $73$ at $z\sim4$, $5$, $6$, and $7$, respectively. See Tables 1 and 3 and Section 3.1 in \citet{2018PASJ...70S..10O} for more details. The sample contains a large number of bright dropout galaxies with $M_{\rm UV}\lesssim-23$, which enables us to study morphological properties of sources at the bright end of galaxy UV LFs. 

With the sample, we select bright dropout galaxies with a number-density excess from the \authorcite{1976ApJ...203..297S} function steep end of UV LFs. \citet{2018PASJ...70S..10O} have obtained two functional forms of galaxy UV LFs by fitting \authorcite{1976ApJ...203..297S} and DPL functions to the number density of dropout galaxies. Based on the best-fit Schechter and DPL functions, we determine a threshold of UV magnitude $M_{\rm UV, th}$ where the number density of the DPL functions is $0.1$ dex larger than that of the Schechter functions. The $M_{\rm UV, th}$ values range from $M_{\rm UV, th}\sim-23$ to $M_{\rm UV, th}\sim-22$ (corresponding to $L_{\rm UV, th}\sim3-5\, L_{\rm UV}^*$) which depend on the redshift of dropout galaxies. In this study, we define sources whose UV magnitude is brighter than $M_{\rm UV, th}$ as ``{\it bright dropout galaxies}". To reduce contaminations in the sample, we exclude 1) $z\sim4$ bright dropout galaxies with $m_{\rm UV}<22.0$ in the HSC-SSP UltraDeep fields because these bright sources are likely to be low-$z$ interlopers (see Section 4.1 of \cite{2018PASJ...70S..10O}), 2) $z\sim4-5$ bright dropout galaxies with $m_{\rm UV}$ where the contamination fractions of low-$z$ interlopers exceed $\sim50$\% (see Section 3.2 of \cite{2018PASJ...70S..10O}), 3) AGN-like bright objects in an $M_{\rm UV}$ range where the AGN fraction is $100$\% estimated with a spectroscopic sample of \citet{2018PASJ...70S..10O}, and 4) AGN-like compact sources with the magnitude difference $m_{\rm PSF}-m_{\rm CModel}<0.15$ \citep{2019ApJ...883..183M}, where $m_{\rm PSF}$ and $m_{\rm CModel}$ and the magnitudes measured with the point spread function (PSF) and the CModel profiles, respectively (see \cite{2018PASJ...70S...4A} for details). Using these selection criteria, we select $6412$ at $z\sim4$, $16$ at $z\sim5$, $94$ at $z\sim6$, and $13$ at $z\sim7$. The total number of the bright dropout galaxies is $6535$. For the photometric sample of dropout galaxies, we regard $z=3.8$, $z=4.9$, $z=5.9$, and $z=6.9$ as each representative redshift which is identical to the average redshift values measured in \citet{2018PASJ...70S..10O}. 

As a control sample, we randomly select ``{\it faint dropout galaxies}" in a UV luminosity of $L_{\rm UV}\sim2.5\, L_{\rm UV}^*$ where $L_{\rm UV}^*$ is the characteristic UV luminosity at $z\sim3$ corresponding to $M_{\rm UV}=-21$ \citep{1999ApJ...519....1S}. The number of the faint dropout galaxies is $880$ at $z\sim4$, $748$ at $z\sim5$, and $10$ at $z\sim6$. Note that faint dropout galaxies with $L_{\rm UV} \sim2.5\, L_{\rm UV}^*$ have not been identified at $z\sim7$ in \citet{2018PASJ...70S..10O}. 

The magnitude, luminosity, and numbers of the sample dropout galaxies are summarized in Table \ref{tab_sample}.

\begin{figure*}
 \begin{center}
  \includegraphics[width=170mm]{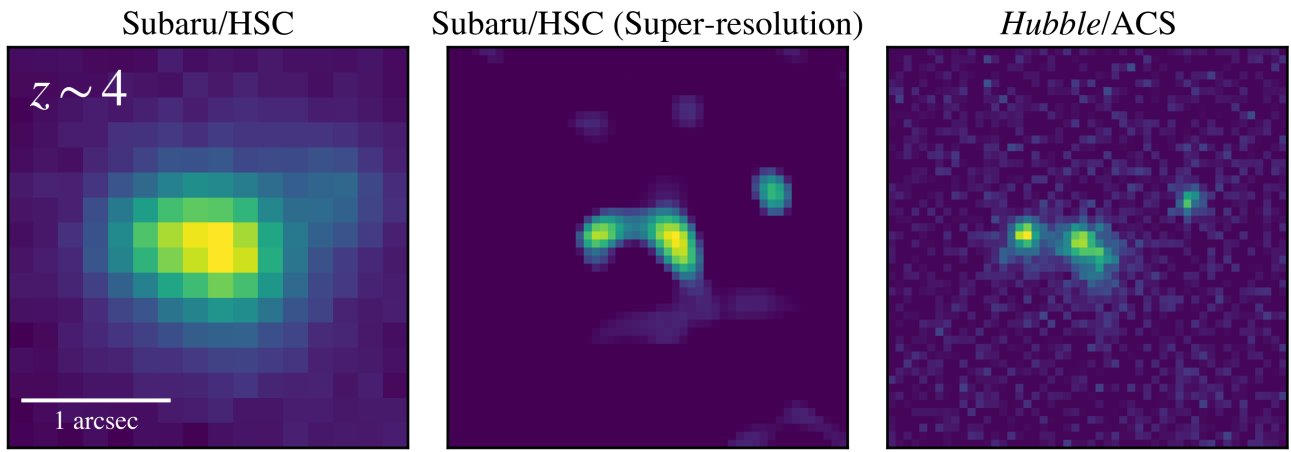}
  \end{center}
   \caption{Images of an example galaxy at $z\sim4$ analyzed in our super-resolution technique. (Left) Original HSC $i$-band image. (Middle) Super-resolved HSC $i$-band image. (Right) {\it Hubble} $I_{814}$-band image. The white horizontal bar indicates $1^{\prime\prime}$. }\label{fig_big_image}
\end{figure*}

\begin{figure*}
 \begin{center}
  \includegraphics[width=110mm]{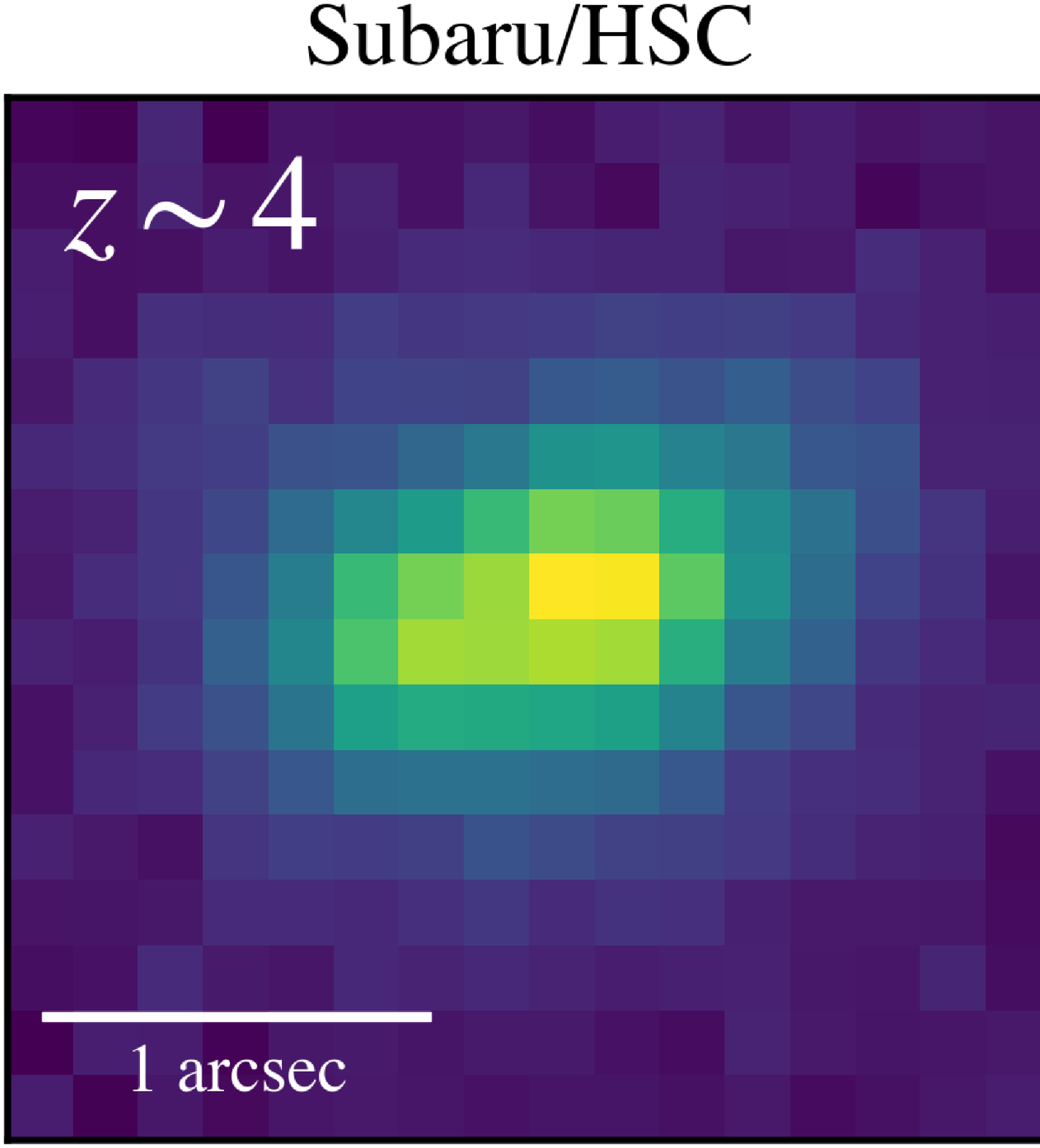}\\
  \includegraphics[width=110mm]{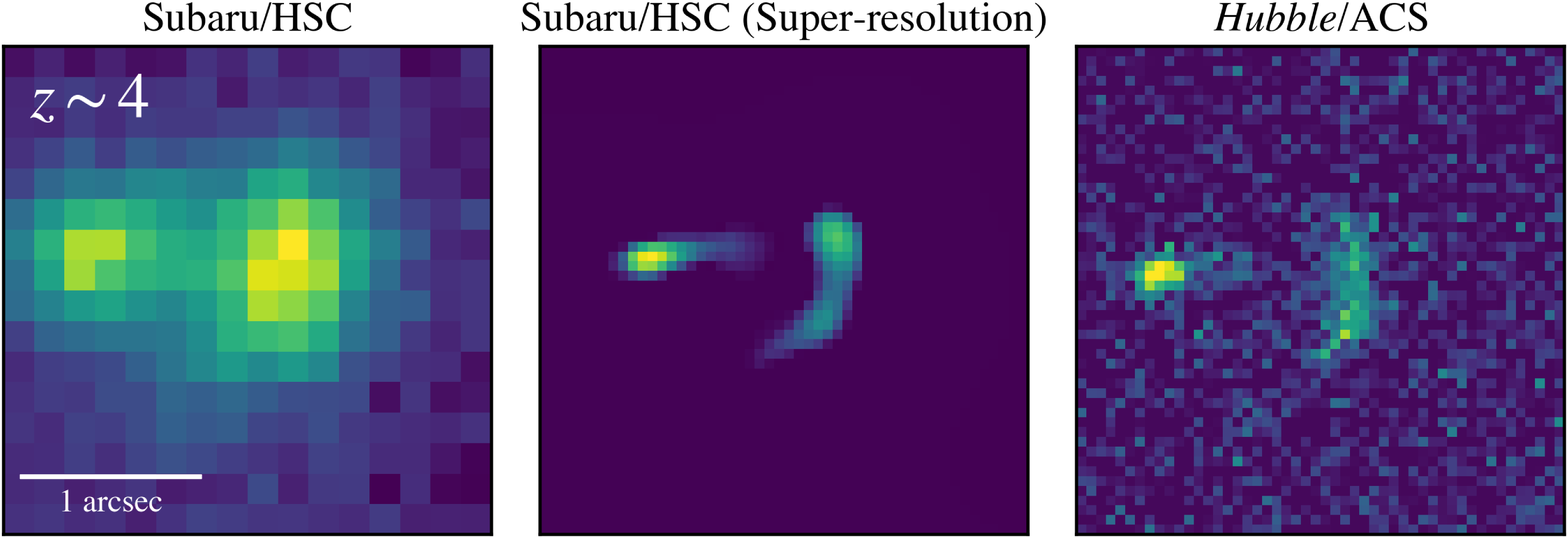}\\
  \includegraphics[width=110mm]{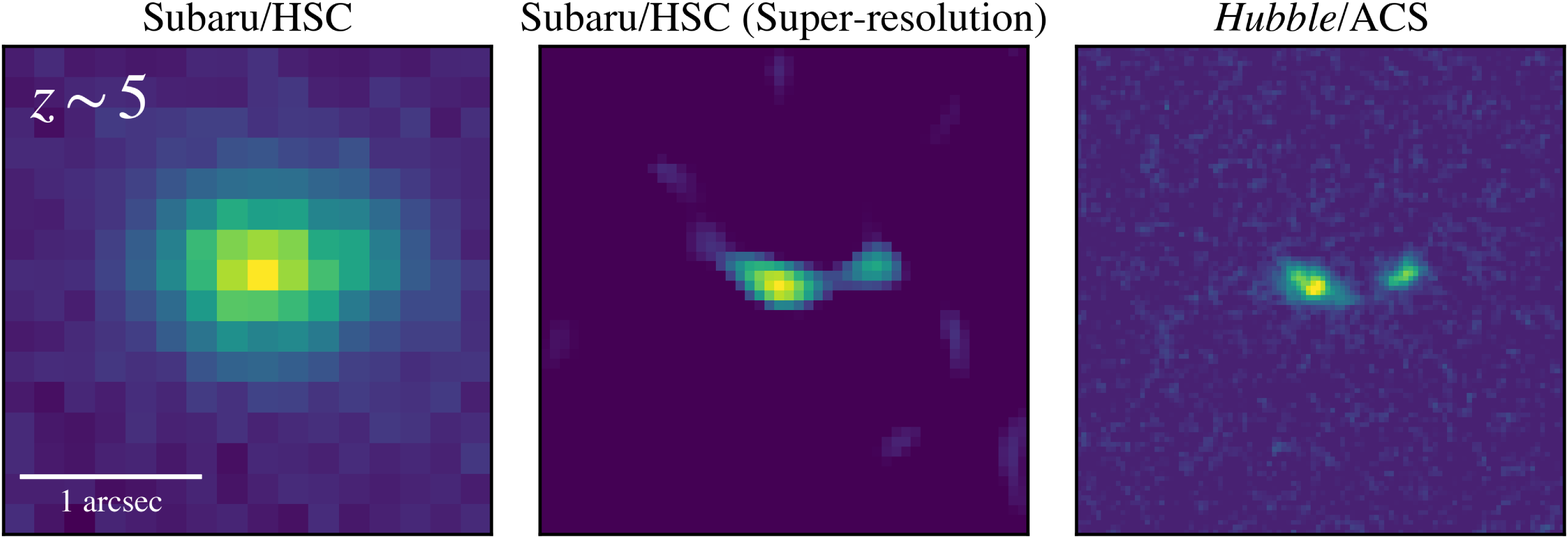}\\
  \includegraphics[width=110mm]{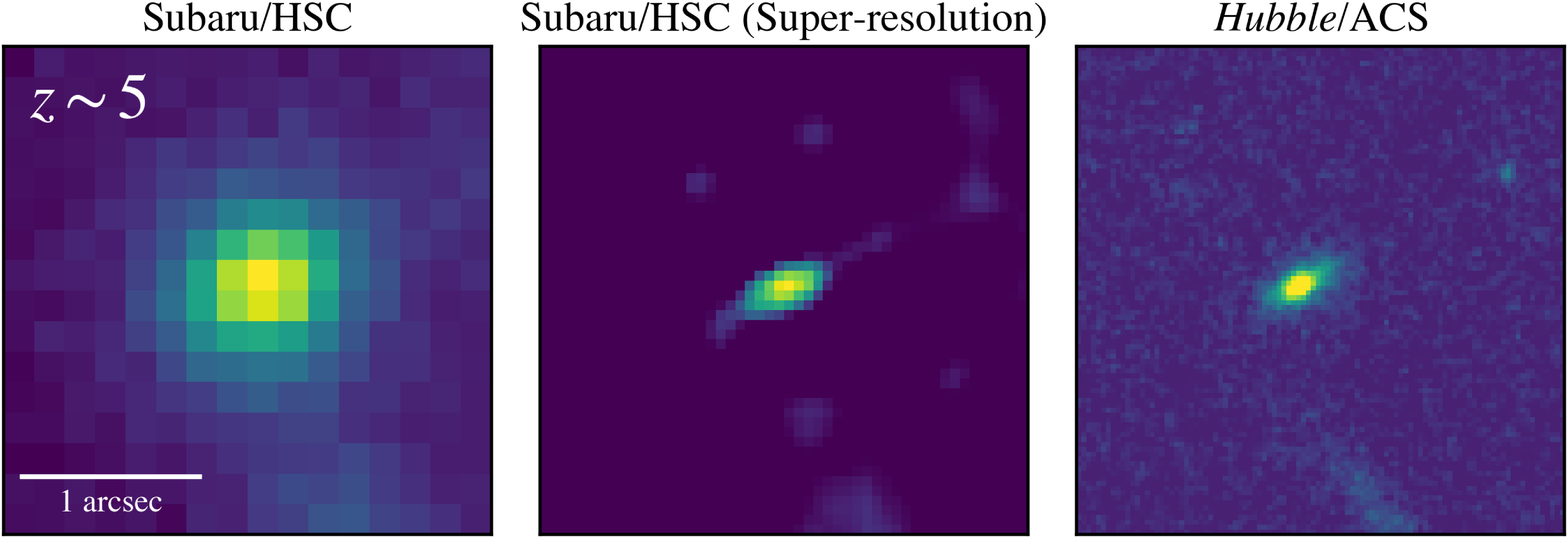}\\
  \includegraphics[width=110mm]{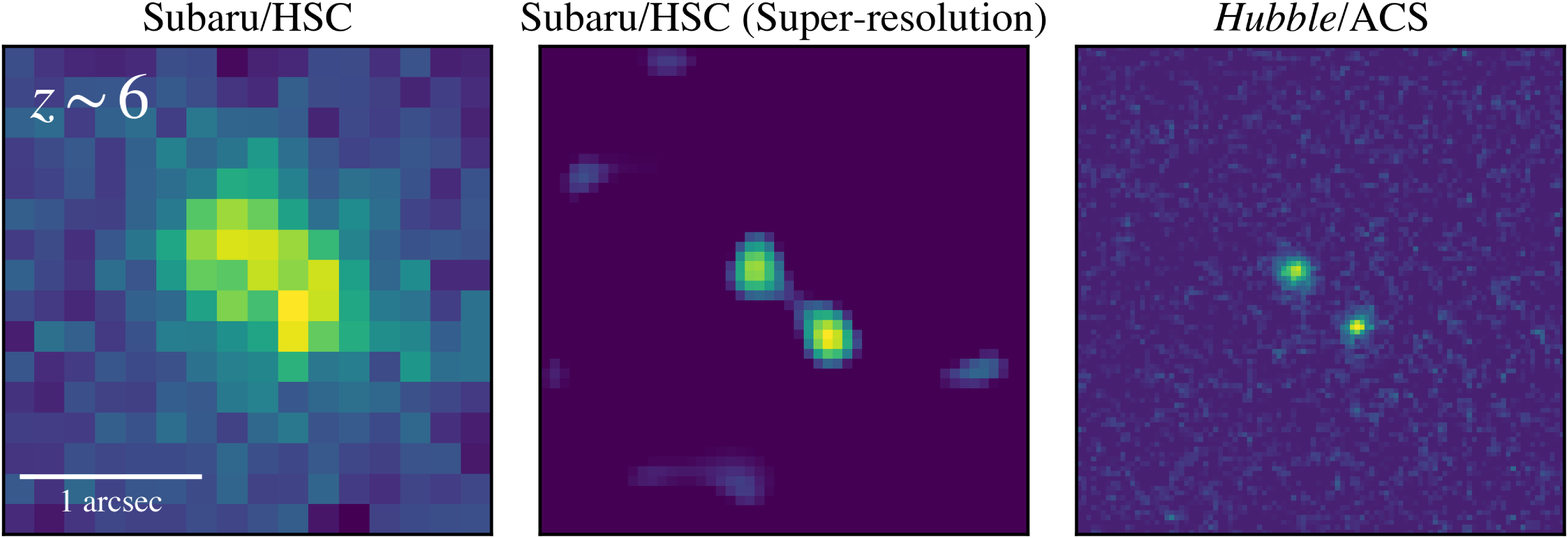}
  \includegraphics[width=110mm]{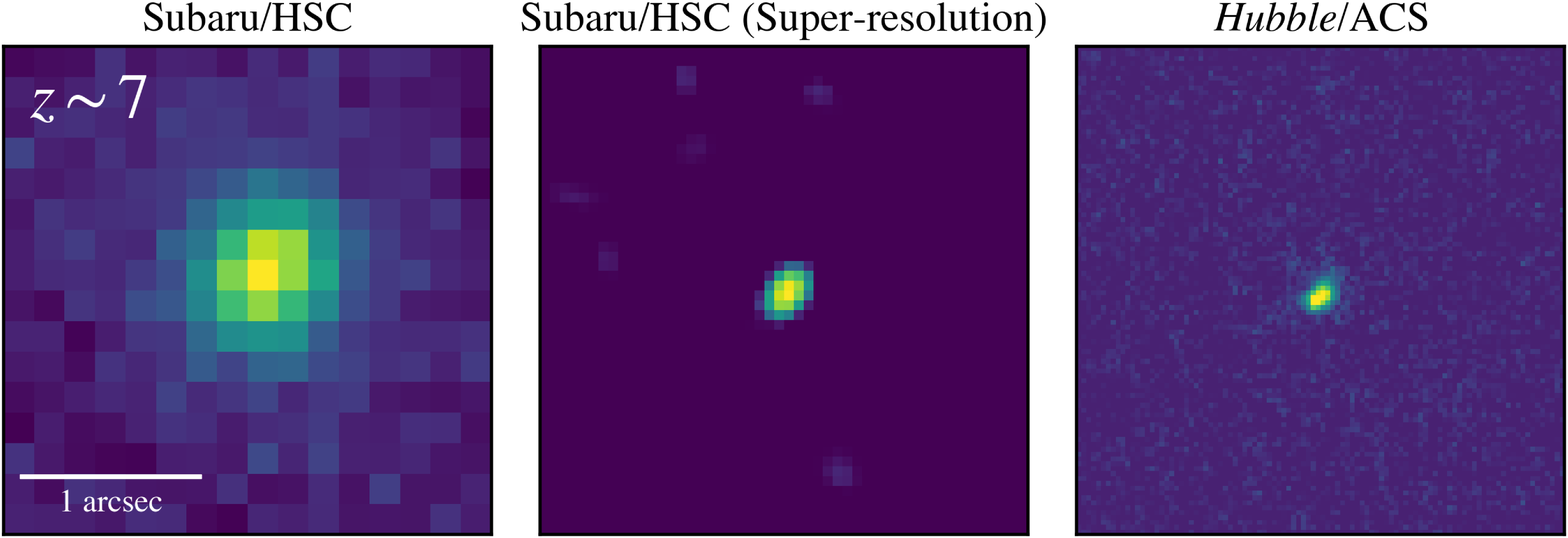}
  \end{center}
   \caption{Same as Figure \ref{fig_big_image}, but for other example galaxies at $z\sim4-7$. Note that the pixel scales of the {\it Hubble} images for $z\sim5-7$ are different from the original. See Section \ref{sec_comp_cont} for details. }\label{fig_image_z47}
\end{figure*}

\section{Analysis}\label{sec_analysis}

\subsection{Super-resolution}\label{sec_super_resolution}

For our super-resolution analysis, we employ the Richardson-Lucy (RL) deconvolution \citep{1972JOSA...62...55R, 1974AJ.....79..745L}. The RL deconvolution is a classical maximum-likelihood algorithm simply assuming that the noise of an observed image $\mathbf{y}$, 

\begin{equation}\label{eq_poisson}
\mathbf{y} = p( \mathbf{Px} ),  
\end{equation}

\noindent follows Poisson statistics $p$ at a specific image pixel $i$, 

\begin{equation}\label{eq_poisson_dist}
p(\mathbf{y}_i|\mathbf{x}) = \frac{(\mathbf{Px})_i^{\mathbf{y}_i} e^{-(\mathbf{Px})_i}}{\mathbf{y}_i!}, 
\end{equation}

\noindent where $\mathbf{P}$ is a PSF matrix, and $\mathbf{x}$ is a vector of a restored image. It is well known that the RL deconvolution yields a super-resolution effect because the cutoff spatial frequency of the image $\mathbf{x}$ gradually increases during the maximum likelihood estimation. While the RL deconvolution algorithm has been widely used in many applications, especially for the image de-noising, this technique has been scarcely applied to studies of high-$z$ galaxies. One advantage to use this algorithm is no requirement of large training samples that are commonly prepared for recent machine learning-based studies. Only observed object and PSF images are needed for the RL deconvolution.

To make the optimization of the RL deconvolution stable and efficient, we exploit the Alternating Direction Method of Multipliers (ADMM) algorithm \citep{MAL-016}. The ADMM algorithm is a powerful tool for solving optimization problems. By introducing auxiliary variables and an augmented Lagrangian, ADMM splits an optimization problem into small subproblems, each of which are easier to handle. The ADMM algorithm enables us to perform more stable and faster optimization than the RL {\tt scikit-image} {\tt Python} function, {\tt skimage.restoration.richardson\_lucy}, which implements the simple multiplicative update rule for the image restoration\footnote{https:\/\/scikit-image.org\/docs\/dev\/api\/skimage.restoration.html\#skimage.restoration.richardson\_lucy}. 

We minimize the following two penalty functions, 

\begin{eqnarray}\label{eq_penalty}
&& \mathop{\rm minimize}\limits_{\{\mathbf{x}\}}\ -\log{\{p(\mathbf{y}|\mathbf{w}_1)\}} + \mathit{I}_{R_+}(\mathbf{w}_2) \\
&& {\rm subject\ to}\ 
\mathbf{A}\mathbf{x} - \mathbf{w} = 0, \,\,\,\,\,
\mathbf{A} = 
\left[
    \begin{array}{c}
      \mathbf{P} \\
      \mathbf{I}
    \end{array}
\right], \,\,\,
\mathbf{w} =
\left[
    \begin{array}{c}
      \mathbf{w}_1 \\
      \mathbf{w}_2
    \end{array}
\right]
\nonumber
\end{eqnarray}

\noindent where the first term $-\log{\{p(\mathbf{y}|\mathbf{w}_1)\}}$ is the log-likelihood estimator for the Poisson noise (Equation \ref{eq_poisson_dist}), the second term $\mathit{I}_{R_+}(\mathbf{w}_2)$ is the indicator function defined as the following Equation (\ref{eq_nonnega}), and $\mathbf{I}$ is the identity matrix. 

\begin{equation}\label{eq_nonnega}
  I_{R_+}(w) = 
  \left\{
    \begin{array}{ll}
      0 & w \in R_+\\
      +\infty & w \notin R_+
    \end{array}
  \right.
\end{equation}

\noindent where $R_+$ is the closed nonempty convex set of the flux non-negativity constraints that force the image flux values to be always positive in the iteration. The $\mathbf{w}_1$ and $\mathbf{w}_2$ vectors are auxiliary variables that are newly introduced for the terms of the Poisson noise and the flux non-negativity, respectively. 

Following the standard ADMM formulation \citep{MAL-016}, we define the Augmented Lagrangian as

\begin{eqnarray}\label{eq_lagrangian}
 L_\rho (\mathbf{x}, \mathbf{w}, \mathbf{h}) &=& -\log{\{p(\mathbf{y}|\mathbf{w}_1)\}} +  \mathit{I}_{R_+}(\mathbf{w}_2) \nonumber \\
 && + \mathbf{h}^T (\mathbf{A}\mathbf{x}-\mathbf{w}) + \frac{\rho}{2} \|\mathbf{A}\mathbf{x}-\mathbf{w}\|_2^2
\end{eqnarray}

\noindent where $\mathbf{h}$ is the Lagrange multiplier, and $\rho$ is a parameter for the rate of the ADMM updates. The first and second terms are the penalty functions in Equation (\ref{eq_penalty}). The third and forth terms are introduced to separate the penalty functions into subproblems. Using the ADMM algorithm, we divide the problem of the RL deconvolution into subproblems, update iteratively the ($\mathbf{x}$, $\mathbf{w}_1$, $\mathbf{w}_2$, $\mathbf{h}$) vectors, and obtain a super-resolution image $\mathbf{x}$ in a method similar to those of, e.g., \citet{5492199}, \citet{Wen:2016vb}, and \citet{2018NatSR...811489I}. Section \ref{sec_appendix} presents details of how to solve each subproblem. Based on a parameter search for $\rho$, we choose $\rho=1\times10^{-3}$ for the stable and fast convergence, although the $\rho$ value does not significantly affect results of the super-resolution. 

For the convergence check during the ADMM iteration, we monitor the total absolute percentage error, 

\begin{equation}\label{eq_tape}
E(n_{\rm iter}) = \sum_{i=1}^{N_{\rm im}} \frac{|P_i * x_i -y_i|}{|y_i|}
\end{equation}

\noindent where $*$ denotes the convolution operator, $i$ is a specific pixel, $n_{\rm iter}$ is the current iteration number, and $N_{\rm im}$ is the total number of image pixels. We force the iteration to stop when the super-resolution image $\mathbf{x}$ meets a convergence criterion of $|{\rm d} E(n_{\rm iter})/ {\rm d}n_{\rm iter}| < 1\times10^{-3}$ at least $10$ times consecutively, or $n_{\rm iter}$ exceeds $N_{\rm iter}=250$. 

A number of recent studies have introduced other regularization terms of, e.g., the total variation \citep{1992PhyD...60..259R} and the L1 norm \citep{10.2307/2346178}. To avoid tuning parameters for such regularization terms, we use only the two penalty functions related to the Poisson noise and the flux non-negativity (Equation \ref{eq_penalty}). These two penalty functions are sufficient to resolve compact and faint components of high-$z$ galaxy mergers (Section \ref{sec_identify_mergers}). While this approach is appropriate for the current use of the images, a further check is needed for other applications, e.g., interpreting color gradients of galaxies at high resolution.

\begin{figure*}
 \begin{center}
  \includegraphics[height=60mm]{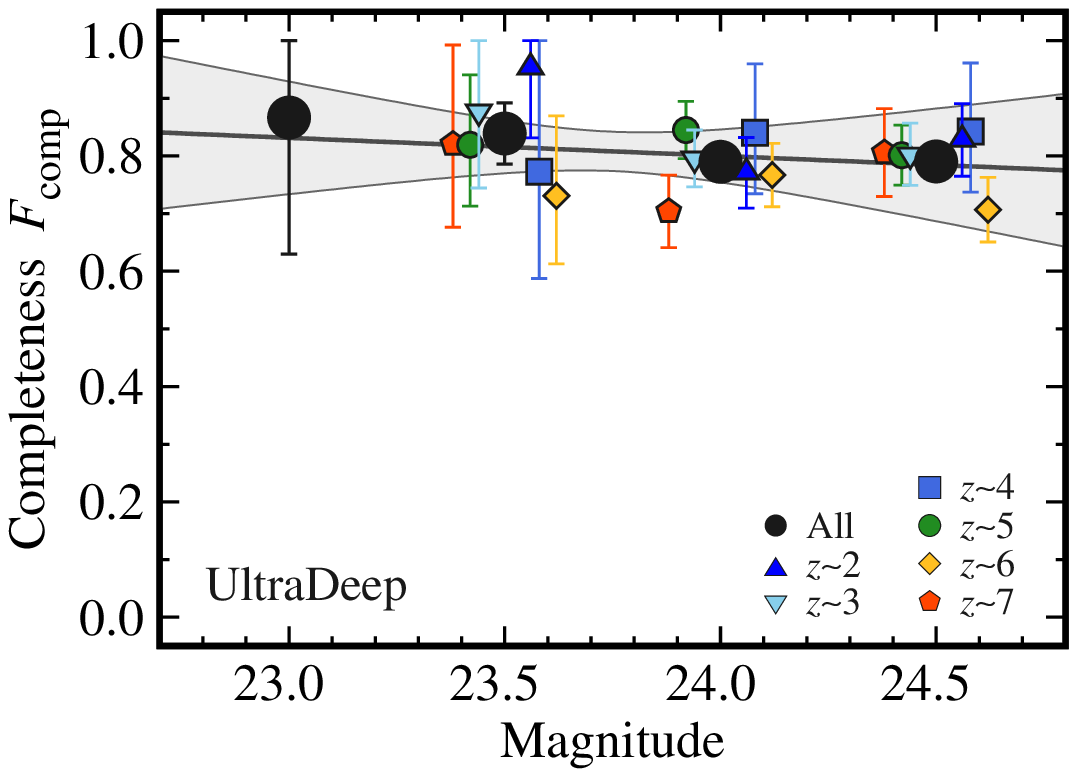}
  \mbox{}
  \includegraphics[height=60mm]{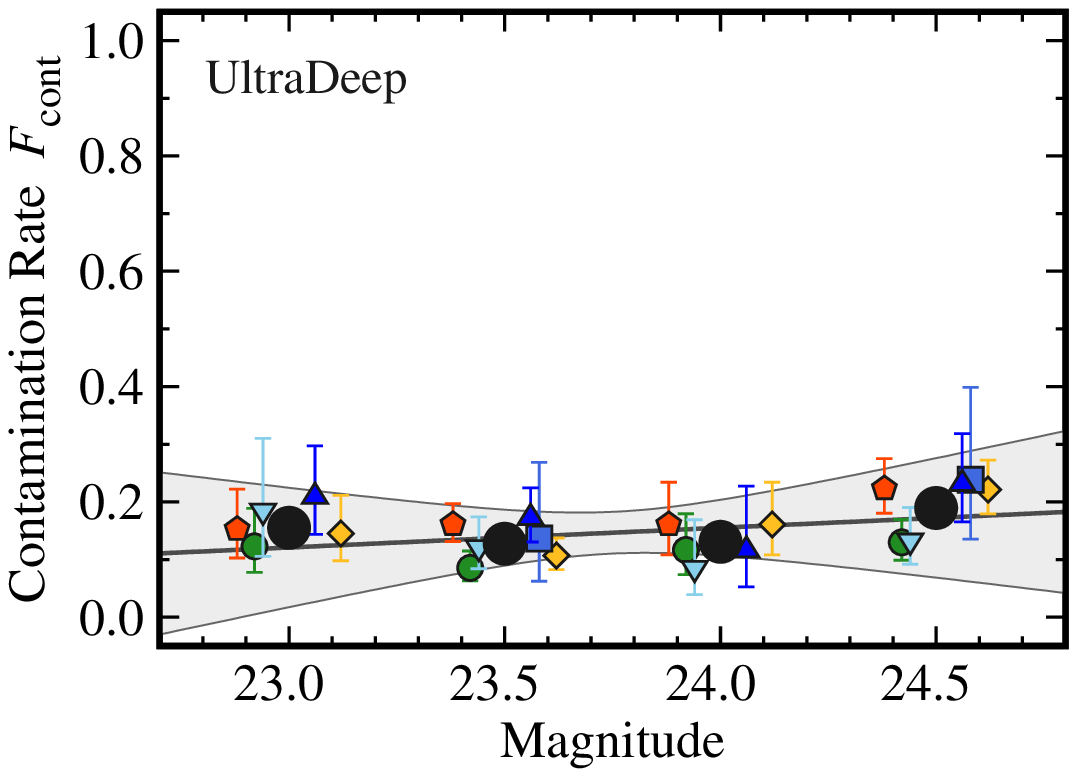}
  \mbox{}\\
  \mbox{}\\
  \includegraphics[height=60mm]{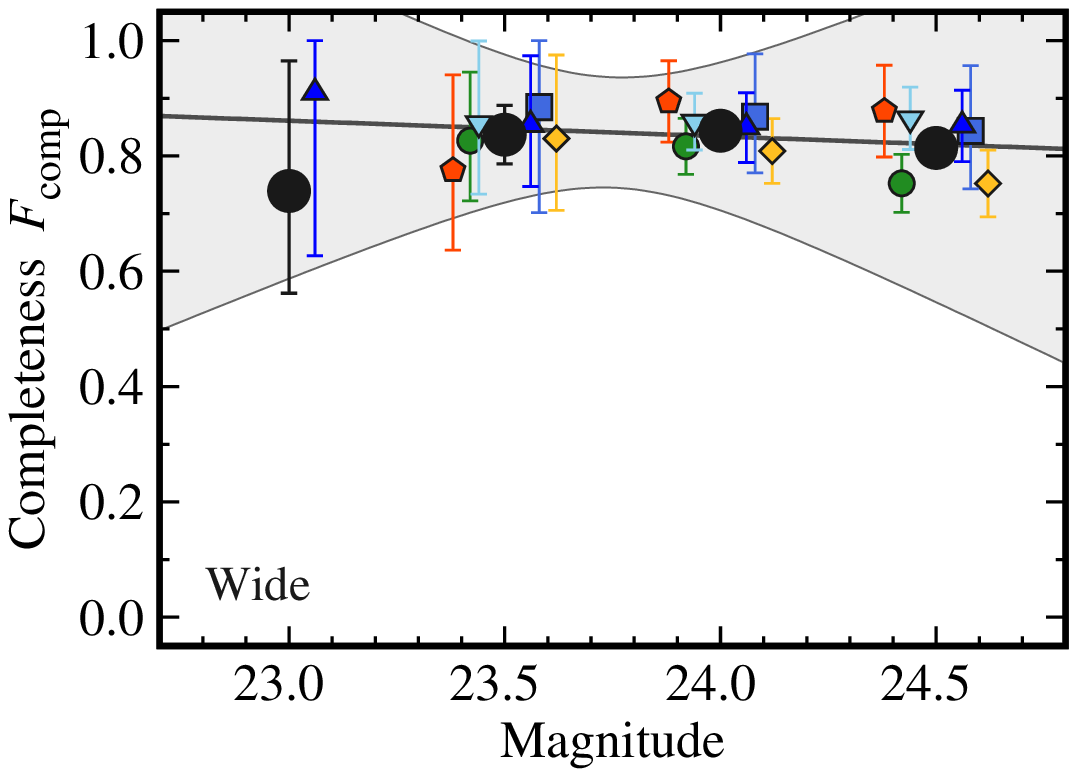}
  \mbox{}
  \includegraphics[height=60mm]{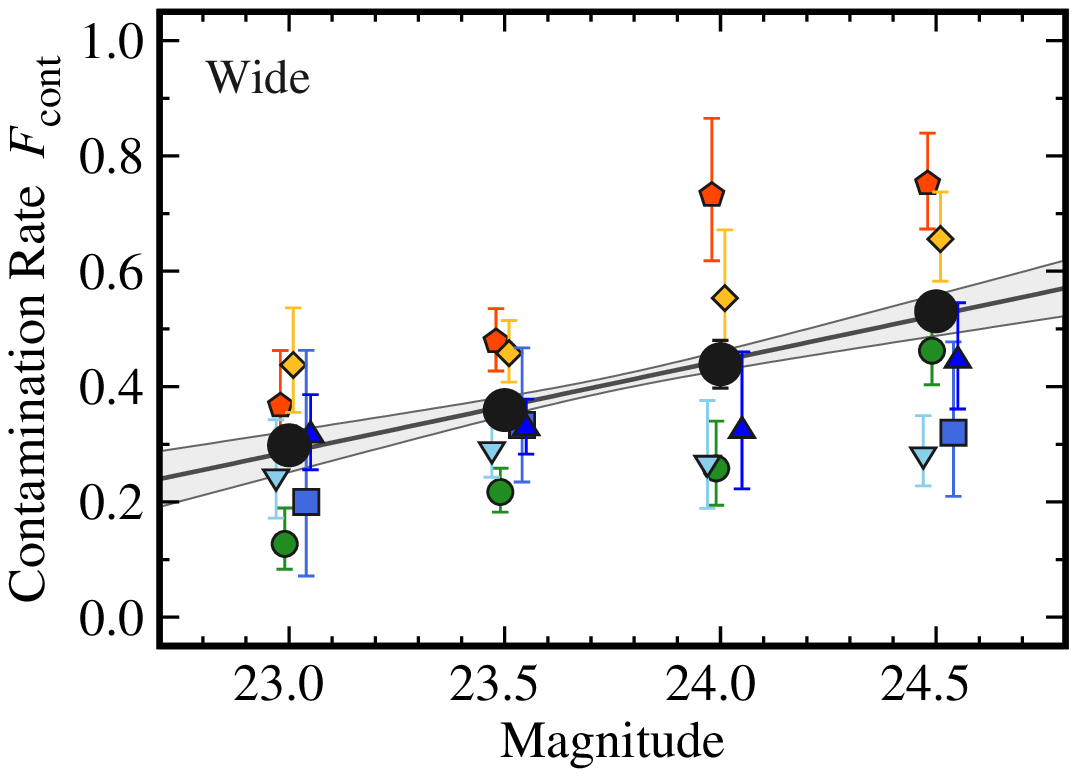}
  \end{center}
   \caption{Completeness (left) and contamination rate (right) as a function of magnitude. The top and bottom panels denote the UltraDeep and Wide fields, respectively. The symbols indicate the completeness and contamination rate estimated for sources at all redshifts (black circles), $z\sim2$ (blue triangles), $z\sim3$ (cyan inverse triangles), $z\sim4$ (blue squares), $z\sim5$ (green circles), $z\sim6$ (yellow diamonds), and $z\sim7$ (red pentagons). The data points are slightly offset along the $x$ axis for clarity. The panels show the completeness and contamination rate estimated with at least $10$ galaxies in each magnitude bin. The error bars of each symbol are calculated based on Poisson statistics from the galaxy number counts. The black solid lines and shaded gray regions depict the best-fit linear functions with $1\sigma$ significance errors to the measurements for all redshifts (i.e., black circles). }\label{fig_completeness_contamination}
\end{figure*}

\subsection{Identification of galaxy mergers}\label{sec_identify_mergers}

Using the super-resolution technique, we analyze the HSC-SSP images to identify galaxy mergers. The super-resolved images are obtained in the following steps. First, we retrieve $36$ pixels $\times$ $36$ pixels ($\sim6^{\prime\prime}\times6^{\prime\prime}$) cutout HSC S20A coadd images at the position of each dropout galaxy \citep{2019PASJ...71..114A}. These HSC-SSP S20A data have been obtained in observations during March 2014 through January 2020, and have been reduced with the {\tt hscPipe} 8.0-8.4 software (\cite{2018PASJ...70S...5B}; see also \cite{2010SPIE.7740E..15A, 2017ASPC..512..279J, 2019ApJ...873..111I}). We also extract PSF images for each galaxy with the {\tt HSC PSF picker} tool\footnote{https://hsc-release.mtk.nao.ac.jp/doc/index.php/tools-2/}. We use images of $i$-band for $z\sim4$ $g$-dropout, $z$-band for $z\sim5$ $r$-dropout, and $y$-band for $z\sim6-7$ $i, z$-dropout galaxies to trace the rest-frame UV continuum emission. The choice of these wavebands is the same as the ones used for galaxy UV LFs of \citet{2018PASJ...70S..10O}. Second, we sub-sample the original pixels of HSC coadd and PSF images and define a new pixel grid to match the pixel scale of {\it Hubble}/ACS (i.e., $0.\!\!^{\prime\prime}06$ per pixel). Here, we linearly interpolate the flux distribution in the new pixel grid. Finally, we super-resolve the HSC coadd images in the technique described in Section \ref{sec_super_resolution}. Figures \ref{fig_big_image} and \ref{fig_image_z47} show examples of super-resolved and de-noised HSC images together with corresponding {\it Hubble} images, which demonstrates the performance of our super-resolution technique. Our super-resolution technique clearly reveals galaxy sub-structures, e.g., candidates of galaxy merger components, in a scale smaller than the PSF FWHM of HSC (i.e., $\sim0.\!\!^{\prime\prime}6-1.\!\!^{\prime\prime}0$). Examples of mis-classified objects are presented in Figure \ref{fig_image_false} in Appendix. In most cases, objects are mis-classified due to the flux ratio uncertainty of individual galaxy components and/or the contamination of faint fake objects in high-resolution images.

From the super-resolved cutout HSC images, we detect sources with the {\tt SExtractor} software \citep{1996A&AS..117..393B}. The detection parameters of {\tt SExtractor} are set to {\tt DETECT\_MINAREA}$=5$, {\tt DETECT\_THRESH}$=2.5$, {\tt ANALYSIS\_THRESH}$=5.0$, {\tt DEBLEND\_NTHRESH}$=16$, {\tt DEBLEND\_MINCONT}$=0.0001$, {\tt PHOT\_AUTOPARAMS}$=2.5, 3.5$, {\tt SATUR\_LEVEL}$=120.0$, {\tt MAG\_GAMMA}$=4.0$, {\tt GAIN}$=2.0$ which are similar to those of \citet{2013ApJS..207...24G} for the {\it Hubble} CANDELS GOODS-S data. 

In the source catalogs constructed with {\tt SExtractor}, we select galaxy mergers. In this study, we focus only on major mergers with similar-mass components, because it is relatively difficult to differentiate galaxy minor mergers from clumpy galaxies (e.g., \cite{2016ApJ...821...72S}). Hereafter, the major merger is referred to as simply the ``galaxy merger", otherwise specified. We define galaxy close-pairs with a flux ratio of $\mu=f_2/f_1\geq1/4$ and a source separation within $d=r_{\rm min}-r_{\rm max}=3-6.5$ kpc as galaxy mergers, where $f_1$ and $f_2$ are the flux for the bright primary and the faint secondary galaxy components, respectively. The criterion of $\mu$ is the same as that of most previous studies on major mergers (see, e.g., Table 5 of \cite{2018MNRAS.475.1549M}, a summary table of previous studies). The criterion of source separation is different from that of previous studies for $z\sim0-3$ (typically $d=5-30$ kpc or $d=14-43$ kpc). The small value of $d=3-6.5$ kpc is adopted for two reasons: i) our criterion of source separation is roughly comparable to that of studies on high-$z$ dropout galaxies (e.g., \cite{2013ApJ...773..153J, 2013AJ....145....4W, 2017MNRAS.466.3612B}), ii) one of our main objectives is to examine the source blending effect on the DPL shape of galaxy UV LFs (see Section \ref{sec_intro}). We aim at resolving high-$z$ galaxy mergers with such a small source separation which are blended at the ground-based resolution.

\subsection{Completeness and Contamination Rate}\label{sec_comp_cont}

We estimate the completeness $F_{\rm comp}$ and contamination rate $F_{\rm cont}$ in identifying galaxy mergers. The completeness $F_{\rm comp}$ is a true positive rate or sensitivity which is defined as 

\begin{equation}\label{eq_completeness}
  F_{\rm comp} = \frac{N_{\rm true}}{N_{\rm real\mathchar`-merger}}
\end{equation}

\noindent where $N_{\rm true}$ is the number of objects that are correctly selected as galaxy mergers in our analysis, and $N_{\rm real\mathchar`-merger}$ is the total number of real mergers. The contamination rate $F_{\rm cont}$ is a false positive rate or fall-out which is defined as 

\begin{equation}\label{eq_contamination}
  F_{\rm cont} = \frac{N_{\rm false}}{N_{\rm isolated}}
\end{equation}

\noindent where $N_{\rm false}$ is the number of isolated galaxies that are incorrectly selected as galaxy mergers in our analysis, and $N_{\rm isolated}$ is the total number of isolated galaxies.\footnote{In object classification studies with machine learning techniques, the contamination rate is often defined as $F_{\rm cont}=N_{\rm false}/(N_{\rm true}+N_{\rm false})$. In this study, we use the definition of $F_{\rm cont}$ which is the same as that of previous studies on high-$z$ galaxy mergers. } As shown in Figures \ref{fig_big_image} and \ref{fig_image_z47}, some galaxy sub-structures are not clearly reproduced, and faint fake objects emerge in the super-resolved HSC images. The $F_{\rm comp}$ and $F_{\rm cont}$ values are essential to estimate the major merger fractions with our super-resolution technique. 

To estimate $F_{\rm comp}$ and $F_{\rm cont}$, we prepare ground-truth images of galaxy mergers and isolated galaxies. The part of the HSC-SSP UltraDeep COSMOS field has been observed with the {\it Hubble}/Advanced Camera for Survey (ACS) $I_{814}$ filter, which enables us to compare the super-resolved HSC images with the {\it Hubble} ones. By using a detection source catalog of \citet{2007ApJS..172..219L} constructed for the {\it Hubble} COSMOS-Wide data and the HSC-SSP {\tt MIZUKI} photometric redshifts \citep{2018PASJ...70S...9T}, we select galaxy mergers and isolated galaxies at $z>2$. Here, we apply the same selection criteria of the flux ratio and source separation as those for the dropout galaxies (Section \ref{sec_identify_mergers}). The numbers of selected galaxy mergers and isolated galaxies are $\sim1,500$ and $\sim1,100$, respectively. In the same manner as that for the dropout galaxies, we analyze the HSC $i$-band images for galaxies at $z\sim2$, $3$, and $4$. 

Because of no wide-area {\it Hubble} data covered by wavelength ranges of HSC $z$ and $y$ bands, we artificially create HSC $z$- and $y$-band images for galaxies at $z\sim5$, $6$, and $7$. First, we shrink the super-resolved HSC $i$-band images of galaxies at $z\sim2, 3$, and $4$. The shrinkage factor is determined from the galaxy size evolution of $r_{\rm e}\sim(1+z)^{-1.2}$ \citep{2015ApJS..219...15S}. In this process of the image shrinkage, we take into account the difference in angular diameter distances between the original and target redshifts. Second, we convolve these shrunk images with the randomly-selected $z$- and $y$-band PSF images. Finally, these PSF-convolved images are embedded into $z$- and $y$-band sky regions in the HSC-SSP UltraDeep and Wide fields to add the real sky background noise. As in the HSC $i$-band images, we estimate $F_{\rm comp}$ and $F_{\rm cont}$ for galaxies at $z\sim5$, $6$, and $7$. 

Figure \ref{fig_completeness_contamination} shows $F_{\rm comp}$ and $F_{\rm cont}$ as a function of magnitude. Even with the relatively low spatial resolution images of the ground-based Subaru telescope, our super-resolution technique successfully identifies galaxy mergers at a high completeness value of $F_{\rm comp}\gtrsim90$\% and a low contamination rate of $F_{\rm cont}\lesssim20$\% at bright magnitude $m\lesssim23$. Given that the dropout galaxies are bright (i.e., $m\sim22-24$, see Table \ref{tab_sample}), the major merger fraction can be estimated at high $F_{\rm comp}$ and relatively low $F_{\rm cont}$ values. Although $F_{\rm comp}$ and $F_{\rm cont}$ do not largely change with magnitude, $F_{\rm comp}$ ($F_{\rm cont}$) slowly decreases (increases) toward faint magnitudes. There is no strong dependence of $F_{\rm comp}$ and $F_{\rm cont}$ on image depths (i.e., UltraDeep or Wide). 

To obtain functional forms of $F_{\rm comp}$ and $F_{\rm cont}$, we fit a linear function to the data points of all-redshift galaxies (i.e., black circles). As shown in Figure \ref{fig_completeness_contamination}, most data points are distributed within the $1\sigma$ error regions of the best-fit linear functions for $F_{\rm comp}$ in the UltraDeep and Wide fields and $F_{\rm cont}$ in the UltraDeep field. Because of this consistency, we assume that these $F_{\rm comp}$ and $F_{\rm cont}$ at any redshifts follow the best-fit linear functions. In contrast, some data points of $F_{\rm cont}$ in the Wide field deviate from the $1\sigma$ error regions of the best-fit linear functions. Due to the dependence of $F_{\rm cont}$ on redshift, we make the best-fit $F_{\rm cont}$ linear function for each redshift bin in the Wide field, in which case the slope are fixed to that of the representative best-fit function. The $F_{\rm comp}$ and $F_{\rm cont}$ functions for the Deep fields are calculated from the depth-weighted average of the results for the UltraDeep and Wide fields.

\subsection{Estimate of Major Merger Fraction}\label{sec_merger_frac}

In this section, we describe how to estimate the major merger fraction. The major merger fraction is calculated as

\begin{equation}\label{eq_merger_frac}
  f_{\rm merger} = \frac{N_{\rm merger}}{N}
\end{equation}

\noindent where $N_{\rm merger}$ and $N$ are the numbers of major merger systems and dropout galaxies, respectively.\footnote{The definition corresponds to {\it the pair fraction} $f_{\rm pair}$ (e.g., \cite{2009ApJ...697.1369B}). } The $N_{\rm merger}$ value is obtained from $N_{\rm selected}$ which is the number of major merger systems selected in the selection criteria of $\mu\geq1/4$ and $d=r_{\rm min}-r_{\rm max}=3-6.5$ kpc (see Section \ref{sec_identify_mergers}). In the same manners as that of \citet{2013MNRAS.431.2661C}, we correct $N_{\rm selected}$ for the incompleteness and contamination in identifying galaxy mergers. Using $F_{\rm comp}$ and $F_{\rm cont}$ estimated in Section \ref{sec_comp_cont}, $N_{\rm merger}$ is derived as 

\begin{equation}\label{eq_correction}
  N_{\rm merger} = \frac{N_{\rm selected} - F_{\rm cont} N}{F_{\rm comp} - F_{\rm cont}}. 
\end{equation}

\noindent The major merger fractions derived in Equations (\ref{eq_merger_frac}) and (\ref{eq_correction}) are presented in Column (2) of Table \ref{tab_merger_frac_muv}. 

In addition to the incompleteness and contamination, we take into account the chance projection of foreground/background sources. Our method of identifying galaxy mergers relies on the projected separation with no redshift information of each galaxy component. Thus, foreground/background sources can be observed, by chance, near isolated galaxies. Following methods of e.g., \citet{2000MNRAS.311..565L}, \citet{2000ApJ...536..153P}, \citet{2009ApJ...697.1369B}, \citet{2009MNRAS.394L..51B}, and \citet{2012ApJ...747...34B}, we correct for the chance projection effect by calculating 

\begin{equation}\label{eq_merger_frac_proj}
  f_{\rm merger} = \frac{N_{\rm merger}-N_{\rm proj}}{N}
\end{equation}

\noindent where $N_{\rm proj}$ is the expected number of chance projection sources in a galaxy sample. The expected number of chance projection sources per galaxy is obtained by multiplying the area of the major merger search annulus $\pi(r_{\rm max}^2-r_{\rm min}^2)$ by the average surface number density of galaxies $S$ in the flux interval $0.25 f_1 \leq f < f_1$. To estimate $S$, we extract extended sources (i.e., galaxies) in the HSC-SSP UltraDeep COSMOS field by setting the {\tt (i, z, y)\_extendedness\_value} flag to $1$ \citep{2018PASJ...70S...8A}. 

Column (3) of Table \ref{tab_merger_frac_muv} summarizes the major merger fractions corrected for all of the incompleteness, the contamination, and the chance projection effect. Note that we have included their uncertainties related to $F_{\rm comp}$, $F_{\rm cont}$, and/or the Poisson error of $N_{\rm proj}$ to $f_{\rm merger}$ in Table \ref{tab_merger_frac_muv}. Even if we take into account the chance projection effect, the correction changes $f_{\rm merger}$ by only $\sim10-20$\%. This is because the small area of search annulus reduces the probability of the chance projection. 

In the next sections, we basically use $f_{\rm merger}$ corrected for all of the incompleteness, the contamination, and the chance projection effect (i.e., Column 3 of Table \ref{tab_merger_frac_muv}), otherwise specified. For the discussion about the source blending effect on galaxy UV LFs (Section \ref{sec_blend}), we adopt $f_{\rm merger}$ in Column (2) of Table \ref{tab_merger_frac_muv}. 

\begin{figure*}
 \begin{center}
  \includegraphics[width=130mm]{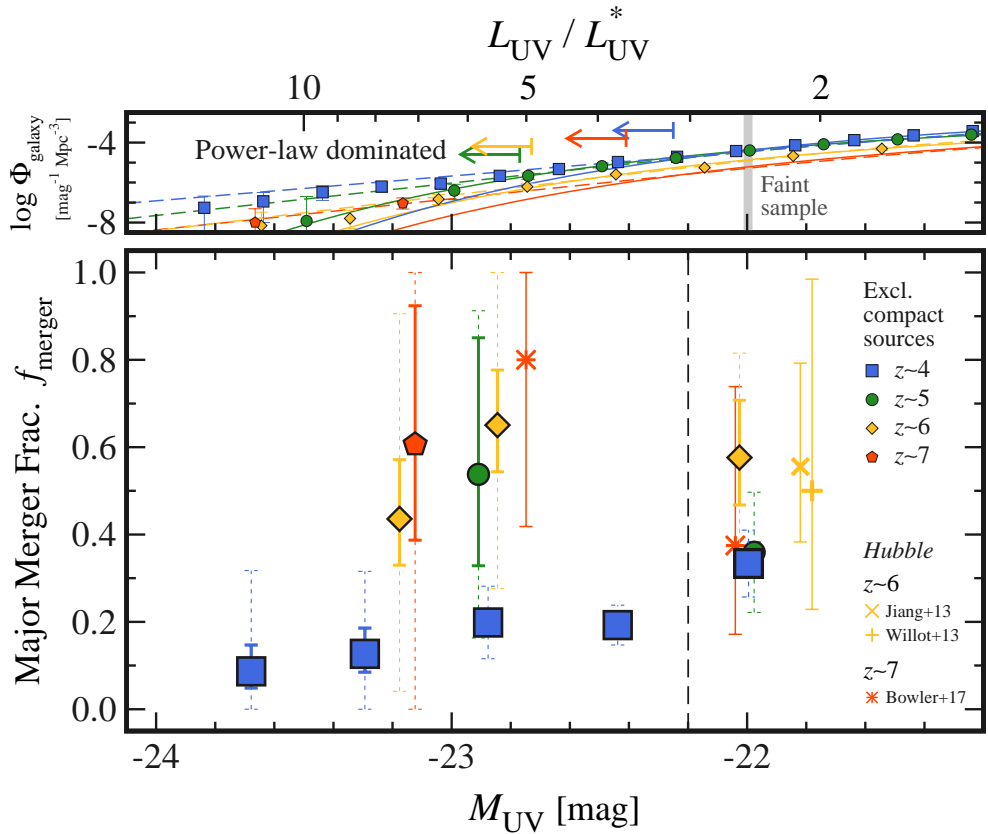}
  \end{center}
   \caption{Major merger fraction as a function of absolute UV magnitude. The filled symbols present dropout galaxies at $z\sim4$ (blue squares), $z\sim5$ (green circles), $z\sim6$ (yellow diamonds), and $z\sim7$ (red pentagons). In the $f_{\rm merger}$ estimates for these filled symbols, compact sources are excluded with a selection criterion of magnitude difference $m_{\rm PSF}-m_{\rm CModel}<0.15$ \citep{2019ApJ...883..183M}. Dropout galaxies in previous {\it Hubble} studies are plotted (yellow cross: $z\sim6$ in \cite{2013ApJ...773..153J}; yellow plus mark: $z\sim6$ in \cite{2013AJ....145....4W}; red asterisks: $z\sim7$ in \cite{2017MNRAS.466.3612B}). The $f_{\rm merger}$ values of \citet{2017MNRAS.466.3612B} are calculated for dropout galaxies in ranges of $M_{\rm UV, th}\lesssim-22.41$  and $M_{\rm UV}=-22.0\pm0.4$ by assuming that sources with multiple distinct components are major mergers. The absolute UV magnitude of the dropout galaxies in \citet{2013ApJ...773..153J} and \citet{2013AJ....145....4W} are assumed to be $M_{\rm UV}\sim-21.8$. The thick solid error bars of ours and the previous {\it Hubble} studies are calculated based on Poisson statistics from the galaxy number counts. In addition to the Poisson error, the thin dashed error bars include the uncertainties related to the incompleteness and contamination in the galaxy merger identification and the galaxy chance projection effect. The vertical dashed line is depicted for the division between the bright and faint galaxy samples. The data points are slightly offset along the $x$ axis for clarity. The top panel indicates the galaxy UV luminosity functions at $z\sim4-7$ from \citet{2018PASJ...70S..10O}. The solid and dashed curves correspond to the best-fit Schechter and DPL functions, respectively, with the color coding the same as for the symbols. The AGN contribution has been subtracted from these galaxy UV luminosity functions. The arrows indicate the UV luminosity regime where the number density of the DPL functions is $0.1$ dex higher than that of the Schechter functions. The vertical gray line denotes the luminosity of the faint galaxy sample, $L_{\rm UV}\sim2.5\, L_{\rm UV}^*$. The top $x$ axis represents the corresponding UV luminosity in units of $L_{\rm UV}^*$. }\label{fig_mag_merger_frac}
\end{figure*}

\section{Results}\label{sec_results}

\begin{table}
  \tbl{Major merger fractions as a function of $M_{\rm UV}$\footnotemark[$*$] }{%
  \begin{tabular}{ccc}
 \hline
 $M_{\rm UV}$ & $f_{\rm merger}$ & $f_{\rm merger}$ \\
 & ($F_{\rm comp}, F_{\rm cont}$) & ($F_{\rm comp}, F_{\rm cont}, N_{\rm proj}$) \\
 & -corrected & -corrected \\
 (mag) &  & \\
 (1) & (2) & (3) \\
  \hline
 \multicolumn{3}{c}{$z\sim4$} \\
$-23.68$ & $0.10_{-0.04}^{+0.06}$ ($_{-0.10}^{+0.23}$) & $0.09_{-0.04}^{+0.06}$ ($_{-0.09}^{+0.23}$) \\ 
$-23.29$ & $0.14_{-0.04}^{+0.06}$ ($_{-0.14}^{+0.19}$) & $0.13_{-0.04}^{+0.06}$ ($_{-0.13}^{+0.19}$) \\ 
$-22.88$ & $0.22_{-0.02}^{+0.02}$ ($_{-0.08}^{+0.08}$) & $0.20_{-0.01}^{+0.01}$ ($_{-0.08}^{+0.08}$) \\ 
$-22.44$ & $0.22_{-0.01}^{+0.01}$ ($_{-0.05}^{+0.05}$) & $0.19_{-0.01}^{+0.01}$ ($_{-0.05}^{+0.05}$) \\ 
\hline
 \multicolumn{3}{c}{$z\sim5$} \\
$-22.87$ & $0.57_{-0.21}^{+0.32}$ ($_{-0.36}^{+0.36}$) & $0.54_{-0.21}^{+0.31}$ ($_{-0.37}^{+0.37}$) \\ 
\hline
 \multicolumn{3}{c}{$z\sim6$} \\
$-23.18$ & $0.49_{-0.11}^{+0.14}$ ($_{-0.39}^{+0.46}$) & $0.44_{-0.11}^{+0.14}$ ($_{-0.39}^{+0.47}$) \\ 
$-22.84$ & $0.70_{-0.11}^{+0.13}$ ($_{-0.37}^{+0.30}$) & $0.65_{-0.11}^{+0.13}$ ($_{-0.37}^{+0.35}$) \\ 
\hline
 \multicolumn{3}{c}{$z\sim7$} \\
$-23.12$ & $0.68_{-0.23}^{+0.32}$ ($_{-0.68}^{+0.32}$) & $0.61_{-0.22}^{+0.32}$ ($_{-0.61}^{+0.39}$) \\ 
 \hline
\end{tabular}}\label{tab_merger_frac_muv}
  \begin{tabnote}
 \footnotemark[$*$](1) Absolute UV magnitude. (2) Major merger fraction corrected for the incompleteness and contamination in the major merger identification. (3) Major merger fraction corrected for the incompleteness and contamination in the major merger identification and the chance projection effect. In the columns (2) and (3), the errors are based on Poisson statistics. In addition to the Poisson error, the values in parenthesis include the uncertainties related to $F_{\rm comp}$, $F_{\rm cont}$, and $N_{\rm proj}$. 
  \end{tabnote}
\end{table}

\begin{table}
  \tbl{Major merger fractions as a function of redshift\footnotemark[$*$] }{%
  \begin{tabular}{cc}
 \hline
 Redshift & $f_{\rm merger}$ \\
 (1) & (2) \\
  \hline
$4$ & $0.20_{-0.02}^{+0.02}$ ($_{-0.09}^{+0.09}$) \\ 
$5$ & $0.54_{-0.23}^{+0.35}$ ($_{-0.41}^{+0.41}$) \\
$6$ & $0.61_{-0.11}^{+0.13}$ ($_{-0.34}^{+0.34}$) \\ 
$7$ & $0.78_{-0.27}^{+0.22}$ ($_{-0.78}^{+0.22}$) \\ 
 \hline
\end{tabular}}\label{tab_merger_frac_z}
  \begin{tabnote}
 \footnotemark[$*$](1) Redshift. (2) Major merger fraction in a narrow UV magnitude range $-23.4<M_{\rm UV}<-22.8$. The $f_{\rm merger}$ values are corrected for the incompleteness and contamination in the major merger identification and the chance projection effect. The errors are based on Poisson statistics. In addition to the Poisson error, the values in parenthesis include the uncertainties related to $F_{\rm comp}$, $F_{\rm cont}$, and $N_{\rm proj}$. 
  \end{tabnote}
\end{table}

\subsection{Major Merger Fraction as a Function of $M_{\rm UV}$}\label{sec_muv_merger_frac}

Figure \ref{fig_mag_merger_frac} presents the major merger fractions as a function of $M_{\rm UV}$ for dropout galaxies at $z\sim4-7$. The combination of the wide-area HSC-SSP data with our super-resolution technique reveals $f_{\rm merger}$ in the number density excess regime of $M_{\rm UV}\lesssim-22.2$ ($L_{\rm UV}\gtrsim3\, L_{\rm UV}^*$) of galaxy UV LFs. Among the redshifts of $z\sim4-7$, the $z\sim4$ galaxy sample covers the widest absolute UV magnitude range of $-24\lesssim M_{\rm UV}\lesssim-22.2$ corresponding to $L_{\rm UV}\sim3-15\, L_{\rm UV}^*$. We find that $f_{\rm merger}$ at $z\sim4$ is almost constant at $\sim10-20$\% in the $M_{\rm UV}$ range, but shows a mild decreasing trend toward bright $M_{\rm UV}$. Albeit with relatively large error bars at higher redshifts due to the small sample size, $f_{\rm merger}$ seem to increase from $z\sim4$ to $z\sim5-7$. The redshift evolution of $f_{\rm merger}$ is examined in details in Section \ref{sec_evolution}. At each redshift, there is no enhancement in $f_{\rm merger}$ for the bright dropout galaxy sample compared to those for the corresponding control faint dropout galaxy sample, by considering the uncertainties. The no enhancement in $f_{\rm merger}$ would imply the major merger is not a dominant source for the bright-end DPL shape of galaxy UV LFs (Section \ref{sec_discuss}). A similar result on $f_{\rm merger}$ is obtained in our machine learning-based study, which will be presented in a companion paper.

We compare $f_{\rm merger}$ estimated for our 6535 bright dropout galaxies with those of previous {\it Hubble} studies. As described below, our $f_{\rm merger}$ estimates are consistent with those of the previous {\it Hubble} studies within the $1\sigma$ uncertainty. Due to the small field of view (FoV) of {\it Hubble}, there have been few statistical studies on the morphology of high-$z$ galaxies with $M_{\rm UV}\lesssim-22$ at sub-structure levels of merger systems (see, e.g., \cite{2006ApJ...652..963R, 2006ApJ...636..592L, 2009MNRAS.397..208C, 2010ApJ...709L..21O, 2015ApJ...804..103K, 2016MNRAS.457..440C} for faint galaxies at $z\gtrsim4$; see also ALMA studies on $f_{\rm merger}$ with [C {\sc ii}]$158\mu m$ line maps, e.g., \cite{2020A&A...643A...1L}). Despite of the small FoV, some studies have examined the morphology of bright dropout galaxies by using the existing {\it Hubble} data of, e.g., CANDELS \citep{2011ApJS..197...35G, 2011ApJS..197...36K} or by conducting multiple {\it Hubble} follow-up observations. \citet{2013ApJ...773..153J} have carried out morphological analyses for $z\sim6$ galaxies with $-22\lesssim M_{\rm UV}\lesssim-19.5$ based on the visual inspection, reporting that 10 out of 18 bright sources with $M_{\rm UV}\leq-21$ are mergers. \citet{2013AJ....145....4W} have checked six bright galaxies at $z\sim6$ detected in the COSMOS {\it Hubble}/ACS data. Half of the six bright galaxies have multiple components which are likely to be galaxy mergers. \citet{2017MNRAS.466.3612B} have visually examined the morphology of 22 $z\sim7$ bright galaxies with $-23.2\lesssim M_{\rm UV}\lesssim-21.2$, nine of which are irregular and/or multiple-component systems. We plot the three previous {\it Hubble} studies on Figure \ref{fig_mag_merger_frac}. Here, the galaxies of \citet{2013ApJ...773..153J} and \citet{2013AJ....145....4W} are assumed to typically have $M_{\rm UV}\sim-21.8$. We divide the galaxies of \citet{2017MNRAS.466.3612B} into the $M_{\rm UV}<-22.41$ bright and $M_{\rm UV}=-22\pm0.4$ faint samples to match the $M_{\rm UV}$ ranges. As shown in Figure \ref{fig_mag_merger_frac}, our $f_{\rm merger}$ estimates at $z\sim6-7$ are consistent with those of these previous {\it Hubble} studies within the uncertainties, ensuring that our super-resolution technique works well in identifying galaxy mergers at high redshifts.

\begin{figure*}
 \begin{center}
  \includegraphics[width=110mm]{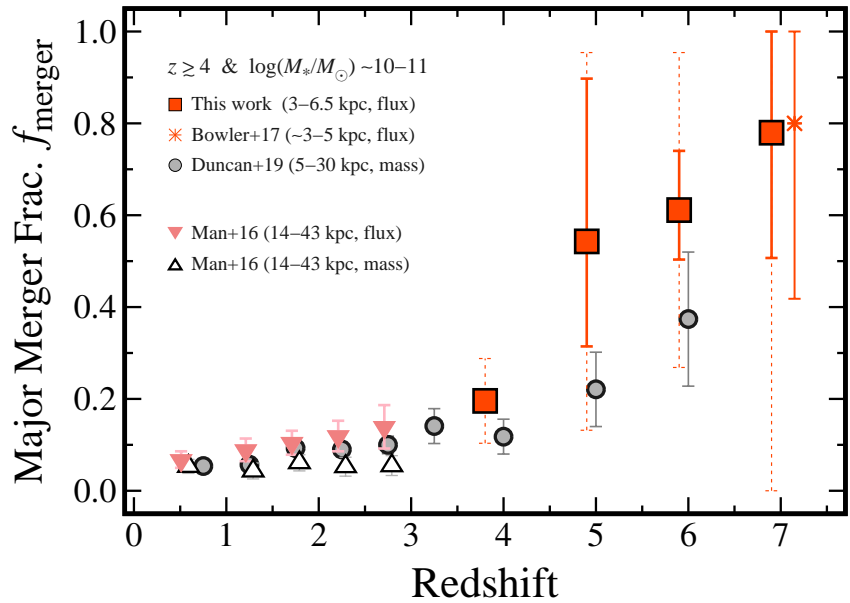}
  \end{center}
   \caption{Redshift evolution of the major merger fraction for massive galaxies. The red squares indicate the bright dropout galaxies with $-23.4 < M_{\rm UV} < -22.8$ in this study. The thick solid error bars are calculated based on Poisson statistics from the galaxy number counts. In addition to the Poisson error, the thin dashed error bars include the uncertainties related to the incompleteness and contamination rate in the galaxy merger identification and the galaxy chance projection effect. The red asterisk represents bright dropout galaxies with $M_{\rm UV}<-22.41$ in \citet{2017MNRAS.466.3612B}. According to the $M_{\rm UV}-M_*$ relation, the stellar mass of these bright dropout galaxies would be typically $\log{(M_*/M_\odot)}\sim10-11$. The gray filled  circles denote galaxies with $\log{(M_*/M_\odot)}>10.3$ in \citet{2019ApJ...876..110D}. The magenta filled inverse triangles and open triangles present $f_{\rm merger}$ for $\log{(M_*/M_\odot)}>10.8$ galaxies in \citet{2016ApJ...830...89M} based on flux and $M_*$ ratio selections, respectively. Other selection criteria (i.e., the source separation and ``flux" or ``mass" ratios) are shown in the legends. See  Section \ref{sec_evolution} for more details. The data points are slightly offset along the $x$ axis for clarity. }\label{fig_z_merger_frac}
\end{figure*}

\begin{figure*}
 \begin{center}
  \includegraphics[width=80mm]{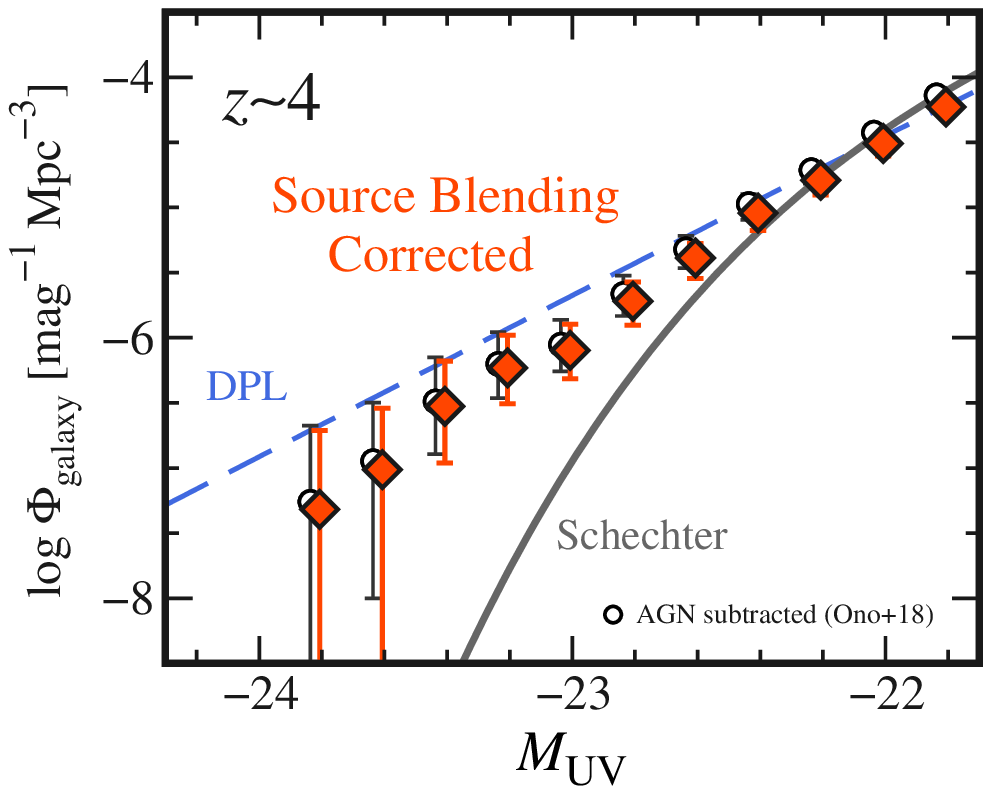}
  \mbox{}
  \includegraphics[width=80mm]{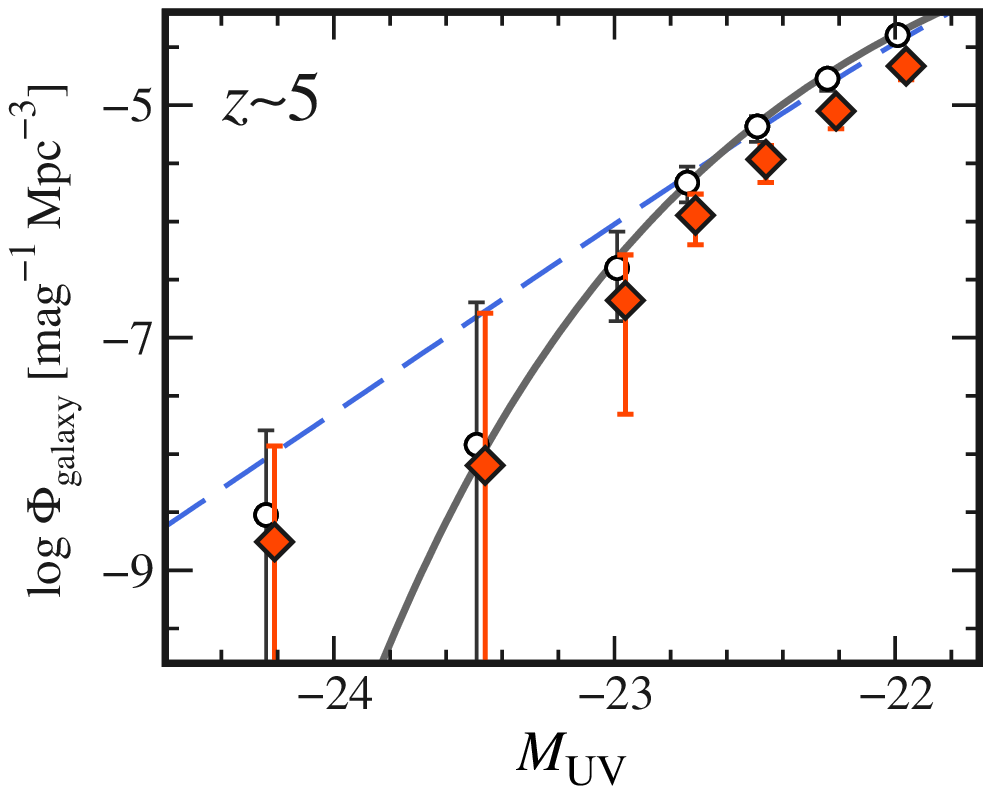}\\ 
  \mbox{} \\
  \includegraphics[width=80mm]{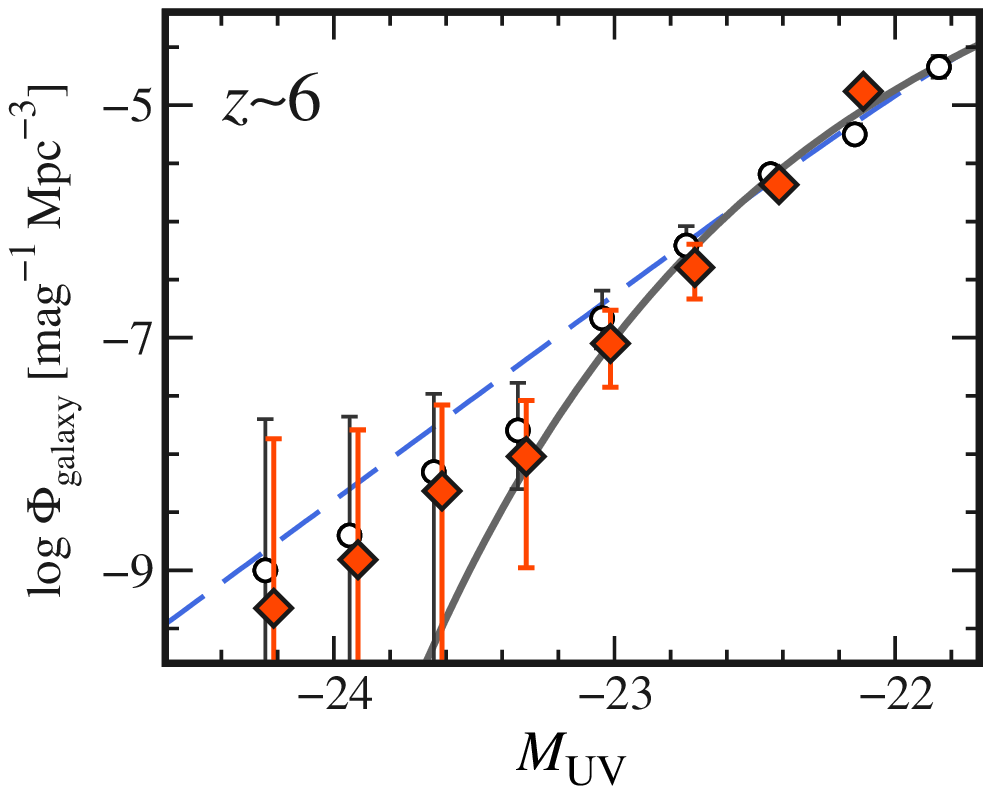}
  \mbox{}
  \includegraphics[width=80mm]{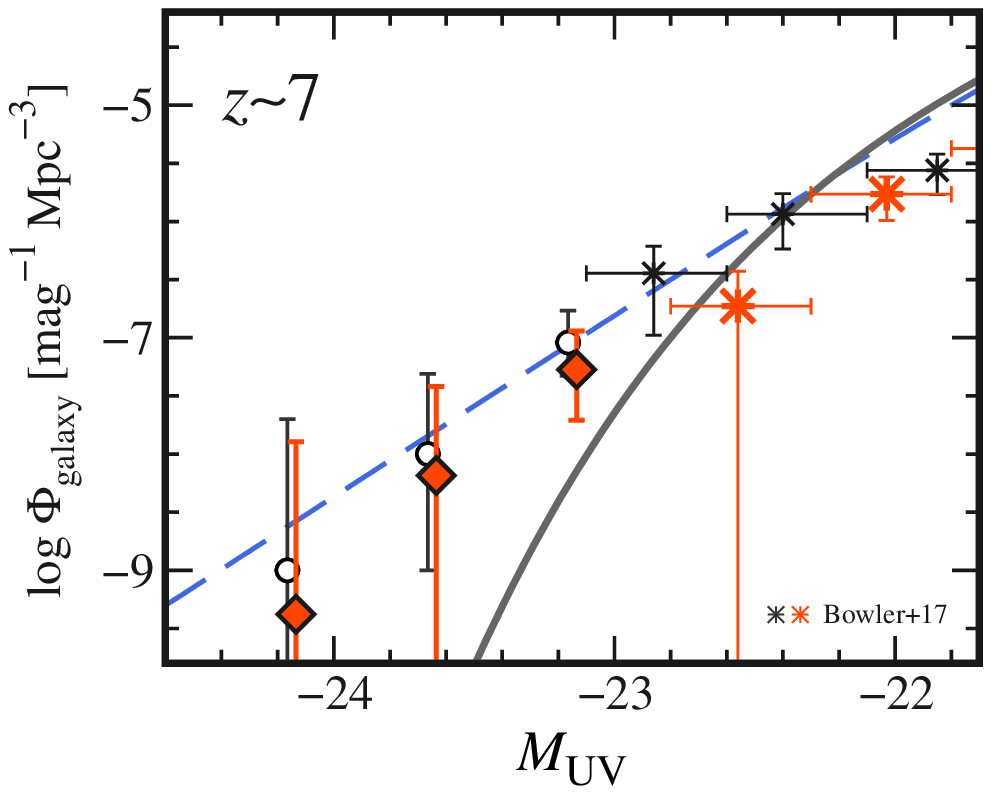}
  \end{center}
   \caption{Galaxy UV LFs at $z\sim4$ (top left), $z\sim5$ (top right), $z\sim6$ (bottom left), and $z\sim7$ (bottom right). The open circles with error bars indicate galaxy UV LFs whose AGN contribution is subtracted \citep{2018PASJ...70S..10O}. The gray solid and blue dashed curves correspond to the best-fit Schechter and DPL functions, respectively \citep{2018PASJ...70S..10O}. The red filled diamonds are galaxy UV LFs that are corrected for the source blending effect. For clarity, the red filled diamonds are slightly offset along the $x$ axis. The error bars of the two brightest data points in all the panels reach the Schechter functions because these lower limits are $\Phi_{\rm galaxy} = 0\,\, {\rm mag}^{-1} {\rm Mpc}^{-3}$. The red large (black small) asterisks at $z\sim7$ represent the galaxy UV LF in \citet{2017MNRAS.466.3612B} that are corrected (not corrected) for the source blending effect. }\label{fig_uvlf_blend}
\end{figure*}

\begin{figure*}
 \begin{center}
  \includegraphics[width=80mm]{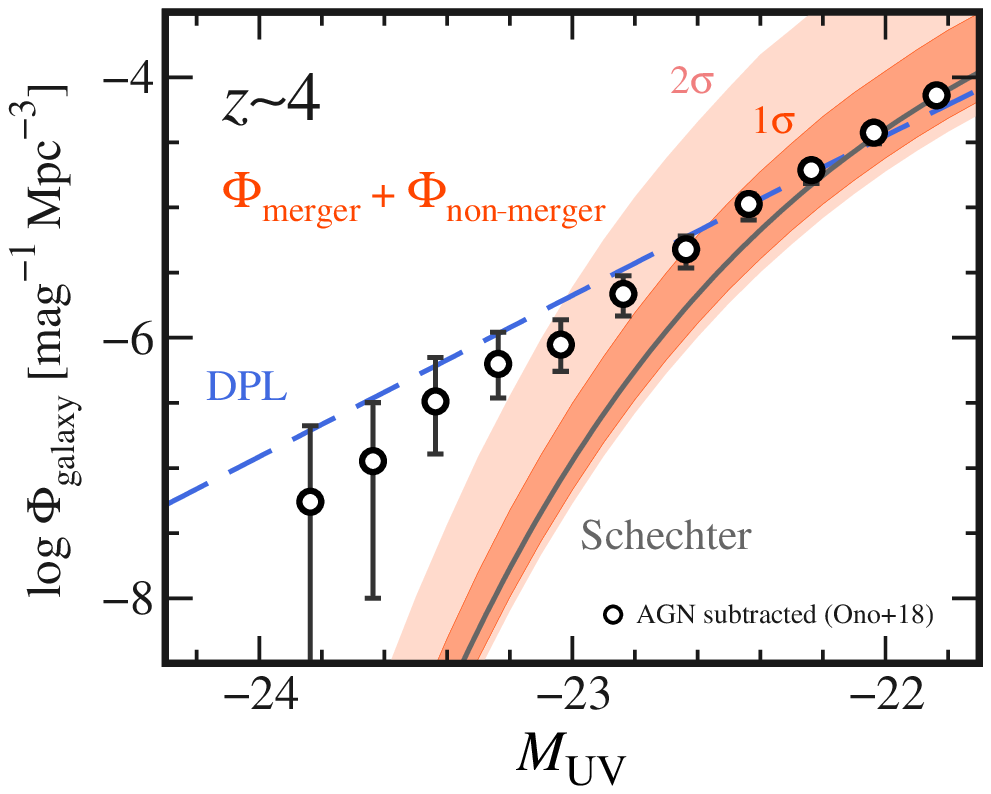}
  \mbox{}
  \includegraphics[width=80mm]{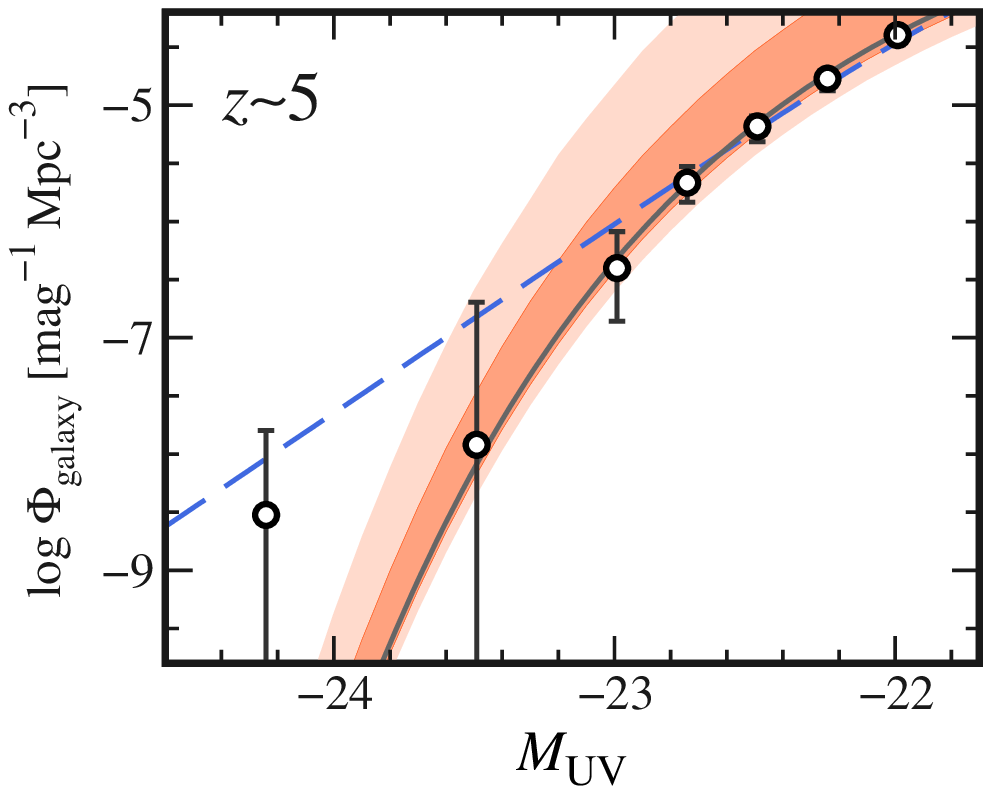}\\ 
  \mbox{} \\
  \includegraphics[width=80mm]{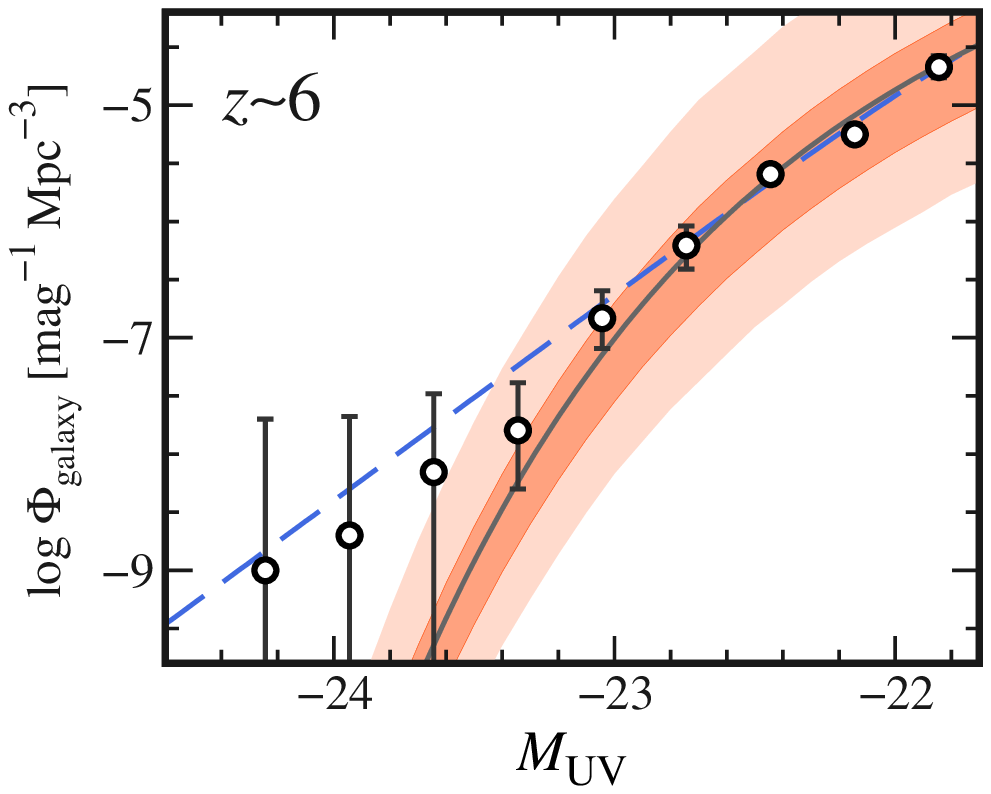}
  \mbox{}
  \includegraphics[width=80mm]{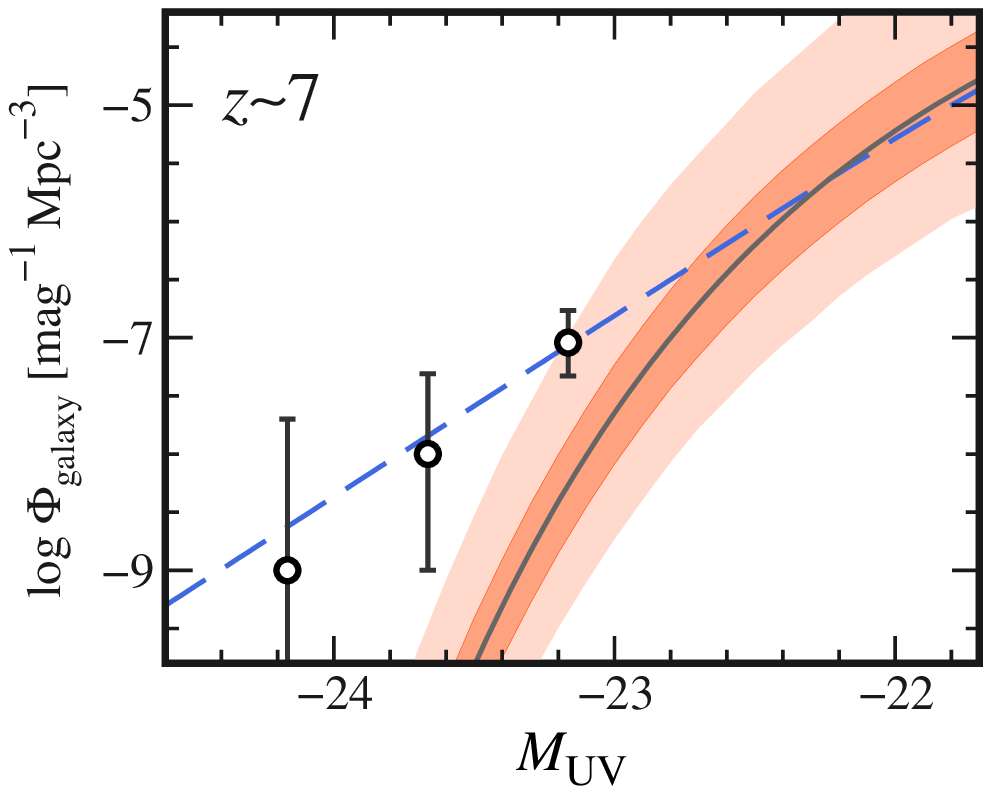}
  \end{center}
   \caption{Same as Figure \ref{fig_uvlf_blend}, but for galaxy UV LFs with the number density error regions allowed by the $f_{\rm merger}$ estimates. The dark-red and light-red areas represent the $1\sigma$ and $2\sigma$ error regions, respectively, of total UV LFs $\Phi_{\rm tot}$ that combine two UV LFs of major mergers $\Phi_{\rm m}$ and non-mergers $\Phi_{\rm nm}$ (see Section \ref{sec_merger_lf} for details). }\label{fig_uvlf}
\end{figure*}

\subsection{Redshift Evolution of Major Merger Fraction}\label{sec_evolution}

We investigate the redshift evolution of the major merger fractions. To reduce the effect that $f_{\rm merger}$ depends on $M_{\rm UV}$, we re-estimate $f_{\rm merger}$ in a relatively narrow $M_{\rm UV}$ range of $-23.4 < M_{\rm UV} < -22.8$. Figure \ref{fig_z_merger_frac} and Table \ref{tab_merger_frac_z} present $f_{\rm merger}$ as a function of redshift. Albeit with huge error bars at $z\gtrsim5$, the major merger fractions tend to increase with increasing redshift, from $f_{\rm merger}\sim20$\% at $z\sim4$ to $\sim50$\% at $z\sim5$, $\sim60$\% at $z\sim6$ and $\sim80$\% at $z\sim7$. 

We compare the redshift evolution of $f_{\rm merger}$ estimated in this work and previous studies in Figure \ref{fig_z_merger_frac}. Although methods to identify major mergers are inhomogeneous between this work and previous studies, the comparison enables us to grasp the rough evolutionary trend. According to the relation between $M_{\rm UV}$ and stellar mass $M_*$ (e.g., \cite{2015ApJS..219...15S, 2016ApJ...821..123H, 2016ApJ...825....5S, 2021arXiv210316571S}), our bright dropout galaxies at $z\sim4-7$ would typically have $\log{(M_*/M_\odot)}\sim10-11$. In this study, we compare our $f_{\rm merger}$ estimates with a $z\sim0-6$ result of \citet{2019ApJ...876..110D} who have investigated comparably massive galaxies with $\log{(M_*/M_\odot)}>10.3$ (see, e.g., \cite{2014A&A...565A..10T, 2017MNRAS.470.3507M, 2020ApJ...895..115F} for studies on less massive galaxies with $\log{M_*/M_\odot}<10$). As in \citet{2019ApJ...876..110D}, our major merger fractions similarly increase with increasing redshift. However, our $f_{\rm merger}$ values appear to be systematically higher than those of \citet{2019ApJ...876..110D}. This systematic offset might be originated from the difference of major merger selection methods. \citet{2016ApJ...830...89M} have estimated $f_{\rm merger}$ based on the flux and $M_*$ ratios for $\log{(M_*/M_\odot)}>10.8$ massive galaxies at $z\sim0-3$. While the $M_*$-ratio based $f_{\rm merger}$ is almost constant at $f_{\rm merger}\sim10$\% in the redshift range of $z\sim0-3$, the flux-ratio based $f_{\rm merger}$ is systematically higher than the $M_*$-ratio based values by a factor of $\sim1.5-2.5$, and steadily increases with increasing redshift. The difference between the flux- and $M_*$-ratio based $f_{\rm merger}$ could be linked to, e.g., mass-to-light ratio and gas fractions of faint galaxy components (e.g., \cite{2009ApJ...697.1369B, 2010MNRAS.402.1693H, 2011ApJ...742..103L, 2018MNRAS.475.1549M}). If the difference of the selection similarly affects $f_{\rm merger}$ at $z\gtrsim4$, our flux-ratio based major merger fractions might be comparable to the $M_*$-ratio based $f_{\rm merger}$ values of \citet{2019ApJ...876..110D}. These results for massive and bright galaxies indicate that the evolutionary trend of $f_{\rm merger}$ at $z\sim0-6$ continues up to a very high redshift of $z\sim7$, providing insights into physical properties of high-$z$ galaxies, e.g., the high gas fraction and active star-formation at high redshifts (e.g., \cite{2018ApJ...853..179T, 2019ApJ...887..235L}). 

Some of recent observational and theoretical studies have reported that $f_{\rm merger}$ increases from $z\sim0$ up to $z\sim2-3$, and subsequently decreases toward high $z$ (e.g., \cite{2017A&A...608A...9V, 2019A&A...631A..87V, 2017MNRAS.464.1659Q, 2020MNRAS.493.1178E, 2021MNRAS.501.3215O}), possibly due to, e.g., the evolving merger time scale $T_{\rm merger}$ (e.g., \cite{2017MNRAS.468..207S}). Galaxies of these previous studies have lower stellar masses of $\log{(M_*/M_\odot)}\sim9-10$ than ours, i.e., $\log{(M_*/M_\odot)}\sim10-11$. Due to the different stellar mass range, these previous studies on low-mass galaxies are not plotted in Figure \ref{fig_z_merger_frac}. If the merger fraction does indeed decline with increasing redshift at $z\gtrsim3$, although this is not indicated by our results, there would be even less indication that it could contribute significantly to shaping the bright-end of the UV luminosity function (see Section \ref{sec_discuss} for more detailed discussion).

\section{Discussion}\label{sec_discuss}

We discuss the implications for the shape of galaxy UV LFs with $f_{\rm merger}$ estimated in Section \ref{sec_muv_merger_frac}. In the following subsections, we check whether the number density excess of galaxy UV LFs can be explained by major mergers in two ways: 1) starting from the DPL function (Section \ref{sec_blend}), and 2) starting from the Schechter function (Section \ref{sec_merger_lf}). For the former and the latter, we consider the source blending effect and the number density enhancement resulting from major mergers, respectively. In this study, we make use of $z\sim4-7$ galaxy UV LFs whose AGN contribution has been subtracted \citep{2018PASJ...70S..10O} for the comparison. In the redshift range of $z\sim4-7$, we mainly discuss the galaxy UV LFs at $z\sim4$ and $z\sim7$ that have shown the significant number density excess.

\subsection{From DPL Function}\label{sec_blend}

First, we consider the source blending effect to check whether the number density excess of galaxy UV LFs can be explained by major mergers, starting from the DPL function. Before the discussion, we obtain functional forms of $f_{\rm merger}$ by fitting a linear function of $f_{\rm merger}(M_{\rm UV}) = a M_{\rm UV} + b$ to the $f_{\rm merger}$ values at $z\sim4$. Here, $f_{\rm merger}$ in Column (2) of Table \ref{tab_merger_frac_muv} is used because it is unnecessary to correct for $N_{\rm proj}$ for the source blending effect. The best-fit linear function at $z\sim4$ is $f_{\rm merger}(M_{\rm UV}) = (0.071\pm0.034) M_{\rm UV} + (1.81\pm0.76)$. For the higher redshifts of $z\gtrsim5$, we fit or scale $f_{\rm merger}(M_{\rm UV}) = 0.071 M_{\rm UV} + b$ to the $f_{\rm merger}$ estimates by fixing the parameter of slope $a=0.071$, because there are few $f_{\rm merger}$ estimates at $z\sim5-7$. The intercepts $b$ at $z\sim(5, 6, 7)$ are $b=(2.19, 2.24, 2.32)$ whose $1\sigma$ errors are fixed to that of $z\sim4$. 

Using $f_{\rm merger}(M_{\rm UV})$, we correct the observed number density for the source blending effect. In a super-resolution image, a major merger with  $L_{\rm UV}$ is separated into a brighter galaxy component with $R L_{\rm UV}$ and a fainter one with $(1-R) L_{\rm UV}$. Here, $R$ is the luminosity ratio of the brighter galaxy component to the blended single source. According to our super-resolution analysis, we find that the average luminosity ratio is $R=0.7$ which does not significantly depend on magnitude and redshift in the entire $M_{\rm UV}$ and $z$ ranges of $-24\lesssim M_{\rm UV}\lesssim-22$ and $z\sim4-7$. By combining $f_{\rm merger}(M_{\rm UV})$ and $R$, we count individual de-blended sources, and reconstruct galaxy UV LFs. To scale the original number density and its errors, we use the number of dropout galaxies in the HSC-SSP Wide field \citep{2018PASJ...70S..10O}. We simply scale the error bars of the galaxy number density based on the change in galaxy number counts before/after the correction for the source blending effect. 

Figure \ref{fig_uvlf_blend} presents the galaxy UV LFs that are corrected for the source blending effect. The number density correction ranges from $\lesssim0.1$ dex at $z\sim4$ to $\sim0.5$ dex at $z\sim7$, which is roughly comparable to those of a {\it Hubble} study at $z\sim7$ \citep{2017MNRAS.466.3612B}. At $z\sim4$, the correction slightly reduces the number density of the galaxy UV LF. However, the error bars of number density do not touch the Schechter function at $-23.4\lesssim M_{\rm UV}\lesssim-22.6$. We find that the correction factor tends to be larger at higher $z$. This trend is originated from high $f_{\rm merger}$ values at high $z$. In a higher $f_{\rm merger}$, a larger fraction of galaxies are de-blended into faint components, which decreases the number density more largely compared to lower redshifts. Even at the highest redshift of $z\sim7$, these correction factors are too small to reach the Schechter functions. In addition, we find some data points of number density ``{\it increase}" from the original values after the correction at a faint magnitude (e.g., at $M_{\rm UV}\sim-22$ at $z\sim6$). A similar trend is found in \citet{2017MNRAS.466.3612B}. This would be because $f_{\rm merger}$ increases for faint galaxies (see the $f_{\rm merger}$ dependence on magnitude in Figure \ref{fig_mag_merger_frac}), and many faint de-blended components in a bright $M_{\rm UV}$ bin move into a faint $M_{\rm UV}$ bin. Thus, we conclude that the source blending effect cannot consistently explain the shape of galaxy UV LFs in a wide magnitude range of $-24\lesssim M_{\rm UV} \lesssim -22$.

\subsection{From Schechter Function}\label{sec_merger_lf}

Next, we consider the number density enhancement caused by major mergers to check whether the number density excess of galaxy UV LFs can be explained by major mergers, starting from the Schechter function. As in Section \ref{sec_blend}, we first obtain functional forms of $f_{\rm merger}(M_{\rm UV}) = a M_{\rm UV} + b$. In contrast to Section \ref{sec_blend}, $f_{\rm merger}$ in Column (3) of Table \ref{tab_merger_frac_muv} is used to estimate the intrinsic number density of major mergers. The best-fit linear function at $z\sim4$ is $f_{\rm merger}(M_{\rm UV}) = (0.055\pm0.036) M_{\rm UV} + (1.43\pm0.81)$. The intercepts $b$ at $z\sim(5, 6, 7)$ are $b=(1.80, 1.83, 1.88)$. 

Using $f_{\rm merger}(M_{\rm UV})$, we construct total UV LFs that combines the contributions from major mergers and non-mergers. For this purpose, we incorporate $f_{\rm merger}(M_{\rm UV})$ into the Schechter function. Here, we use the Schechter function that is described as a function of $M_{\rm UV}$,

\begin{eqnarray}\label{eq_Schechter_MUV}
\Phi_{\rm Sch}(M_{\rm UV})
	&=& \frac{\ln 10}{2.5} \phi^\ast 10^{-0.4 (M_{\rm UV} - M_{\rm UV}^\ast) (\alpha +1)} \nonumber \\
	&& \times \exp \left( - 10^{-0.4 (M_{\rm UV} - M_{\rm UV}^\ast)} \right), 
\end{eqnarray}

\noindent where $\phi^*$ is the normalization factor in the number density, $M_{\rm UV}^\ast$ is the characteristic UV magnitude, and $\alpha$ is the faint-end slope. We apply the best-fit Schechter parameters ($\phi^*, M_{\rm UV}^\ast, \alpha$) derived in \citet{2018PASJ...70S..10O}. From the major merger fraction defined as, 

\begin{equation}\label{eq_fmerger_density}
f_{\rm merger} \equiv \frac{\Phi_{\rm m}}{\Phi_{\rm tot}} \equiv \frac{\Phi_{\rm m}}{\Phi_{\rm m}+\Phi_{\rm nm}}, 
\end{equation} 

\noindent the total UV LF $\Phi_{\rm tot}$ is derived as, 

\begin{equation}\label{eq_mod_Schechter}
\Phi_{\rm tot} = \Phi_{\rm m} + \Phi_{\rm nm} = \frac{1}{1-f_{\rm merger}} \Phi_{\rm nm}, 
\end{equation} 

\noindent where $\Phi_{\rm m}$ and $\Phi_{\rm nm}$ are the UV LFs of major mergers and non-mergers, respectively. All the variables of $\Phi_{\rm tot}, \Phi_{\rm m}, \Phi_{\rm nm}$, and $f_{\rm merger}$ are as a function of $M_{\rm UV}$, which is omitted to avoid the complication of these equations. Because non-merger UV LFs have not been well constrained at high redshifts in previous studies, $\Phi_{\rm nm}$ is assumed to be $\Phi_{\rm nm} = \{ 1-f_{\rm merger}(M_{\rm UV}=-22) \} \Phi_{\rm Sch}$ where $f_{\rm merger}(M_{\rm UV}=-22)$ is the major merger fractions estimated for the faint galaxy sample at $M_{\rm UV}=-22$. The assumption of $\Phi_{\rm nm}$ would be too simplistic. In this assumption, it is difficult to reproduce the DPL shape with the $\Phi_{\rm tot}$ function unless the $f_{\rm merger}$ errors are included. The objective of this analysis is to investigate whether the observed number density of galaxy UV LFs can be reproduced by varying $\Phi_{\rm tot}$ within the $f_{\rm merger}$ uncertainty.

Figure \ref{fig_uvlf} shows the total UV LFs, $\Phi_{\rm tot}$. The $1\sigma$ and $2\sigma$ error regions of $\Phi_{\rm tot}$ are depicted based on Monte-Carlo simulations in which $\Phi_{\rm tot}$ is varied within the $f_{\rm merger}(M_{\rm UV})$ uncertainty. The number density of $\Phi_{\rm tot}$ is matched to that of the Schechter functions at $M_{\rm UV}=-22$ for the comparison of the bright-end shapes of UV LFs. We find that $\Phi_{\rm tot}$ is consistent with the observed number density in the galaxy UV LFs within the $2\sigma$ uncertainty at a faint UV magnitude of $M_{\rm UV}\gtrsim-23.2$ ($L_{\rm UV}\lesssim8\, L_{\rm UV}^*$) at $z\sim4$ and $M_{\rm UV}\gtrsim-23.6$ ($L_{\rm UV}\lesssim10\, L_{\rm UV}^*$) at $z\sim7$, indicating that the DPL shape is partly explained by major mergers. However, an offset in the number density from the total UV LFs still remains at the very bright end of $M_{\rm UV}\lesssim-23.6$ ($L_{\rm UV}\gtrsim10\, L_{\rm UV}^*$) at $\sim2\sigma$ significance. 

\begin{figure}
 \begin{center}
  \includegraphics[width=83mm]{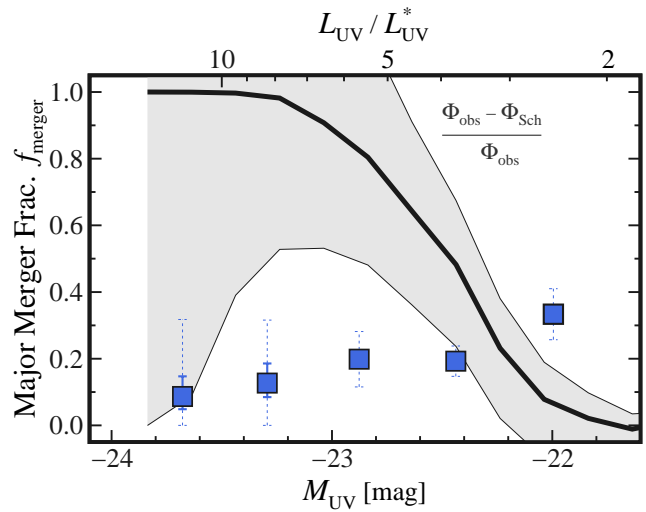}
  \end{center}
   \caption{Same as Figure \ref{fig_mag_merger_frac}, but for the comparison of the major merger fractions at $z\sim4$ estimated from our super-resolution techniques (blue filled squares) and the galaxy number densities (black solid line). The shaded region indicates the $1\sigma$ uncertainty of $f_{\rm merger}$ from the galaxy number densities, which is calculated with the errors in $\Phi_{\rm obs}$. See Section \ref{sec_merger_lf} for details.}\label{fig_mag_merger_frac_model}
\end{figure}

Because the assumption of $\Phi_{\rm nm}$ is over-simplified, we take another approach. We compare our $f_{\rm merger}$ estimates with $f_{\rm merger}$ calculated from the galaxy number densities. Instead of Equation (\ref{eq_fmerger_density}), we calculate the major merger fraction from the galaxy number densities as

\begin{equation}\label{eq_fmerger_density2}
f_{\rm merger} \equiv \frac{\Phi_{\rm m}}{\Phi_{\rm tot}} = \frac{\Phi_{\rm obs}-\Phi_{\rm Sch}}{\Phi_{\rm obs}}, 
\end{equation} 

\noindent where $\Phi_{\rm obs}$ is the observed galaxy UV LFs (i.e., the open circles in Figures \ref{fig_uvlf_blend} and \ref{fig_uvlf}). In Equation (\ref{eq_fmerger_density2}), we make an assumption that all of the number-density excess are originated from the galaxy major mergers, $\Phi_{\rm m} = \Phi_{\rm obs}-\Phi_{\rm Sch}$. Again, we apply the best-fit Schechter parameters of \citet{2018PASJ...70S..10O} to $\Phi_{\rm Sch}$. Figure \ref{fig_mag_merger_frac_model} presents $f_{\rm merger}$ estimated from our super-resolution analysis and from the galaxy number densities at $z\sim4$ where the dropout galaxy sample is sufficiently large for the comparison of the $f_{\rm merger}$ trend. The major merger fraction estimated with Equation (\ref{eq_fmerger_density2}) increases with decreasing $M_{\rm UV}$, although the uncertainty is large at $M_{\rm UV}<-23.5$. The increasing trend is in tension with $f_{\rm merger}$ estimated from our super-resolution analysis. This result would support the conclusion that the number-density excess of galaxy UV LFs cannot be explained only by major mergers.

\mbox{}\\ 

The discussion in Sections \ref{sec_merger_lf} and \ref{sec_blend} indicates that the two effects of the source blending effect and the number density enhancement resulting from major mergers can partly explain the DPL shape at $L_{\rm UV}\sim3-10\, L_{\rm UV}^*$. However, the DPL shape cannot be explained at the very bright end of $L_{\rm UV}\gtrsim10\, L_{\rm UV}^*$. These results would suggest that galaxy major mergers are not enough to reproduce the DPL shape of galaxy UV LFs. This implication has already been obtained from the result of the no $f_{\rm merger}$ enhancement at bright $M_{\rm UV}$ (Section \ref{sec_muv_merger_frac} and Figure \ref{fig_mag_merger_frac}). The discussion in this section supports more quantitatively the small impacts of galaxy major mergers on the bright-end DPL shape. In addition to galaxy major mergers, other physical mechanisms could contribute to the number density excess of galaxy UV LFs. Possible candidates for these physical mechanisms include 1) insignificant mass quenching and inefficient star-formation/AGN feedback effects in the early Universe (e.g., \cite{1977ApJ...215..483B, 2006MNRAS.365...11C, 2010ApJ...721..193P, 2019ApJ...878..114R}), 2) low dust obscuration at high-$z$ and the bright end \citep{2015MNRAS.452.1817B, 2020MNRAS.493.2059B, 2021arXiv210800830V}, 3) gravitational lensing magnification (e.g., \cite{2011Natur.469..181W, 2011ApJ...742...15T, 2015ApJ...805...79M, 2015MNRAS.450.1224B}), 4) hidden AGN activity missed by shallow spectroscopic follow-up observations (e.g., \cite{2021ApJ...910L..11K}). See also \citet{2021arXiv210801090H} for detailed discussion about these physical mechanisms. To reveal the physical mechanisms related to the number density excess of galaxy UV LFs, it would be important to investigate, e.g., kinematics of inter-stellar/circum-galactic media, dust abundance, and high ionization lines for bright galaxies, and to compare the contributions from galaxy major mergers and other physical processes.  

For future observational studies on galaxy major mergers at $z\gtrsim4$, analyses with deep IR imaging and spectroscopic data are essential to more accurately estimate $f_{\rm merger}$ at high redshifts. Our method to identify major mergers is not based on the stellar-mass ratio, but just the flux ratio. Although some of our main discussion need only the flux ratio-based $f_{\rm merger}$ (e.g., the source blending effect on galaxy UV LFs), the stellar mass should be measured. Deep IR imaging data of, e.g., the {\it James Webb Space Telescope} would be useful to reveal the intrinsic redshift evolution of major merger fractions. In addition, the spectroscopic identification of galaxy pairs is required for more precise $f_{\rm merger}$ estimates without the contamination of chance projection. 

\section{Summary and Conclusions}\label{sec_summary}

We estimate the major merger fractions, $f_{\rm merger}$, for the bright dropout galaxies at $z\sim4-7$ using the HSC-SSP data and the super-resolution technique. Our super-resolution technique improves the spatial resolution of the ground-based HSC images, from $\sim0.\!\!^{\prime\prime}6-1.\!\!^{\prime\prime}0$ to $\lesssim0.\!\!^{\prime\prime}1$, which is comparable to that of the {\it Hubble Space Telescope} (Figures \ref{fig_big_image} and \ref{fig_image_z47}). The combination of the survey capability of HSC and the resolving power allows us to investigate morphological properties of rare galaxy populations at high redshifts. The comparison between the {\it Hubble} and super-resolution HSC images suggests that we are able to identify $z\sim4-7$ bright major mergers at a high completeness value of $\gtrsim90$\% and a relatively low contamination rate of $\sim20$\% (Figure \ref{fig_completeness_contamination}). We apply the super-resolution technique to $6535$ very bright dropout galaxies in a high UV luminosity regime of $L_{\rm UV}\sim3-15\, L_{\rm UV}^*$ (corresponding to $-24\lesssim M_{\rm UV}\lesssim-22$) where the number density of galaxies and low-luminosity AGNs is comparable. Our major findings in this study are as follows. 

\begin{itemize}
  \item After the correction for the analysis biases (e.g., the incompleteness of major merger identification), the major merger fractions for the bright dropout galaxies are estimated to be $f_{\rm merger}\sim10-20$\% at $z\sim4$ and $\sim50-70$\% at $z\sim5-7$. We find no $f_{\rm merger}$ enhancement at $L_{\rm UV}\sim3-15\, L_{\rm UV}^*$ compared to those of the control sample of faint dropout galaxies with $L_{\rm UV}\sim2.5\, L_{\rm UV}^*$ (Figure \ref{fig_mag_merger_frac}), implying that the major merger is not a dominant source for the bright-end DPL shape of galaxy UV LFs. 
  
  \item In a relatively narrow $M_{\rm UV}$ range of $-23.4 < M_{\rm UV} < -22.8$, the major merger fractions show a gradual $f_{\rm merger}$ increase toward high redshift, from $f_{\rm merger}\sim20$\% at $z\sim4$ to $\sim50$\% at $z\sim5$, $\sim60$\% at $z\sim6$ and $\sim80$\% at $z\sim7$, which is similar to the redshift evolution in $f_{\rm merger}$ found in previous {\it Hubble} studies (Figure \ref{fig_z_merger_frac}). The slight difference in $f_{\rm merger}$ between this work and the previous studies might be interpreted by the difference of merger identification methods, i.e., based on the flux or stellar-mass ratios of galaxy components in merger systems. This result might indicate that the evolutionary trend of $f_{\rm merger}$ continues up to a very high redshift of $z\sim7$, providing insights into physical properties of high-$z$ galaxies.   
  
  \item Based on the $f_{\rm merger}$ estimates, we take into account two effects related to major mergers: 1) the source blending effects (Figure \ref{fig_uvlf_blend}), and 2) the number density enhancement resulting from major mergers (Figure \ref{fig_uvlf}) to check whether the bright-end DPL shape of galaxy UV LFs can be reproduced. While these two effects partly explain the DPL shape at $L_{\rm UV}\sim3-10\, L_{\rm UV}^*$, the DPL shape cannot be explained at the very bright end of $L_{\rm UV}\gtrsim10\, L_{\rm UV}^*$, even after the AGN contribution is subtracted. The results support scenarios in which other additional mechanisms such as insignificant mass quenching, low dust obscuration, and gravitational lensing magnification contribute to the DPL shape of galaxy UV LFs at high redshifts. 
  
\end{itemize} 

\begin{ack}

We thank Akio K. Inoue, Yuki Isobe, Ken Mawatari, Yoshiaki Ono, Marcin Sawicki, Aswin P. Vijayan,  Gordon Wetzstein, and the anonymous referee for valuable comments and suggestions that have improved our paper. This work was supported by JSPS KAKENHI Grant Number 20K14508. The Cosmic Dawn Center is funded by the Danish National Research Foundation under grant No. 140. S.F. acknowledges support from the European Research Council (ERC) Consolidator Grant funding scheme (project ConTExt, grant No. 648179). This project has received funding from the European Union's Horizon 2020 research and innovation program under the Marie Sk\l odowska-Curie grant agreement No. 847523 `INTERACTIONS'.

 The HSC collaboration includes the astronomical communities of Japan and Taiwan, and Princeton University. The HSC instrumentation and software were developed by the National Astronomical Observatory of Japan (NAOJ), the Kavli Institute for the Physics and Mathematics of the Universe (Kavli IPMU), the University of Tokyo, the High Energy Accelerator Research Organization (KEK), the Academia Sinica Institute for Astronomy and Astrophysics in Taiwan (ASIAA), and Princeton University. Funding was contributed by the FIRST program from Japanese Cabinet Office, the Ministry of Education, Culture, Sports, Science and Technology (MEXT), the Japan Society for the Promotion of Science (JSPS), Japan Science and Technology Agency (JST), the Toray Science Foundation, NAOJ, Kavli IPMU, KEK, ASIAA, and Princeton University. 

This paper makes use of software developed for the Large Synoptic Survey Telescope. We thank the LSST Project for making their code available as free software at  http://dm.lsst.org. 

This paper is based on data collected at the Subaru Telescope and retrieved from the HSC data archive system, which is operated by Subaru Telescope and Astronomy Data Center (ADC) at NAOJ. We are honored and grateful for the opportunity of observing the Universe from Maunakea, which has the cultural, historical and natural significance in Hawaii. Data analysis was in part carried out with the cooperation of Center for Computational Astrophysics (CfCA), (CfCA), NAOJ.

The Pan-STARRS1 Surveys (PS1) have been made possible through contributions of the Institute for Astronomy, the University of Hawaii, the Pan-STARRS Project Office, the Max-Planck Society and its participating institutes, the Max Planck Institute for Astronomy, Heidelberg and the Max Planck Institute for Extraterrestrial Physics, Garching, The Johns Hopkins University, Durham University, the University of Edinburgh, Queen's University Belfast, the Harvard-Smithsonian Center for Astrophysics, the Las Cumbres Observatory Global Telescope Network Incorporated, the National Central University of Taiwan, the Space Telescope Science Institute, the National Aeronautics and Space Administration under Grant No. NNX08AR22G issued through the Planetary Science Division of the NASA Science Mission Directorate, the National Science Foundation under Grant No. AST-1238877, the University of Maryland, and Eotvos Lorand University (ELTE) and the Los Alamos National Laboratory.  

\end{ack}

\bibliographystyle{aa}
\bibliography{bib_shibuya}

\begin{thebibliography}{111}
\expandafter\ifx\csname natexlab\endcsname\relax\def\natexlab#1{#1}\fi

\bibitem[{{Adams} {et~al.}(2020){Adams}, {Bowler}, {Jarvis}, {H{\"a}u{\ss}ler},
  {McLure}, {Bunker}, {Dunlop}, \& {Verma}}]{2020MNRAS.494.1771A}
{Adams}, N.~J., {Bowler}, R.~A.~A., {Jarvis}, M.~J., {et~al.} 2020, \mnras,
  494, 1771

\bibitem[{{Aihara} {et~al.}(2019){Aihara}, {AlSayyad}, {Ando}, {Armstrong},
  {Bosch}, {Egami}, {Furusawa}, {Furusawa}, {Goulding}, {Harikane}, {Hikage},
  {Ho}, {Hsieh}, {Huang}, {Ikeda}, {Imanishi}, {Ito}, {Iwata}, {Jaelani},
  {Kakuma}, {Kawana}, {Kikuta}, {Kobayashi}, {Koike}, {Komiyama}, {Li},
  {Liang}, {Lin}, {Luo}, {Lupton}, {Lust}, {MacArthur}, {Matsuoka}, {Mineo},
  {Miyatake}, {Miyazaki}, {More}, {Murata}, {Namiki}, {Nishizawa}, {Oguri},
  {Okabe}, {Okamoto}, {Okura}, {Ono}, {Onodera}, {Onoue}, {Osato}, {Ouchi},
  {Shibuya}, {Strauss}, {Sugiyama}, {Suto}, {Takada}, {Takagi}, {Takata},
  {Takita}, {Tanaka}, {Terai}, {Toba}, {Uchiyama}, {Utsumi}, {Wang}, {Wang}, \&
  {Yamada}}]{2019PASJ...71..114A}
{Aihara}, H., {AlSayyad}, Y., {Ando}, M., {et~al.} 2019, \pasj, 71, 114

\bibitem[{{Aihara} {et~al.}(2018{\natexlab{a}}){Aihara}, {Arimoto},
  {Armstrong}, {Arnouts}, {Bahcall}, {Bickerton}, {Bosch}, {Bundy}, {Capak},
  {Chan}, {Chiba}, {Coupon}, {Egami}, {Enoki}, {Finet}, {Fujimori}, {Fujimoto},
  {Furusawa}, {Furusawa}, {Goto}, {Goulding}, {Greco}, {Greene}, {Gunn},
  {Hamana}, {Harikane}, {Hashimoto}, {Hattori}, {Hayashi}, {Hayashi},
  {He{\l}miniak}, {Higuchi}, {Hikage}, {Ho}, {Hsieh}, {Huang}, {Huang},
  {Ikeda}, {Imanishi}, {Inoue}, {Iwasawa}, {Iwata}, {Jaelani}, {Jian},
  {Kamata}, {Karoji}, {Kashikawa}, {Katayama}, {Kawanomoto}, {Kayo}, {Koda},
  {Koike}, {Kojima}, {Komiyama}, {Konno}, {Koshida}, {Koyama}, {Kusakabe},
  {Leauthaud}, {Lee}, {Lin}, {Lin}, {Lupton}, {Mand elbaum}, {Matsuoka},
  {Medezinski}, {Mineo}, {Miyama}, {Miyatake}, {Miyazaki}, {Momose}, {More},
  {More}, {Moritani}, {Moriya}, {Morokuma}, {Mukae}, {Murata}, {Murayama},
  {Nagao}, {Nakata}, {Niida}, {Niikura}, {Nishizawa}, {Obuchi}, {Oguri},
  {Oishi}, {Okabe}, {Okamoto}, {Okura}, {Ono}, {Onodera}, {Onoue}, {Osato},
  {Ouchi}, {Price}, {Pyo}, {Sako}, {Sawicki}, {Shibuya}, {Shimasaku},
  {Shimono}, {Shirasaki}, {Silverman}, {Simet}, {Speagle}, {Spergel},
  {Strauss}, {Sugahara}, {Sugiyama}, {Suto}, {Suyu}, {Suzuki}, {Tait},
  {Takada}, {Takata}, {Tamura}, {Tanaka}, {Tanaka}, {Tanaka}, {Tanaka},
  {Terai}, {Terashima}, {Toba}, {Tominaga}, {Toshikawa}, {Turner}, {Uchida},
  {Uchiyama}, {Umetsu}, {Uraguchi}, {Urata}, {Usuda}, {Utsumi}, {Wang}, {Wang},
  {Wong}, {Yabe}, {Yamada}, {Yamanoi}, {Yasuda}, {Yeh}, {Yonehara}, \&
  {Yuma}}]{2018PASJ...70S...4A}
{Aihara}, H., {Arimoto}, N., {Armstrong}, R., {et~al.} 2018{\natexlab{a}},
  \pasj, 70, S4

\bibitem[{{Aihara} {et~al.}(2018{\natexlab{b}}){Aihara}, {Armstrong},
  {Bickerton}, {Bosch}, {Coupon}, {Furusawa}, {Hayashi}, {Ikeda}, {Kamata},
  {Karoji}, {Kawanomoto}, {Koike}, {Komiyama}, {Lang}, {Lupton}, {Mineo},
  {Miyatake}, {Miyazaki}, {Morokuma}, {Obuchi}, {Oishi}, {Okura}, {Price},
  {Takata}, {Tanaka}, {Tanaka}, {Tanaka}, {Uchida}, {Uraguchi}, {Utsumi},
  {Wang}, {Yamada}, {Yamanoi}, {Yasuda}, {Arimoto}, {Chiba}, {Finet},
  {Fujimori}, {Fujimoto}, {Furusawa}, {Goto}, {Goulding}, {Gunn}, {Harikane},
  {Hattori}, {Hayashi}, {He{\l}miniak}, {Higuchi}, {Hikage}, {Ho}, {Hsieh},
  {Huang}, {Huang}, {Imanishi}, {Iwata}, {Jaelani}, {Jian}, {Kashikawa},
  {Katayama}, {Kojima}, {Konno}, {Koshida}, {Kusakabe}, {Leauthaud}, {Lee},
  {Lin}, {Lin}, {Mandelbaum}, {Matsuoka}, {Medezinski}, {Miyama}, {Momose},
  {More}, {More}, {Mukae}, {Murata}, {Murayama}, {Nagao}, {Nakata}, {Niida},
  {Niikura}, {Nishizawa}, {Oguri}, {Okabe}, {Ono}, {Onodera}, {Onoue}, {Ouchi},
  {Pyo}, {Shibuya}, {Shimasaku}, {Simet}, {Speagle}, {Spergel}, {Strauss},
  {Sugahara}, {Sugiyama}, {Suto}, {Suzuki}, {Tait}, {Takada}, {Terai}, {Toba},
  {Turner}, {Uchiyama}, {Umetsu}, {Urata}, {Usuda}, {Yeh}, \&
  {Yuma}}]{2018PASJ...70S...8A}
{Aihara}, H., {Armstrong}, R., {Bickerton}, S., {et~al.} 2018{\natexlab{b}},
  \pasj, 70, S8

\bibitem[{{Akiyama} {et~al.}(2018){Akiyama}, {He}, {Ikeda}, {Niida}, {Nagao},
  {Bosch}, {Coupon}, {Enoki}, {Imanishi}, {Kashikawa}, {Kawaguchi}, {Komiyama},
  {Lee}, {Matsuoka}, {Miyazaki}, {Nishizawa}, {Oguri}, {Ono}, {Onoue}, {Ouchi},
  {Schulze}, {Silverman}, {Tanaka}, {Tanaka}, {Terashima}, {Toba}, \&
  {Ueda}}]{2018PASJ...70S..34A}
{Akiyama}, M., {He}, W., {Ikeda}, H., {et~al.} 2018, \pasj, 70, S34

\bibitem[{{Axelrod} {et~al.}(2010){Axelrod}, {Kantor}, {Lupton}, \&
  {Pierfederici}}]{2010SPIE.7740E..15A}
{Axelrod}, T., {Kantor}, J., {Lupton}, R.~H., \& {Pierfederici}, F. 2010, in
  Society of Photo-Optical Instrumentation Engineers (SPIE) Conference Series,
  Vol. 7740, Software and Cyberinfrastructure for Astronomy, ed. N.~M.
  {Radziwill} \& A.~{Bridger}, 774015

\bibitem[{{Barone-Nugent} {et~al.}(2015){Barone-Nugent}, {Wyithe}, {Trenti},
  {Treu}, {Oesch}, {Bouwens}, {Illingworth}, \&
  {Schmidt}}]{2015MNRAS.450.1224B}
{Barone-Nugent}, R.~L., {Wyithe}, J.~S.~B., {Trenti}, M., {et~al.} 2015,
  \mnras, 450, 1224

\bibitem[{{Behroozi} {et~al.}(2015){Behroozi}, {Zhu}, {Ferguson}, {Hearin},
  {Lotz}, {Silk}, {Kassin}, {Lu}, {Croton}, {Somerville}, \&
  {Watson}}]{2015MNRAS.450.1546B}
{Behroozi}, P.~S., {Zhu}, G., {Ferguson}, H.~C., {et~al.} 2015, \mnras, 450,
  1546

\bibitem[{{Bertin} \& {Arnouts}(1996)}]{1996A&AS..117..393B}
{Bertin}, E. \& {Arnouts}, S. 1996, \aaps, 117, 393

\bibitem[{{Binney}(1977)}]{1977ApJ...215..483B}
{Binney}, J. 1977, \apj, 215, 483

\bibitem[{{Bluck} {et~al.}(2009){Bluck}, {Conselice}, {Bouwens}, {Daddi},
  {Dickinson}, {Papovich}, \& {Yan}}]{2009MNRAS.394L..51B}
{Bluck}, A. F.~L., {Conselice}, C.~J., {Bouwens}, R.~J., {et~al.} 2009, \mnras,
  394, L51

\bibitem[{{Bluck} {et~al.}(2012){Bluck}, {Conselice}, {Buitrago},
  {Gr{\"u}tzbauch}, {Hoyos}, {Mortlock}, \& {Bauer}}]{2012ApJ...747...34B}
{Bluck}, A. F.~L., {Conselice}, C.~J., {Buitrago}, F., {et~al.} 2012, \apj,
  747, 34

\bibitem[{{Bosch} {et~al.}(2018){Bosch}, {Armstrong}, {Bickerton}, {Furusawa},
  {Ikeda}, {Koike}, {Lupton}, {Mineo}, {Price}, {Takata}, {Tanaka}, {Yasuda},
  {AlSayyad}, {Becker}, {Coulton}, {Coupon}, {Garmilla}, {Huang}, {Krughoff},
  {Lang}, {Leauthaud}, {Lim}, {Lust}, {MacArthur}, {Mandelbaum}, {Miyatake},
  {Miyazaki}, {Murata}, {More}, {Okura}, {Owen}, {Swinbank}, {Strauss},
  {Yamada}, \& {Yamanoi}}]{2018PASJ...70S...5B}
{Bosch}, J., {Armstrong}, R., {Bickerton}, S., {et~al.} 2018, \pasj, 70, S5

\bibitem[{{Bouwens} {et~al.}(2015){Bouwens}, {Illingworth}, {Oesch}, {Trenti},
  {Labb{\'e}}, {Bradley}, {Carollo}, {van Dokkum}, {Gonzalez}, {Holwerda},
  {Franx}, {Spitler}, {Smit}, \& {Magee}}]{2015ApJ...803...34B}
{Bouwens}, R.~J., {Illingworth}, G.~D., {Oesch}, P.~A., {et~al.} 2015, \apj,
  803, 34

\bibitem[{{Bouwens} {et~al.}(2021){Bouwens}, {Oesch}, {Stefanon},
  {Illingworth}, {Labbe}, {Reddy}, {Atek}, {Montes}, {Naidu}, {Nanayakkara},
  {Nelson}, \& {Wilkins}}]{2021arXiv210207775B}
{Bouwens}, R.~J., {Oesch}, P.~A., {Stefanon}, M., {et~al.} 2021, arXiv
  e-prints, arXiv:2102.07775

\bibitem[{{Bowler} {et~al.}(2021){Bowler}, {Adams}, {Jarvis}, \&
  {H{\"a}u{\ss}ler}}]{2021MNRAS.502..662B}
{Bowler}, R.~A.~A., {Adams}, N.~J., {Jarvis}, M.~J., \& {H{\"a}u{\ss}ler}, B.
  2021, \mnras, 502, 662

\bibitem[{{Bowler} {et~al.}(2012){Bowler}, {Dunlop}, {McLure}, {McCracken},
  {Milvang-Jensen}, {Furusawa}, {Fynbo}, {Le F{\`e}vre}, {Holt}, {Ideue},
  {Ihara}, {Rogers}, \& {Taniguchi}}]{2012MNRAS.426.2772B}
{Bowler}, R.~A.~A., {Dunlop}, J.~S., {McLure}, R.~J., {et~al.} 2012, \mnras,
  426, 2772

\bibitem[{{Bowler} {et~al.}(2015){Bowler}, {Dunlop}, {McLure}, {McCracken},
  {Milvang-Jensen}, {Furusawa}, {Taniguchi}, {Le F{\`e}vre}, {Fynbo}, {Jarvis},
  \& {H{\"a}u{\ss}ler}}]{2015MNRAS.452.1817B}
{Bowler}, R.~A.~A., {Dunlop}, J.~S., {McLure}, R.~J., {et~al.} 2015, \mnras,
  452, 1817

\bibitem[{{Bowler} {et~al.}(2017){Bowler}, {Dunlop}, {McLure}, \&
  {McLeod}}]{2017MNRAS.466.3612B}
{Bowler}, R.~A.~A., {Dunlop}, J.~S., {McLure}, R.~J., \& {McLeod}, D.~J. 2017,
  \mnras, 466, 3612

\bibitem[{{Bowler} {et~al.}(2014){Bowler}, {Dunlop}, {McLure}, {Rogers},
  {McCracken}, {Milvang-Jensen}, {Furusawa}, {Fynbo}, {Taniguchi}, {Afonso},
  {Bremer}, \& {Le F{\`e}vre}}]{2014MNRAS.440.2810B}
{Bowler}, R.~A.~A., {Dunlop}, J.~S., {McLure}, R.~J., {et~al.} 2014, \mnras,
  440, 2810

\bibitem[{{Bowler} {et~al.}(2020){Bowler}, {Jarvis}, {Dunlop}, {McLure},
  {McLeod}, {Adams}, {Milvang-Jensen}, \& {McCracken}}]{2020MNRAS.493.2059B}
{Bowler}, R.~A.~A., {Jarvis}, M.~J., {Dunlop}, J.~S., {et~al.} 2020, \mnras,
  493, 2059

\bibitem[{Boyd {et~al.}(2011)Boyd, Parikh, Chu, Peleato, \& Eckstein}]{MAL-016}
Boyd, S., Parikh, N., Chu, E., Peleato, B., \& Eckstein, J. 2011, Foundations
  and Trends{\textregistered} in Machine Learning, 3, 1

\bibitem[{{Bradley} {et~al.}(2012){Bradley}, {Trenti}, {Oesch}, {Stiavelli},
  {Treu}, {Bouwens}, {Shull}, {Holwerda}, \& {Pirzkal}}]{2012ApJ...760..108B}
{Bradley}, L.~D., {Trenti}, M., {Oesch}, P.~A., {et~al.} 2012, \apj, 760, 108

\bibitem[{{Bundy} {et~al.}(2009){Bundy}, {Fukugita}, {Ellis}, {Targett},
  {Belli}, \& {Kodama}}]{2009ApJ...697.1369B}
{Bundy}, K., {Fukugita}, M., {Ellis}, R.~S., {et~al.} 2009, \apj, 697, 1369

\bibitem[{{Cibinel} {et~al.}(2019){Cibinel}, {Daddi}, {Sargent}, {Le Floc'h},
  {Liu}, {Bournaud}, {Oesch}, {Amram}, {Calabr{\`o}}, {Duc}, {Pannella},
  {Puglisi}, {Perret}, {Elbaz}, \& {Kokorev}}]{2019MNRAS.485.5631C}
{Cibinel}, A., {Daddi}, E., {Sargent}, M.~T., {et~al.} 2019, \mnras, 485, 5631

\bibitem[{{Conselice} \& {Arnold}(2009)}]{2009MNRAS.397..208C}
{Conselice}, C.~J. \& {Arnold}, J. 2009, \mnras, 397, 208

\bibitem[{{Cotini} {et~al.}(2013){Cotini}, {Ripamonti}, {Caccianiga}, {Colpi},
  {Della Ceca}, {Mapelli}, {Severgnini}, \& {Segreto}}]{2013MNRAS.431.2661C}
{Cotini}, S., {Ripamonti}, E., {Caccianiga}, A., {et~al.} 2013, \mnras, 431,
  2661

\bibitem[{{Croton} {et~al.}(2006){Croton}, {Springel}, {White}, {De Lucia},
  {Frenk}, {Gao}, {Jenkins}, {Kauffmann}, {Navarro}, \&
  {Yoshida}}]{2006MNRAS.365...11C}
{Croton}, D.~J., {Springel}, V., {White}, S. D.~M., {et~al.} 2006, \mnras, 365,
  11

\bibitem[{{Curtis-Lake} {et~al.}(2016){Curtis-Lake}, {McLure}, {Dunlop},
  {Rogers}, {Targett}, {Dekel}, {Ellis}, {Faber}, {Ferguson}, {Grogin},
  {Kocevski}, {Koekemoer}, {Lai}, {M{\'a}rmol-Queralt{\'o}}, \&
  {Robertson}}]{2016MNRAS.457..440C}
{Curtis-Lake}, E., {McLure}, R.~J., {Dunlop}, J.~S., {et~al.} 2016, \mnras,
  457, 440

\bibitem[{{Duncan} {et~al.}(2019){Duncan}, {Conselice}, {Mundy}, {Bell},
  {Donley}, {Galametz}, {Guo}, {Grogin}, {Hathi}, {Kartaltepe}, {Kocevski},
  {Koekemoer}, {P{\'e}rez-Gonz{\'a}lez}, {Mantha}, {Snyder}, \&
  {Stefanon}}]{2019ApJ...876..110D}
{Duncan}, K., {Conselice}, C.~J., {Mundy}, C., {et~al.} 2019, \apj, 876, 110

\bibitem[{{Ellis} {et~al.}(2013){Ellis}, {McLure}, {Dunlop}, {Robertson},
  {Ono}, {Schenker}, {Koekemoer}, {Bowler}, {Ouchi}, {Rogers}, {Curtis-Lake},
  {Schneider}, {Charlot}, {Stark}, {Furlanetto}, \&
  {Cirasuolo}}]{2013ApJ...763L...7E}
{Ellis}, R.~S., {McLure}, R.~J., {Dunlop}, J.~S., {et~al.} 2013, \apjl, 763, L7

\bibitem[{{Ellison} {et~al.}(2013){Ellison}, {Mendel}, {Patton}, \&
  {Scudder}}]{2013MNRAS.435.3627E}
{Ellison}, S.~L., {Mendel}, J.~T., {Patton}, D.~R., \& {Scudder}, J.~M. 2013,
  \mnras, 435, 3627

\bibitem[{{Ellison} {et~al.}(2008){Ellison}, {Patton}, {Simard}, \&
  {McConnachie}}]{2008AJ....135.1877E}
{Ellison}, S.~L., {Patton}, D.~R., {Simard}, L., \& {McConnachie}, A.~W. 2008,
  \aj, 135, 1877

\bibitem[{{Endsley} {et~al.}(2020){Endsley}, {Behroozi}, {Stark}, {Williams},
  {Robertson}, {Rieke}, {Gottl{\"o}ber}, \& {Yepes}}]{2020MNRAS.493.1178E}
{Endsley}, R., {Behroozi}, P., {Stark}, D.~P., {et~al.} 2020, \mnras, 493, 1178

\bibitem[{{Ferreira} {et~al.}(2020){Ferreira}, {Conselice}, {Duncan}, {Cheng},
  {Griffiths}, \& {Whitney}}]{2020ApJ...895..115F}
{Ferreira}, L., {Conselice}, C.~J., {Duncan}, K., {et~al.} 2020, \apj, 895, 115

\bibitem[{{Figueiredo} \& {Bioucas-Dias}(2010)}]{5492199}
{Figueiredo}, M. A.~T. \& {Bioucas-Dias}, J.~M. 2010, IEEE Transactions on
  Image Processing, 19, 3133

\bibitem[{{Finkelstein} {et~al.}(2021){Finkelstein}, {Bagley}, {Song},
  {Larson}, {Papovich}, {Dickinson}, {Finkelstein}, {Koekemoer}, {Pirzkal},
  {Somerville}, {Yung}, {Behroozi}, {Ferguson}, {Giavalisco}, {Grogin},
  {Hathi}, {Hutchison}, {Jung}, {Kocevski}, {Kawinwanichakij}, {Rojas-Ruiz},
  {Ryan}, {Snyder}, \& {Tacchella}}]{2021arXiv210613813F}
{Finkelstein}, S.~L., {Bagley}, M., {Song}, M., {et~al.} 2021, arXiv e-prints,
  arXiv:2106.13813

\bibitem[{{Finkelstein} {et~al.}(2015){Finkelstein}, {Ryan}, {Papovich},
  {Dickinson}, {Song}, {Somerville}, {Ferguson}, {Salmon}, {Giavalisco},
  {Koekemoer}, {Ashby}, {Behroozi}, {Castellano}, {Dunlop}, {Faber}, {Fazio},
  {Fontana}, {Grogin}, {Hathi}, {Jaacks}, {Kocevski}, {Livermore}, {McLure},
  {Merlin}, {Mobasher}, {Newman}, {Rafelski}, {Tilvi}, \&
  {Willner}}]{2015ApJ...810...71F}
{Finkelstein}, S.~L., {Ryan}, Russell~E., J., {Papovich}, C., {et~al.} 2015,
  \apj, 810, 71

\bibitem[{{Furusawa} {et~al.}(2018){Furusawa}, {Koike}, {Takata}, {Okura},
  {Miyatake}, {Lupton}, {Bickerton}, {Price}, {Bosch}, {Yasuda}, {Mineo},
  {Yamada}, {Miyazaki}, {Nakata}, {Koshida}, {Komiyama}, {Utsumi},
  {Kawanomoto}, {Jeschke}, {Noumaru}, {Schubert}, {Iwata}, {Finet},
  {Fujiyoshi}, {Tajitsu}, {Terai}, \& {Lee}}]{2018PASJ...70S...3F}
{Furusawa}, H., {Koike}, M., {Takata}, T., {et~al.} 2018, \pasj, 70, S3

\bibitem[{{Grogin} {et~al.}(2011){Grogin}, {Kocevski}, {Faber}, {Ferguson},
  {Koekemoer}, {Riess}, {Acquaviva}, {Alexander}, {Almaini}, {Ashby}, {Barden},
  {Bell}, {Bournaud}, {Brown}, {Caputi}, {Casertano}, {Cassata}, {Castellano},
  {Challis}, {Chary}, {Cheung}, {Cirasuolo}, {Conselice}, {Roshan Cooray},
  {Croton}, {Daddi}, {Dahlen}, {Dav{\'e}}, {de Mello}, {Dekel}, {Dickinson},
  {Dolch}, {Donley}, {Dunlop}, {Dutton}, {Elbaz}, {Fazio}, {Filippenko},
  {Finkelstein}, {Fontana}, {Gardner}, {Garnavich}, {Gawiser}, {Giavalisco},
  {Grazian}, {Guo}, {Hathi}, {H{\"a}ussler}, {Hopkins}, {Huang}, {Huang},
  {Jha}, {Kartaltepe}, {Kirshner}, {Koo}, {Lai}, {Lee}, {Li}, {Lotz}, {Lucas},
  {Madau}, {McCarthy}, {McGrath}, {McIntosh}, {McLure}, {Mobasher},
  {Moustakas}, {Mozena}, {Nandra}, {Newman}, {Niemi}, {Noeske}, {Papovich},
  {Pentericci}, {Pope}, {Primack}, {Rajan}, {Ravindranath}, {Reddy}, {Renzini},
  {Rix}, {Robaina}, {Rodney}, {Rosario}, {Rosati}, {Salimbeni}, {Scarlata},
  {Siana}, {Simard}, {Smidt}, {Somerville}, {Spinrad}, {Straughn}, {Strolger},
  {Telford}, {Teplitz}, {Trump}, {van der Wel}, {Villforth}, {Wechsler},
  {Weiner}, {Wiklind}, {Wild}, {Wilson}, {Wuyts}, {Yan}, \&
  {Yun}}]{2011ApJS..197...35G}
{Grogin}, N.~A., {Kocevski}, D.~D., {Faber}, S.~M., {et~al.} 2011, \apjs, 197,
  35

\bibitem[{{Guo} {et~al.}(2013){Guo}, {Ferguson}, {Giavalisco}, {Barro},
  {Willner}, {Ashby}, {Dahlen}, {Donley}, {Faber}, {Fontana}, {Galametz},
  {Grazian}, {Huang}, {Kocevski}, {Koekemoer}, {Koo}, {McGrath}, {Peth},
  {Salvato}, {Wuyts}, {Castellano}, {Cooray}, {Dickinson}, {Dunlop}, {Fazio},
  {Gardner}, {Gawiser}, {Grogin}, {Hathi}, {Hsu}, {Lee}, {Lucas}, {Mobasher},
  {Nandra}, {Newman}, \& {van der Wel}}]{2013ApJS..207...24G}
{Guo}, Y., {Ferguson}, H.~C., {Giavalisco}, M., {et~al.} 2013, \apjs, 207, 24

\bibitem[{{Harikane} {et~al.}(2021){Harikane}, {Ono}, {Ouchi}, {Liu},
  {Sawicki}, {Shibuya}, {Behroozi}, {He}, {Shimasaku}, {Arnouts}, {Coupon},
  {Fujimoto}, {Gwyn}, {Huang}, {Inoue}, {Kashikawa}, {Komiyama}, {Matsuoka}, \&
  {Willott}}]{2021arXiv210801090H}
{Harikane}, Y., {Ono}, Y., {Ouchi}, M., {et~al.} 2021, arXiv e-prints,
  arXiv:2108.01090

\bibitem[{{Harikane} {et~al.}(2016){Harikane}, {Ouchi}, {Ono}, {More}, {Saito},
  {Lin}, {Coupon}, {Shimasaku}, {Shibuya}, {Price}, {Lin}, {Hsieh}, {Ishigaki},
  {Komiyama}, {Silverman}, {Takata}, {Tamazawa}, \&
  {Toshikawa}}]{2016ApJ...821..123H}
{Harikane}, Y., {Ouchi}, M., {Ono}, Y., {et~al.} 2016, \apj, 821, 123

\bibitem[{{Harikane} {et~al.}(2018){Harikane}, {Ouchi}, {Ono}, {Saito},
  {Behroozi}, {More}, {Shimasaku}, {Toshikawa}, {Lin}, {Akiyama}, {Coupon},
  {Komiyama}, {Konno}, {Lin}, {Miyazaki}, {Nishizawa}, {Shibuya}, \&
  {Silverman}}]{2018PASJ...70S..11H}
{Harikane}, Y., {Ouchi}, M., {Ono}, Y., {et~al.} 2018, \pasj, 70, S11

\bibitem[{{Hopkins} {et~al.}(2010){Hopkins}, {Younger}, {Hayward}, {Narayanan},
  \& {Hernquist}}]{2010MNRAS.402.1693H}
{Hopkins}, P.~F., {Younger}, J.~D., {Hayward}, C.~C., {Narayanan}, D., \&
  {Hernquist}, L. 2010, \mnras, 402, 1693

\bibitem[{{Ikoma} {et~al.}(2018){Ikoma}, {Broxton}, {Kudo}, \&
  {Wetzstein}}]{2018NatSR...811489I}
{Ikoma}, H., {Broxton}, M., {Kudo}, T., \& {Wetzstein}, G. 2018, Scientific
  Reports, 8, 11489

\bibitem[{{Ivezi{\'c}} {et~al.}(2019){Ivezi{\'c}}, {Kahn}, {Tyson}, {Abel},
  {Acosta}, {Allsman}, {Alonso}, {AlSayyad}, {Anderson}, {Andrew}, {Angel},
  {Angeli}, {Ansari}, {Antilogus}, {Araujo}, {Armstrong}, {Arndt}, {Astier},
  {Aubourg}, {Auza}, {Axelrod}, {Bard}, {Barr}, {Barrau}, {Bartlett}, {Bauer},
  {Bauman}, {Baumont}, {Bechtol}, {Bechtol}, {Becker}, {Becla}, {Beldica},
  {Bellavia}, {Bianco}, {Biswas}, {Blanc}, {Blazek}, {Blandford}, {Bloom},
  {Bogart}, {Bond}, {Booth}, {Borgland}, {Borne}, {Bosch}, {Boutigny},
  {Brackett}, {Bradshaw}, {Brandt}, {Brown}, {Bullock}, {Burchat}, {Burke},
  {Cagnoli}, {Calabrese}, {Callahan}, {Callen}, {Carlin}, {Carlson},
  {Chandrasekharan}, {Charles-Emerson}, {Chesley}, {Cheu}, {Chiang}, {Chiang},
  {Chirino}, {Chow}, {Ciardi}, {Claver}, {Cohen-Tanugi}, {Cockrum}, {Coles},
  {Connolly}, {Cook}, {Cooray}, {Covey}, {Cribbs}, {Cui}, {Cutri}, {Daly},
  {Daniel}, {Daruich}, {Daubard}, {Daues}, {Dawson}, {Delgado}, {Dellapenna},
  {de Peyster}, {de Val-Borro}, {Digel}, {Doherty}, {Dubois},
  {Dubois-Felsmann}, {Durech}, {Economou}, {Eifler}, {Eracleous}, {Emmons},
  {Fausti Neto}, {Ferguson}, {Figueroa}, {Fisher-Levine}, {Focke}, {Foss},
  {Frank}, {Freemon}, {Gangler}, {Gawiser}, {Geary}, {Gee}, {Geha}, {Gessner},
  {Gibson}, {Gilmore}, {Glanzman}, {Glick}, {Goldina}, {Goldstein}, {Goodenow},
  {Graham}, {Gressler}, {Gris}, {Guy}, {Guyonnet}, {Haller}, {Harris},
  {Hascall}, {Haupt}, {Hernandez}, {Herrmann}, {Hileman}, {Hoblitt}, {Hodgson},
  {Hogan}, {Howard}, {Huang}, {Huffer}, {Ingraham}, {Innes}, {Jacoby}, {Jain},
  {Jammes}, {Jee}, {Jenness}, {Jernigan}, {Jevremovi{\'c}}, {Johns}, {Johnson},
  {Johnson}, {Jones}, {Juramy-Gilles}, {Juri{\'c}}, {Kalirai}, {Kallivayalil},
  {Kalmbach}, {Kantor}, {Karst}, {Kasliwal}, {Kelly}, {Kessler}, {Kinnison},
  {Kirkby}, {Knox}, {Kotov}, {Krabbendam}, {Krughoff}, {Kub{\'a}nek},
  {Kuczewski}, {Kulkarni}, {Ku}, {Kurita}, {Lage}, {Lambert}, {Lange},
  {Langton}, {Le Guillou}, {Levine}, {Liang}, {Lim}, {Lintott}, {Long},
  {Lopez}, {Lotz}, {Lupton}, {Lust}, {MacArthur}, {Mahabal}, {Mandelbaum},
  {Markiewicz}, {Marsh}, {Marshall}, {Marshall}, {May}, {McKercher}, {McQueen},
  {Meyers}, {Migliore}, {Miller}, {Mills}, {Miraval}, {Moeyens}, {Moolekamp},
  {Monet}, {Moniez}, {Monkewitz}, {Montgomery}, {Morrison}, {Mueller},
  {Muller}, {Mu{\~n}oz Arancibia}, {Neill}, {Newbry}, {Nief}, {Nomerotski},
  {Nordby}, {O'Connor}, {Oliver}, {Olivier}, {Olsen}, {O'Mullane}, {Ortiz},
  {Osier}, {Owen}, {Pain}, {Palecek}, {Parejko}, {Parsons}, {Pease},
  {Peterson}, {Peterson}, {Petravick}, {Libby Petrick}, {Petry},
  {Pierfederici}, {Pietrowicz}, {Pike}, {Pinto}, {Plante}, {Plate}, {Plutchak},
  {Price}, {Prouza}, {Radeka}, {Rajagopal}, {Rasmussen}, {Regnault}, {Reil},
  {Reiss}, {Reuter}, {Ridgway}, {Riot}, {Ritz}, {Robinson}, {Roby}, {Roodman},
  {Rosing}, {Roucelle}, {Rumore}, {Russo}, {Saha}, {Sassolas}, {Schalk},
  {Schellart}, {Schindler}, {Schmidt}, {Schneider}, {Schneider}, {Schoening},
  {Schumacher}, {Schwamb}, {Sebag}, {Selvy}, {Sembroski}, {Seppala}, {Serio},
  {Serrano}, {Shaw}, {Shipsey}, {Sick}, {Silvestri}, {Slater}, {Smith},
  {Smith}, {Sobhani}, {Soldahl}, {Storrie-Lombardi}, {Stover}, {Strauss},
  {Street}, {Stubbs}, {Sullivan}, {Sweeney}, {Swinbank}, {Szalay}, {Takacs},
  {Tether}, {Thaler}, {Thayer}, {Thomas}, {Thornton}, {Thukral}, {Tice},
  {Trilling}, {Turri}, {Van Berg}, {Vanden Berk}, {Vetter}, {Virieux},
  {Vucina}, {Wahl}, {Walkowicz}, {Walsh}, {Walter}, {Wang}, {Wang}, {Warner},
  {Wiecha}, {Willman}, {Winters}, {Wittman}, {Wolff}, {Wood-Vasey}, {Wu},
  {Xin}, {Yoachim}, \& {Zhan}}]{2019ApJ...873..111I}
{Ivezi{\'c}}, {\v{Z}}., {Kahn}, S.~M., {Tyson}, J.~A., {et~al.} 2019, \apj,
  873, 111

\bibitem[{{Jiang} {et~al.}(2013){Jiang}, {Egami}, {Fan}, {Windhorst}, {Cohen},
  {Dav{\'e}}, {Finlator}, {Kashikawa}, {Mechtley}, {Ouchi}, \&
  {Shimasaku}}]{2013ApJ...773..153J}
{Jiang}, L., {Egami}, E., {Fan}, X., {et~al.} 2013, \apj, 773, 153

\bibitem[{{Juri{\'c}} {et~al.}(2017){Juri{\'c}}, {Kantor}, {Lim}, {Lupton},
  {Dubois-Felsmann}, {Jenness}, {Axelrod}, {Aleksi{\'c}}, {Allsman},
  {AlSayyad}, {Alt}, {Armstrong}, {Basney}, {Becker}, {Becla}, {Biswas},
  {Bosch}, {Boutigny}, {Kind}, {Ciardi}, {Connolly}, {Daniel}, {Daues},
  {Economou}, {Chiang}, {Fausti}, {Fisher-Levine}, {Freemon}, {Gris},
  {Hernandez}, {Hoblitt}, {Ivezi{\'c}}, {Jammes}, {Jevremovi{\'c}}, {Jones},
  {Kalmbach}, {Kasliwal}, {Krughoff}, {Lurie}, {Lust}, {MacArthur}, {Melchior},
  {Moeyens}, {Nidever}, {Owen}, {Parejko}, {Peterson}, {Petravick},
  {Pietrowicz}, {Price}, {Reiss}, {Shaw}, {Sick}, {Slater}, {Strauss},
  {Sullivan}, {Swinbank}, {Van Dyk}, {Vuj{\v{c}}i{\'c}}, {Withers}, \&
  {Yoachim}}]{2017ASPC..512..279J}
{Juri{\'c}}, M., {Kantor}, J., {Lim}, K.~T., {et~al.} 2017, in Astronomical
  Society of the Pacific Conference Series, Vol. 512, Astronomical Data
  Analysis Software and Systems XXV, ed. N.~P.~F. {Lorente}, K.~{Shortridge},
  \& R.~{Wayth}, 279

\bibitem[{{Kawamata} {et~al.}(2015){Kawamata}, {Ishigaki}, {Shimasaku},
  {Oguri}, \& {Ouchi}}]{2015ApJ...804..103K}
{Kawamata}, R., {Ishigaki}, M., {Shimasaku}, K., {Oguri}, M., \& {Ouchi}, M.
  2015, \apj, 804, 103

\bibitem[{{Kawanomoto} {et~al.}(2018){Kawanomoto}, {Uraguchi}, {Komiyama},
  {Miyazaki}, {Furusawa}, {Finet}, {Hattori}, {Wang}, {Yasuda}, \&
  {Suzuki}}]{2018PASJ...70...66K}
{Kawanomoto}, S., {Uraguchi}, F., {Komiyama}, Y., {et~al.} 2018, \pasj, 70, 66

\bibitem[{{Kim} \& {Im}(2021)}]{2021ApJ...910L..11K}
{Kim}, Y. \& {Im}, M. 2021, \apjl, 910, L11

\bibitem[{{Koekemoer} {et~al.}(2011){Koekemoer}, {Faber}, {Ferguson}, {Grogin},
  {Kocevski}, {Koo}, {Lai}, {Lotz}, {Lucas}, {McGrath}, {Ogaz}, {Rajan},
  {Riess}, {Rodney}, {Strolger}, {Casertano}, {Castellano}, {Dahlen},
  {Dickinson}, {Dolch}, {Fontana}, {Giavalisco}, {Grazian}, {Guo}, {Hathi},
  {Huang}, {van der Wel}, {Yan}, {Acquaviva}, {Alexander}, {Almaini}, {Ashby},
  {Barden}, {Bell}, {Bournaud}, {Brown}, {Caputi}, {Cassata}, {Challis},
  {Chary}, {Cheung}, {Cirasuolo}, {Conselice}, {Roshan Cooray}, {Croton},
  {Daddi}, {Dav{\'e}}, {de Mello}, {de Ravel}, {Dekel}, {Donley}, {Dunlop},
  {Dutton}, {Elbaz}, {Fazio}, {Filippenko}, {Finkelstein}, {Frazer}, {Gardner},
  {Garnavich}, {Gawiser}, {Gruetzbauch}, {Hartley}, {H{\"a}ussler},
  {Herrington}, {Hopkins}, {Huang}, {Jha}, {Johnson}, {Kartaltepe},
  {Khostovan}, {Kirshner}, {Lani}, {Lee}, {Li}, {Madau}, {McCarthy},
  {McIntosh}, {McLure}, {McPartland}, {Mobasher}, {Moreira}, {Mortlock},
  {Moustakas}, {Mozena}, {Nandra}, {Newman}, {Nielsen}, {Niemi}, {Noeske},
  {Papovich}, {Pentericci}, {Pope}, {Primack}, {Ravindranath}, {Reddy},
  {Renzini}, {Rix}, {Robaina}, {Rosario}, {Rosati}, {Salimbeni}, {Scarlata},
  {Siana}, {Simard}, {Smidt}, {Snyder}, {Somerville}, {Spinrad}, {Straughn},
  {Telford}, {Teplitz}, {Trump}, {Vargas}, {Villforth}, {Wagner}, {Wandro},
  {Wechsler}, {Weiner}, {Wiklind}, {Wild}, {Wilson}, {Wuyts}, \&
  {Yun}}]{2011ApJS..197...36K}
{Koekemoer}, A.~M., {Faber}, S.~M., {Ferguson}, H.~C., {et~al.} 2011, \apjs,
  197, 36

\bibitem[{{Komiyama} {et~al.}(2018){Komiyama}, {Obuchi}, {Nakaya}, {Kamata},
  {Kawanomoto}, {Utsumi}, {Miyazaki}, {Uraguchi}, {Furusawa}, {Morokuma},
  {Uchida}, {Miyatake}, {Mineo}, {Fujimori}, {Aihara}, {Karoji}, {Gunn}, \&
  {Wang}}]{2018PASJ...70S...2K}
{Komiyama}, Y., {Obuchi}, Y., {Nakaya}, H., {et~al.} 2018, \pasj, 70, S2

\bibitem[{{Le F{\`e}vre} {et~al.}(2000){Le F{\`e}vre}, {Abraham}, {Lilly},
  {Ellis}, {Brinchmann}, {Schade}, {Tresse}, {Colless}, {Crampton},
  {Glazebrook}, {Hammer}, \& {Broadhurst}}]{2000MNRAS.311..565L}
{Le F{\`e}vre}, O., {Abraham}, R., {Lilly}, S.~J., {et~al.} 2000, \mnras, 311,
  565

\bibitem[{{Le F{\`e}vre} {et~al.}(2020){Le F{\`e}vre}, {B{\'e}thermin},
  {Faisst}, {Jones}, {Capak}, {Cassata}, {Silverman}, {Schaerer}, {Yan},
  {Amorin}, {Bardelli}, {Boquien}, {Cimatti}, {Dessauges-Zavadsky},
  {Giavalisco}, {Hathi}, {Fudamoto}, {Fujimoto}, {Ginolfi}, {Gruppioni},
  {Hemmati}, {Ibar}, {Koekemoer}, {Khusanova}, {Lagache}, {Lemaux}, {Loiacono},
  {Maiolino}, {Mancini}, {Narayanan}, {Morselli}, {M{\'e}ndez-Hern{\`a}ndez},
  {Oesch}, {Pozzi}, {Romano}, {Riechers}, {Scoville}, {Talia}, {Tasca},
  {Thomas}, {Toft}, {Vallini}, {Vergani}, {Walter}, {Zamorani}, \&
  {Zucca}}]{2020A&A...643A...1L}
{Le F{\`e}vre}, O., {B{\'e}thermin}, M., {Faisst}, A., {et~al.} 2020, \aap,
  643, A1

\bibitem[{{Leauthaud} {et~al.}(2007){Leauthaud}, {Massey}, {Kneib}, {Rhodes},
  {Johnston}, {Capak}, {Heymans}, {Ellis}, {Koekemoer}, {Le F{\`e}vre},
  {Mellier}, {R{\'e}fr{\'e}gier}, {Robin}, {Scoville}, {Tasca}, {Taylor}, \&
  {Van Waerbeke}}]{2007ApJS..172..219L}
{Leauthaud}, A., {Massey}, R., {Kneib}, J.-P., {et~al.} 2007, \apjs, 172, 219

\bibitem[{{Liu} {et~al.}(2019){Liu}, {Schinnerer}, {Groves}, {Magnelli},
  {Lang}, {Leslie}, {Jim{\'e}nez-Andrade}, {Riechers}, {Popping}, {Magdis},
  {Daddi}, {Sargent}, {Gao}, {Fudamoto}, {Oesch}, \&
  {Bertoldi}}]{2019ApJ...887..235L}
{Liu}, D., {Schinnerer}, E., {Groves}, B., {et~al.} 2019, \apj, 887, 235

\bibitem[{{Lotz} {et~al.}(2011){Lotz}, {Jonsson}, {Cox}, {Croton}, {Primack},
  {Somerville}, \& {Stewart}}]{2011ApJ...742..103L}
{Lotz}, J.~M., {Jonsson}, P., {Cox}, T.~J., {et~al.} 2011, \apj, 742, 103

\bibitem[{{Lotz} {et~al.}(2006){Lotz}, {Madau}, {Giavalisco}, {Primack}, \&
  {Ferguson}}]{2006ApJ...636..592L}
{Lotz}, J.~M., {Madau}, P., {Giavalisco}, M., {Primack}, J., \& {Ferguson},
  H.~C. 2006, \apj, 636, 592

\bibitem[{{Lucy}(1974)}]{1974AJ.....79..745L}
{Lucy}, L.~B. 1974, \aj, 79, 745

\bibitem[{{Man} {et~al.}(2016){Man}, {Zirm}, \& {Toft}}]{2016ApJ...830...89M}
{Man}, A. W.~S., {Zirm}, A.~W., \& {Toft}, S. 2016, \apj, 830, 89

\bibitem[{{Mantha} {et~al.}(2018){Mantha}, {McIntosh}, {Brennan}, {Ferguson},
  {Kodra}, {Newman}, {Rafelski}, {Somerville}, {Conselice}, {Cook}, {Hathi},
  {Koo}, {Lotz}, {Simmons}, {Straughn}, {Snyder}, {Wuyts}, {Bell}, {Dekel},
  {Kartaltepe}, {Kocevski}, {Koekemoer}, {Lee}, {Lucas}, {Pacifici}, {Peth},
  {Barro}, {Dahlen}, {Finkelstein}, {Fontana}, {Galametz}, {Grogin}, {Guo},
  {Mobasher}, {Nayyeri}, {P{\'e}rez-Gonz{\'a}lez}, {Pforr}, {Santini},
  {Stefanon}, \& {Wiklind}}]{2018MNRAS.475.1549M}
{Mantha}, K.~B., {McIntosh}, D.~H., {Brennan}, R., {et~al.} 2018, \mnras, 475,
  1549

\bibitem[{{Mason} {et~al.}(2015){Mason}, {Treu}, {Schmidt}, {Collett},
  {Trenti}, {Marshall}, {Barone-Nugent}, {Bradley}, {Stiavelli}, \&
  {Wyithe}}]{2015ApJ...805...79M}
{Mason}, C.~A., {Treu}, T., {Schmidt}, K.~B., {et~al.} 2015, \apj, 805, 79

\bibitem[{{Matsuoka} {et~al.}(2019){Matsuoka}, {Iwasawa}, {Onoue}, {Kashikawa},
  {Strauss}, {Lee}, {Imanishi}, {Nagao}, {Akiyama}, {Asami}, {Bosch},
  {Furusawa}, {Goto}, {Gunn}, {Harikane}, {Ikeda}, {Izumi}, {Kawaguchi},
  {Kato}, {Kikuta}, {Kohno}, {Komiyama}, {Koyama}, {Lupton}, {Minezaki},
  {Miyazaki}, {Murayama}, {Niida}, {Nishizawa}, {Noboriguchi}, {Oguri}, {Ono},
  {Ouchi}, {Price}, {Sameshima}, {Schulze}, {Silverman}, {Sugiyama}, {Tait},
  {Takada}, {Takata}, {Tanaka}, {Tang}, {Toba}, {Utsumi}, {Wang}, \&
  {Yamashita}}]{2019ApJ...883..183M}
{Matsuoka}, Y., {Iwasawa}, K., {Onoue}, M., {et~al.} 2019, \apj, 883, 183

\bibitem[{{McLeod} {et~al.}(2016){McLeod}, {McLure}, \&
  {Dunlop}}]{2016MNRAS.459.3812M}
{McLeod}, D.~J., {McLure}, R.~J., \& {Dunlop}, J.~S. 2016, \mnras, 459, 3812

\bibitem[{{McLure} {et~al.}(2013){McLure}, {Dunlop}, {Bowler}, {Curtis-Lake},
  {Schenker}, {Ellis}, {Robertson}, {Koekemoer}, {Rogers}, {Ono}, {Ouchi},
  {Charlot}, {Wild}, {Stark}, {Furlanetto}, {Cirasuolo}, \&
  {Targett}}]{2013MNRAS.432.2696M}
{McLure}, R.~J., {Dunlop}, J.~S., {Bowler}, R.~A.~A., {et~al.} 2013, \mnras,
  432, 2696

\bibitem[{{Miyazaki} {et~al.}(2018){Miyazaki}, {Komiyama}, {Kawanomoto}, {Doi},
  {Furusawa}, {Hamana}, {Hayashi}, {Ikeda}, {Kamata}, {Karoji}, {Koike},
  {Kurakami}, {Miyama}, {Morokuma}, {Nakata}, {Namikawa}, {Nakaya}, {Nariai},
  {Obuchi}, {Oishi}, {Okada}, {Okura}, {Tait}, {Takata}, {Tanaka}, {Tanaka},
  {Terai}, {Tomono}, {Uraguchi}, {Usuda}, {Utsumi}, {Yamada}, {Yamanoi},
  {Aihara}, {Fujimori}, {Mineo}, {Miyatake}, {Oguri}, {Uchida}, {Tanaka},
  {Yasuda}, {Takada}, {Murayama}, {Nishizawa}, {Sugiyama}, {Chiba}, {Futamase},
  {Wang}, {Chen}, {Ho}, {Liaw}, {Chiu}, {Ho}, {Lai}, {Lee}, {Jeng}, {Iwamura},
  {Armstrong}, {Bickerton}, {Bosch}, {Gunn}, {Lupton}, {Loomis}, {Price},
  {Smith}, {Strauss}, {Turner}, {Suzuki}, {Miyazaki}, {Muramatsu}, {Yamamoto},
  {Endo}, {Ezaki}, {Ito}, {Kawaguchi}, {Sofuku}, {Taniike}, {Akutsu}, {Dojo},
  {Kasumi}, {Matsuda}, {Imoto}, {Miwa}, {Suzuki}, {Takeshi}, \&
  {Yokota}}]{2018PASJ...70S...1M}
{Miyazaki}, S., {Komiyama}, Y., {Kawanomoto}, S., {et~al.} 2018, \pasj, 70, S1

\bibitem[{{Miyazaki} {et~al.}(2012){Miyazaki}, {Komiyama}, {Nakaya}, {Kamata},
  {Doi}, {Hamana}, {Karoji}, {Furusawa}, {Kawanomoto}, {Morokuma}, {Ishizuka},
  {Nariai}, {Tanaka}, {Uraguchi}, {Utsumi}, {Obuchi}, {Okura}, {Oguri},
  {Takata}, {Tomono}, {Kurakami}, {Namikawa}, {Usuda}, {Yamanoi}, {Terai},
  {Uekiyo}, {Yamada}, {Koike}, {Aihara}, {Fujimori}, {Mineo}, {Miyatake},
  {Yasuda}, {Nishizawa}, {Saito}, {Tanaka}, {Uchida}, {Katayama}, {Wang},
  {Chen}, {Lupton}, {Loomis}, {Bickerton}, {Price}, {Gunn}, {Suzuki},
  {Miyazaki}, {Muramatsu}, {Yamamoto}, {Endo}, {Ezaki}, {Itoh}, {Miwa},
  {Yokota}, {Matsuda}, {Ebinuma}, \& {Takeshi}}]{2012SPIE.8446E..0ZM}
{Miyazaki}, S., {Komiyama}, Y., {Nakaya}, H., {et~al.} 2012, in Society of
  Photo-Optical Instrumentation Engineers (SPIE) Conference Series, Vol. 8446,
  Ground-based and Airborne Instrumentation for Astronomy IV, ed. I.~S.
  {McLean}, S.~K. {Ramsay}, \& H.~{Takami}, 84460Z

\bibitem[{{Mundy} {et~al.}(2017){Mundy}, {Conselice}, {Duncan}, {Almaini},
  {H{\"a}u{\ss}ler}, \& {Hartley}}]{2017MNRAS.470.3507M}
{Mundy}, C.~J., {Conselice}, C.~J., {Duncan}, K.~J., {et~al.} 2017, \mnras,
  470, 3507

\bibitem[{{Oesch} {et~al.}(2010{\natexlab{a}}){Oesch}, {Bouwens}, {Carollo},
  {Illingworth}, {Magee}, {Trenti}, {Stiavelli}, {Franx}, {Labb{\'e}}, \& {van
  Dokkum}}]{2010ApJ...725L.150O}
{Oesch}, P.~A., {Bouwens}, R.~J., {Carollo}, C.~M., {et~al.}
  2010{\natexlab{a}}, \apjl, 725, L150

\bibitem[{{Oesch} {et~al.}(2010{\natexlab{b}}){Oesch}, {Bouwens}, {Carollo},
  {Illingworth}, {Trenti}, {Stiavelli}, {Magee}, {Labb{\'e}}, \&
  {Franx}}]{2010ApJ...709L..21O}
{Oesch}, P.~A., {Bouwens}, R.~J., {Carollo}, C.~M., {et~al.}
  2010{\natexlab{b}}, \apjl, 709, L21

\bibitem[{{Oesch} {et~al.}(2013){Oesch}, {Bouwens}, {Illingworth}, {Labb{\'e}},
  {Franx}, {van Dokkum}, {Trenti}, {Stiavelli}, {Gonzalez}, \&
  {Magee}}]{2013ApJ...773...75O}
{Oesch}, P.~A., {Bouwens}, R.~J., {Illingworth}, G.~D., {et~al.} 2013, \apj,
  773, 75

\bibitem[{{Oke}(1974)}]{1974ApJS...27...21O}
{Oke}, J.~B. 1974, \apjs, 27, 21

\bibitem[{{Oke} \& {Gunn}(1983)}]{1983ApJ...266..713O}
{Oke}, J.~B. \& {Gunn}, J.~E. 1983, \apj, 266, 713

\bibitem[{{O'Leary} {et~al.}(2021){O'Leary}, {Moster}, {Naab}, \&
  {Somerville}}]{2021MNRAS.501.3215O}
{O'Leary}, J.~A., {Moster}, B.~P., {Naab}, T., \& {Somerville}, R.~S. 2021,
  \mnras, 501, 3215

\bibitem[{{Ono} {et~al.}(2018){Ono}, {Ouchi}, {Harikane}, {Toshikawa}, {Rauch},
  {Yuma}, {Sawicki}, {Shibuya}, {Shimasaku}, {Oguri}, {Willott}, {Akhlaghi},
  {Akiyama}, {Coupon}, {Kashikawa}, {Komiyama}, {Konno}, {Lin}, {Matsuoka},
  {Miyazaki}, {Nagao}, {Nakajima}, {Silverman}, {Tanaka}, {Taniguchi}, \&
  {Wang}}]{2018PASJ...70S..10O}
{Ono}, Y., {Ouchi}, M., {Harikane}, Y., {et~al.} 2018, \pasj, 70, S10

\bibitem[{{Ostriker} \& {Tremaine}(1975)}]{1975ApJ...202L.113O}
{Ostriker}, J.~P. \& {Tremaine}, S.~D. 1975, \apjl, 202, L113

\bibitem[{{Parsa} {et~al.}(2016){Parsa}, {Dunlop}, {McLure}, \&
  {Mortlock}}]{2016MNRAS.456.3194P}
{Parsa}, S., {Dunlop}, J.~S., {McLure}, R.~J., \& {Mortlock}, A. 2016, \mnras,
  456, 3194

\bibitem[{{Patton} {et~al.}(2000){Patton}, {Carlberg}, {Marzke}, {Pritchet},
  {da Costa}, \& {Pellegrini}}]{2000ApJ...536..153P}
{Patton}, D.~R., {Carlberg}, R.~G., {Marzke}, R.~O., {et~al.} 2000, \apj, 536,
  153

\bibitem[{{Peng} {et~al.}(2010){Peng}, {Lilly}, {Kova{\v{c}}}, {Bolzonella},
  {Pozzetti}, {Renzini}, {Zamorani}, {Ilbert}, {Knobel}, {Iovino}, {Maier},
  {Cucciati}, {Tasca}, {Carollo}, {Silverman}, {Kampczyk}, {de Ravel},
  {Sanders}, {Scoville}, {Contini}, {Mainieri}, {Scodeggio}, {Kneib}, {Le
  F{\`e}vre}, {Bardelli}, {Bongiorno}, {Caputi}, {Coppa}, {de la Torre},
  {Franzetti}, {Garilli}, {Lamareille}, {Le Borgne}, {Le Brun}, {Mignoli},
  {Perez Montero}, {Pello}, {Ricciardelli}, {Tanaka}, {Tresse}, {Vergani},
  {Welikala}, {Zucca}, {Oesch}, {Abbas}, {Barnes}, {Bordoloi}, {Bottini},
  {Cappi}, {Cassata}, {Cimatti}, {Fumana}, {Hasinger}, {Koekemoer},
  {Leauthaud}, {Maccagni}, {Marinoni}, {McCracken}, {Memeo}, {Meneux}, {Nair},
  {Porciani}, {Presotto}, \& {Scaramella}}]{2010ApJ...721..193P}
{Peng}, Y.-j., {Lilly}, S.~J., {Kova{\v{c}}}, K., {et~al.} 2010, \apj, 721, 193

\bibitem[{{Qu} {et~al.}(2017){Qu}, {Helly}, {Bower}, {Theuns}, {Crain},
  {Frenk}, {Furlong}, {McAlpine}, {Schaller}, {Schaye}, \&
  {White}}]{2017MNRAS.464.1659Q}
{Qu}, Y., {Helly}, J.~C., {Bower}, R.~G., {et~al.} 2017, \mnras, 464, 1659

\bibitem[{{Ravindranath} {et~al.}(2006){Ravindranath}, {Giavalisco},
  {Ferguson}, {Conselice}, {Katz}, {Weinberg}, {Lotz}, {Dickinson}, {Fall},
  {Mobasher}, \& {Papovich}}]{2006ApJ...652..963R}
{Ravindranath}, S., {Giavalisco}, M., {Ferguson}, H.~C., {et~al.} 2006, \apj,
  652, 963

\bibitem[{{Ren} {et~al.}(2019){Ren}, {Trenti}, \&
  {Mason}}]{2019ApJ...878..114R}
{Ren}, K., {Trenti}, M., \& {Mason}, C.~A. 2019, \apj, 878, 114

\bibitem[{{Richardson}(1972)}]{1972JOSA...62...55R}
{Richardson}, W.~H. 1972, Journal of the Optical Society of America
  (1917-1983), 62, 55

\bibitem[{{Rudin} {et~al.}(1992){Rudin}, {Osher}, \&
  {Fatemi}}]{1992PhyD...60..259R}
{Rudin}, L.~I., {Osher}, S., \& {Fatemi}, E. 1992, Physica D Nonlinear
  Phenomena, 60, 259

\bibitem[{{Sawicki} {et~al.}(2020){Sawicki}, {Arcila-Osejo}, {Golob},
  {Moutard}, {Arnouts}, \& {Cheema}}]{2020MNRAS.494.1366S}
{Sawicki}, M., {Arcila-Osejo}, L., {Golob}, A., {et~al.} 2020, \mnras, 494,
  1366

\bibitem[{{Schechter}(1976)}]{1976ApJ...203..297S}
{Schechter}, P. 1976, \apj, 203, 297

\bibitem[{{Schenker} {et~al.}(2013){Schenker}, {Robertson}, {Ellis}, {Ono},
  {McLure}, {Dunlop}, {Koekemoer}, {Bowler}, {Ouchi}, {Curtis-Lake}, {Rogers},
  {Schneider}, {Charlot}, {Stark}, {Furlanetto}, \&
  {Cirasuolo}}]{2013ApJ...768..196S}
{Schenker}, M.~A., {Robertson}, B.~E., {Ellis}, R.~S., {et~al.} 2013, \apj,
  768, 196

\bibitem[{{Shibuya} {et~al.}(2015){Shibuya}, {Ouchi}, \&
  {Harikane}}]{2015ApJS..219...15S}
{Shibuya}, T., {Ouchi}, M., \& {Harikane}, Y. 2015, \apjs, 219, 15

\bibitem[{{Shibuya} {et~al.}(2016){Shibuya}, {Ouchi}, {Kubo}, \&
  {Harikane}}]{2016ApJ...821...72S}
{Shibuya}, T., {Ouchi}, M., {Kubo}, M., \& {Harikane}, Y. 2016, \apj, 821, 72

\bibitem[{{Snyder} {et~al.}(2017){Snyder}, {Lotz}, {Rodriguez-Gomez},
  {Guimar{\~a}es}, {Torrey}, \& {Hernquist}}]{2017MNRAS.468..207S}
{Snyder}, G.~F., {Lotz}, J.~M., {Rodriguez-Gomez}, V., {et~al.} 2017, \mnras,
  468, 207

\bibitem[{{Song} {et~al.}(2016){Song}, {Finkelstein}, {Ashby}, {Grazian}, {Lu},
  {Papovich}, {Salmon}, {Somerville}, {Dickinson}, {Duncan}, {Faber}, {Fazio},
  {Ferguson}, {Fontana}, {Guo}, {Hathi}, {Lee}, {Merlin}, \&
  {Willner}}]{2016ApJ...825....5S}
{Song}, M., {Finkelstein}, S.~L., {Ashby}, M. L.~N., {et~al.} 2016, \apj, 825,
  5

\bibitem[{{Stefanon} {et~al.}(2021){Stefanon}, {Bouwens}, {Labb{\'e}},
  {Illingworth}, {Gonzalez}, \& {Oesch}}]{2021arXiv210316571S}
{Stefanon}, M., {Bouwens}, R.~J., {Labb{\'e}}, I., {et~al.} 2021, arXiv
  e-prints, arXiv:2103.16571

\bibitem[{{Stefanon} {et~al.}(2017){Stefanon}, {Labb{\'e}}, {Bouwens},
  {Brammer}, {Oesch}, {Franx}, {Fynbo}, {Milvang-Jensen}, {Muzzin},
  {Illingworth}, {Le F{\`e}vre}, {Caputi}, {Holwerda}, {McCracken}, {Smit}, \&
  {Magee}}]{2017ApJ...851...43S}
{Stefanon}, M., {Labb{\'e}}, I., {Bouwens}, R.~J., {et~al.} 2017, \apj, 851, 43

\bibitem[{{Stefanon} {et~al.}(2019){Stefanon}, {Labb{\'e}}, {Bouwens}, {Oesch},
  {Ashby}, {Caputi}, {Franx}, {Fynbo}, {Illingworth}, {Le F{\`e}vre},
  {Marchesini}, {McCracken}, {Milvang-Jensen}, {Muzzin}, \& {van
  Dokkum}}]{2019ApJ...883...99S}
{Stefanon}, M., {Labb{\'e}}, I., {Bouwens}, R.~J., {et~al.} 2019, \apj, 883, 99

\bibitem[{{Steidel} {et~al.}(1999){Steidel}, {Adelberger}, {Giavalisco},
  {Dickinson}, \& {Pettini}}]{1999ApJ...519....1S}
{Steidel}, C.~C., {Adelberger}, K.~L., {Giavalisco}, M., {Dickinson}, M., \&
  {Pettini}, M. 1999, \apj, 519, 1

\bibitem[{{Stevans} {et~al.}(2018){Stevans}, {Finkelstein}, {Wold},
  {Kawinwanichakij}, {Papovich}, {Sherman}, {Ciardullo}, {Florez}, {Gronwall},
  {Jogee}, {Somerville}, \& {Yung}}]{2018ApJ...863...63S}
{Stevans}, M.~L., {Finkelstein}, S.~L., {Wold}, I., {et~al.} 2018, \apj, 863,
  63

\bibitem[{{Tacconi} {et~al.}(2018){Tacconi}, {Genzel}, {Saintonge}, {Combes},
  {Garc{\'\i}a-Burillo}, {Neri}, {Bolatto}, {Contini}, {F{\"o}rster Schreiber},
  {Lilly}, {Lutz}, {Wuyts}, {Accurso}, {Boissier}, {Boone}, {Bouch{\'e}},
  {Bournaud}, {Burkert}, {Carollo}, {Cooper}, {Cox}, {Feruglio}, {Freundlich},
  {Herrera-Camus}, {Juneau}, {Lippa}, {Naab}, {Renzini}, {Salome}, {Sternberg},
  {Tadaki}, {{\"U}bler}, {Walter}, {Weiner}, \& {Weiss}}]{2018ApJ...853..179T}
{Tacconi}, L.~J., {Genzel}, R., {Saintonge}, A., {et~al.} 2018, \apj, 853, 179

\bibitem[{{Takahashi} {et~al.}(2011){Takahashi}, {Oguri}, {Sato}, \&
  {Hamana}}]{2011ApJ...742...15T}
{Takahashi}, R., {Oguri}, M., {Sato}, M., \& {Hamana}, T. 2011, \apj, 742, 15

\bibitem[{{Tanaka} {et~al.}(2018){Tanaka}, {Coupon}, {Hsieh}, {Mineo},
  {Nishizawa}, {Speagle}, {Furusawa}, {Miyazaki}, \&
  {Murayama}}]{2018PASJ...70S...9T}
{Tanaka}, M., {Coupon}, J., {Hsieh}, B.-C., {et~al.} 2018, \pasj, 70, S9

\bibitem[{{Tasca} {et~al.}(2014){Tasca}, {Le F{\`e}vre}, {L{\'o}pez-Sanjuan},
  {Wang}, {Cassata}, {Garilli}, {Ilbert}, {Le Brun}, {Lemaux}, {Maccagni},
  {Tresse}, {Bardelli}, {Contini}, {Charlot}, {Cucciati}, {Fontana},
  {Giavalisco}, {Kneib}, {Salvato}, {Taniguchi}, {Vergani}, {Zamorani}, \&
  {Zucca}}]{2014A&A...565A..10T}
{Tasca}, L.~A.~M., {Le F{\`e}vre}, O., {L{\'o}pez-Sanjuan}, C., {et~al.} 2014,
  \aap, 565, A10

\bibitem[{Tibshirani(1996)}]{10.2307/2346178}
Tibshirani, R. 1996, Journal of the Royal Statistical Society. Series B
  (Methodological), 58, 267

\bibitem[{{Toshikawa} {et~al.}(2018){Toshikawa}, {Uchiyama}, {Kashikawa},
  {Ouchi}, {Overzier}, {Ono}, {Harikane}, {Ishikawa}, {Kodama}, {Matsuda},
  {Lin}, {Onoue}, {Tanaka}, {Nagao}, {Akiyama}, {Komiyama}, {Goto}, \&
  {Lee}}]{2018PASJ...70S..12T}
{Toshikawa}, J., {Uchiyama}, H., {Kashikawa}, N., {et~al.} 2018, \pasj, 70, S12

\bibitem[{{Ventou} {et~al.}(2017){Ventou}, {Contini}, {Bouch{\'e}}, {Epinat},
  {Brinchmann}, {Bacon}, {Inami}, {Lam}, {Drake}, {Garel}, {Michel-Dansac},
  {Pello}, {Steinmetz}, {Weilbacher}, {Wisotzki}, \&
  {Carollo}}]{2017A&A...608A...9V}
{Ventou}, E., {Contini}, T., {Bouch{\'e}}, N., {et~al.} 2017, \aap, 608, A9

\bibitem[{{Ventou} {et~al.}(2019){Ventou}, {Contini}, {Bouch{\'e}}, {Epinat},
  {Brinchmann}, {Inami}, {Richard}, {Schroetter}, {Soucail}, {Steinmetz}, \&
  {Weilbacher}}]{2019A&A...631A..87V}
{Ventou}, E., {Contini}, T., {Bouch{\'e}}, N., {et~al.} 2019, \aap, 631, A87

\bibitem[{{Vijayan} {et~al.}(2021){Vijayan}, {Wilkins}, {Lovell}, {Thomas},
  {Camps}, {Baes}, {Trayford}, {Kuusisto}, \& {Roper}}]{2021arXiv210800830V}
{Vijayan}, A.~P., {Wilkins}, S.~M., {Lovell}, C.~C., {et~al.} 2021, arXiv
  e-prints, arXiv:2108.00830

\bibitem[{Wen {et~al.}(2016)Wen, Chan, \& Zeng}]{Wen:2016vb}
Wen, Y., Chan, R.~H., \& Zeng, T. 2016, Science China Mathematics, 59, 141

\bibitem[{{Willott} {et~al.}(2013){Willott}, {McLure}, {Hibon}, {Bielby},
  {McCracken}, {Kneib}, {Ilbert}, {Bonfield}, {Bruce}, \&
  {Jarvis}}]{2013AJ....145....4W}
{Willott}, C.~J., {McLure}, R.~J., {Hibon}, P., {et~al.} 2013, \aj, 145, 4

\bibitem[{{Wong} {et~al.}(2011){Wong}, {Blanton}, {Burles}, {Coil}, {Cool},
  {Eisenstein}, {Moustakas}, {Zhu}, \& {Arnouts}}]{2011ApJ...728..119W}
{Wong}, K.~C., {Blanton}, M.~R., {Burles}, S.~M., {et~al.} 2011, \apj, 728, 119

\bibitem[{{Wyithe} {et~al.}(2011){Wyithe}, {Yan}, {Windhorst}, \&
  {Mao}}]{2011Natur.469..181W}
{Wyithe}, J. S.~B., {Yan}, H., {Windhorst}, R.~A., \& {Mao}, S. 2011, \nat,
  469, 181

\end{thebibliography}

\section{Appendix}\label{sec_appendix}

\subsection{Solutions of ADMM Subproblems}\label{sec_subproblem}

In this appendix, we briefly explain how to solve Equation (\ref{eq_penalty}) with the Augmented Lagrangian (Equation \ref{eq_lagrangian}). Based on the standard ADMM procedure \citep{MAL-016}, we divide the problem of the RL deconvolution into subproblems for the PSF convolution, the Poisson noise, and the flux non-negativity. Using the proximal operator $\mathbf{prox}$, we iteratively update the $\mathbf{x}$, $\mathbf{w}_1$, $\mathbf{w}_2$, and $\mathbf{h}$ (i.e., $\mathbf{u}$) vectors as follows,  

\begin{eqnarray}
\mathbf{x} \leftarrow \mathbf{prox}_{q}(\mathbf{v}) &=& \mathop{\rm arg~min}\limits_{\{\mathbf{x}\}} L_\rho (\mathbf{x}, \mathbf{w}, \mathbf{h}) \nonumber \\ \label{eq_iter_x}
 &=& \mathop{\rm arg~min}\limits_{\{\mathbf{x}\}} \frac{1}{2} \|\mathbf{Ax}-\mathbf{w}\|_2^2, \nonumber \\
 && \mathbf{v} = \mathbf{w} - \mathbf{u} \\ 
\mathbf{w}_1 \leftarrow \mathbf{prox}_{p, \rho}(\mathbf{v}) &=& \mathop{\rm arg~min}\limits_{\{\mathbf{w}_1\}} L_\rho (\mathbf{x}, \mathbf{w}, \mathbf{h}) \nonumber \\ \label{eq_iter_w1}
 &=& \mathop{\rm arg~min}\limits_{\{\mathbf{w}_1\}} -\log{p(\mathbf{y}|\mathbf{w}_1)} + \frac{\rho}{2} \|\mathbf{v}-\mathbf{w}_1\|_2^2, \nonumber \\
 && \mathbf{v} = \mathbf{Px} + \mathbf{u}_1 \\ 
\mathbf{w}_2 \leftarrow \mathbf{prox}_{i, \rho}(\mathbf{v}) &=& \mathop{\rm arg~min}\limits_{\{\mathbf{w}_2\}} L_\rho (\mathbf{x}, \mathbf{w}, \mathbf{h}) \nonumber \\ \label{eq_iter_w2}
 &=& \mathop{\rm arg~min}\limits_{\{\mathbf{w}_2\}} \mathit{I}_{R_+}(\mathbf{w}_2) + \frac{\rho}{2} \|\mathbf{v}-\mathbf{w}_2\|_2^2, \nonumber \\
 && \mathbf{v} = \mathbf{x} + \mathbf{u}_2 \\ 
\left[
    \begin{array}{c}
      \mathbf{u_1} \\
      \mathbf{u_2}
    \end{array}
\right]
 &\leftarrow& \mathbf{u} + \mathbf{Ax} - \mathbf{w} \label{eq_iter_u}
\end{eqnarray}

\noindent where $\mathbf{u}$ is the scaled Lagrange multiplier $\mathbf{u} = (1/\rho)\mathbf{h}$. In the following paragraphs, we derive the proximal operators for the $\mathbf{x}$, $\mathbf{w}_1$, and $\mathbf{w}_2$ updates in Equations (\ref{eq_iter_x})-(\ref{eq_iter_w2}). 

\paragraph*{For the $\mathbf{x}$ update: } The proximal operator for the $\mathbf{x}$ update is formulated as the solution of the quadratic problem, 

\begin{eqnarray}
 \mathbf{prox}_{q} (\mathbf{v}) &=& \mathop{\rm arg~min}\limits_{\{\mathbf{x}\}} 
\frac{1}{2} 
\left|\left|
\left[
    \begin{array}{c}
      \mathbf{P} \\
      \mathbf{I}
    \end{array}
\right]
x - 
\left[
    \begin{array}{c}
      \mathbf{v}_1 \\
      \mathbf{v}_2
    \end{array}
\right]
\right|\right|_2^2 \nonumber \\
&=& (\mathbf{P}^T\mathbf{P}+\mathbf{I})^{-1}(\mathbf{P}^T\mathbf{v}_1+\mathbf{v}_2) \nonumber \\
&=& \frac{c\ast v_1+v_2}{c\ast c+1} 
\end{eqnarray}

\noindent where $c$ is the PSF convolution kernel. 

\paragraph*{For the $\mathbf{w}_1$ update: } Next, we derive the proximal operator for the $\mathbf{w}_1$ update. Using Equation (\ref{eq_poisson}), the objective function in Equation (\ref{eq_iter_w1}) is rewritten as 

\begin{equation}
 J(\mathbf{w}_1) = -\log{(\mathbf{w}_1)}^T \mathbf{y} + \mathbf{w}_1^T \mathbf{1} + \frac{\rho}{2} (\mathbf{w}_1-\mathbf{v})^T(\mathbf{w}_1-\mathbf{v}) 
\end{equation}

\noindent We equate the gradient of the objective function $J(\mathbf{w}_1)$ to zero, 

\begin{eqnarray}
  \frac{\partial J}{\partial \mathbf{w}_{1}} &=& -\frac{\mathbf{y}}{\mathbf{w}_{1}}+1+\rho(\mathbf{w}_{1}-\mathbf{v}_{1}) \nonumber \\
 &=& \mathbf{w}^2_{1} + \frac{1-\rho\mathbf{v}}{\rho}\mathbf{w}_{1}-\frac{\mathbf{y}}{\rho} = 0. 
\end{eqnarray}

\noindent By solving this classical root-finding problem of the quadratic equation, we formulate the the proximal operator as, 

\begin{equation}
   \mathbf{prox}_{p, \rho} (\mathbf{v}) = \frac{1-\rho\mathbf{v}}{2\rho}+\sqrt{\left(\frac{1-\rho\mathbf{v}}{2\rho}\right)^2+\frac{\mathbf{y}}{\rho}}. 
\end{equation}

\paragraph*{For the $\mathbf{w}_2$ update: } The proximal operator for the $\mathbf{w}_2$ updates can be formulated as 

\begin{eqnarray}
 \mathbf{prox}_{i, \rho} (\mathbf{v}) &=& \mathop{\rm arg~min}\limits_{\{\mathbf{w}_2\}} I_{R_+}(\mathbf{w}_2) + \frac{\rho}{2} \|\mathbf{v}-\mathbf{w}_2\|_2^2 \nonumber \\
 &=& \mathop{\rm arg~min}\limits_{\{\mathbf{w}_2\}} \|\mathbf{v}-\mathbf{w}_2\|_2^2 \nonumber \\
 &=& \Pi_{R_+} (\mathbf{v})
\end{eqnarray}

\noindent where $\Pi_{R_+}$ is defined as follows, 

\begin{equation}
  \Pi_{R_+}(\mathbf{v}_j) = 
  \left\{
    \begin{array}{ll}
      0 & \mathbf{v}_j < 0 \\
      \mathbf{v}_j & \mathbf{v}_j \geq 0
    \end{array}
  \right.
\end{equation}

\mbox{}

After the convergence of the iteration in Equations (\ref{eq_iter_x})-(\ref{eq_iter_u}), we obtain the restored image $\mathbf{x}$.

\subsection{Examples of Mis-classified Galaxies}\label{sec_subproblem}

\begin{figure}
 \begin{center}
  \includegraphics[width=80mm]{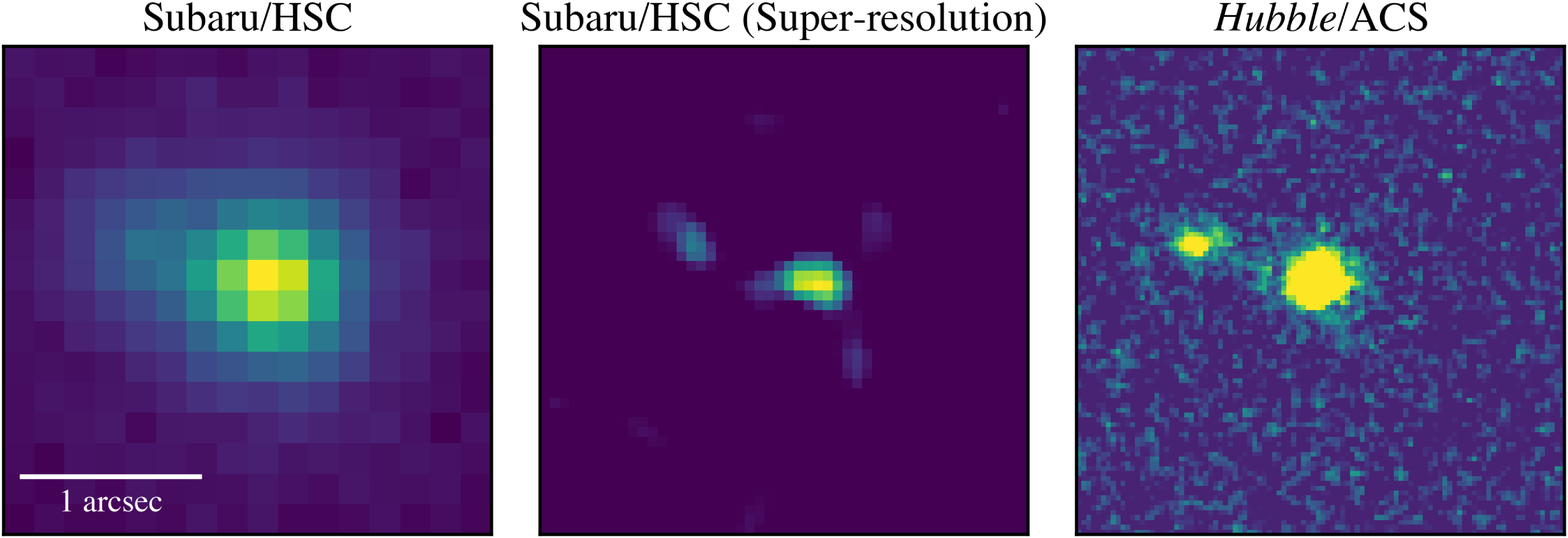}\\
  \includegraphics[width=80mm]{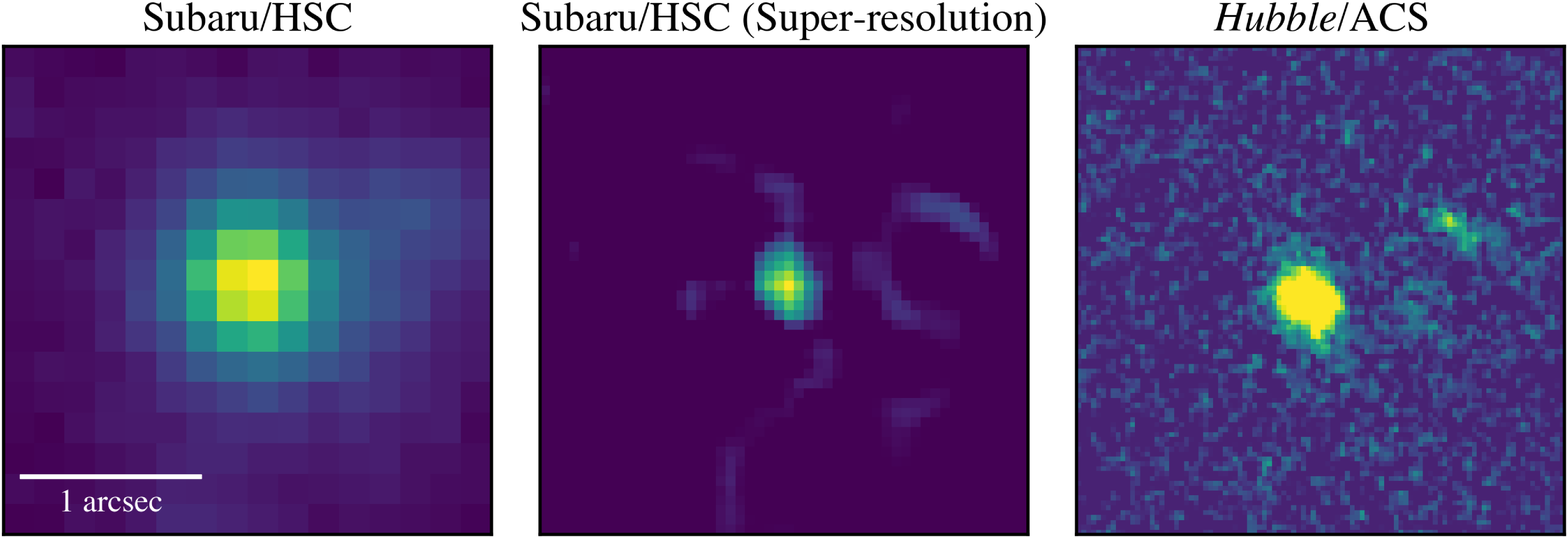}\\
  \includegraphics[width=80mm]{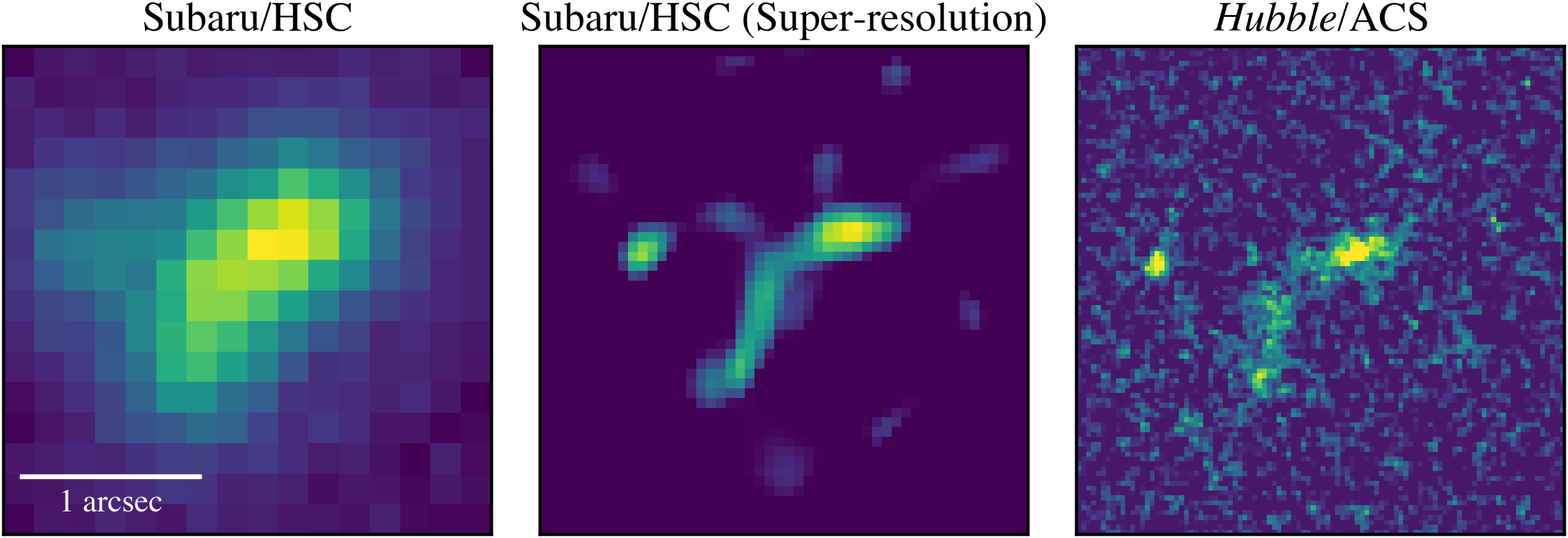}\\
  \mbox{}\\
  \includegraphics[width=80mm]{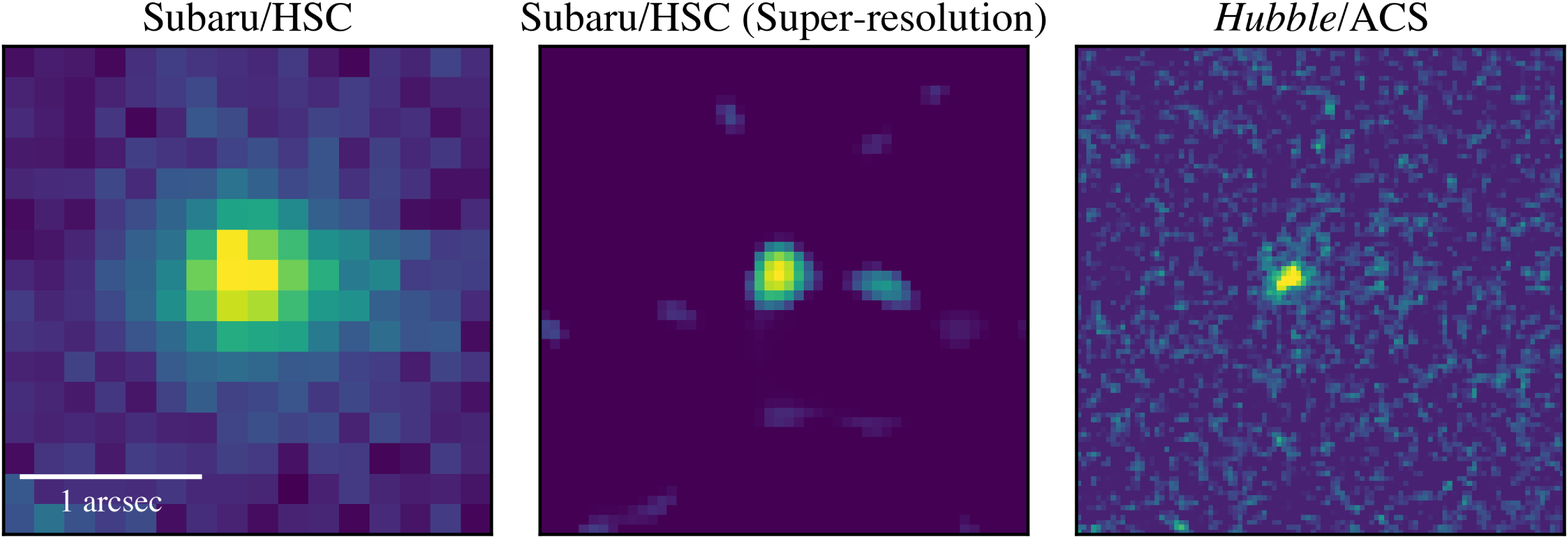}\\
  \includegraphics[width=80mm]{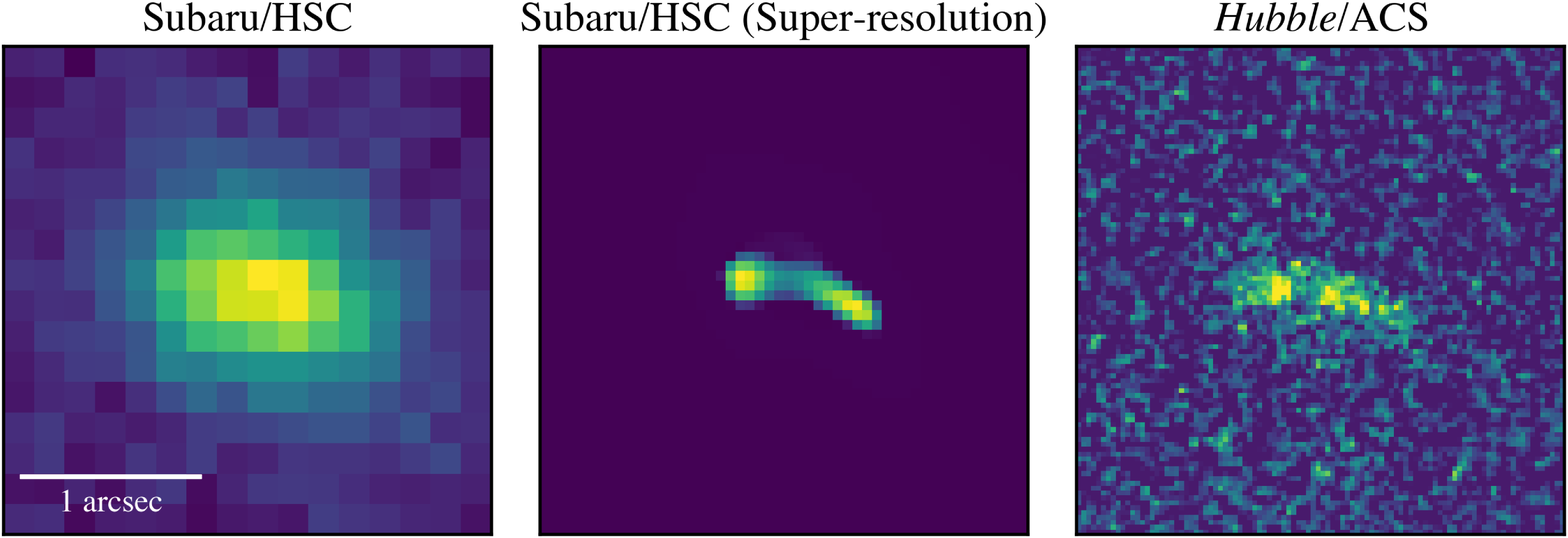}\\ 
  \includegraphics[width=80mm]{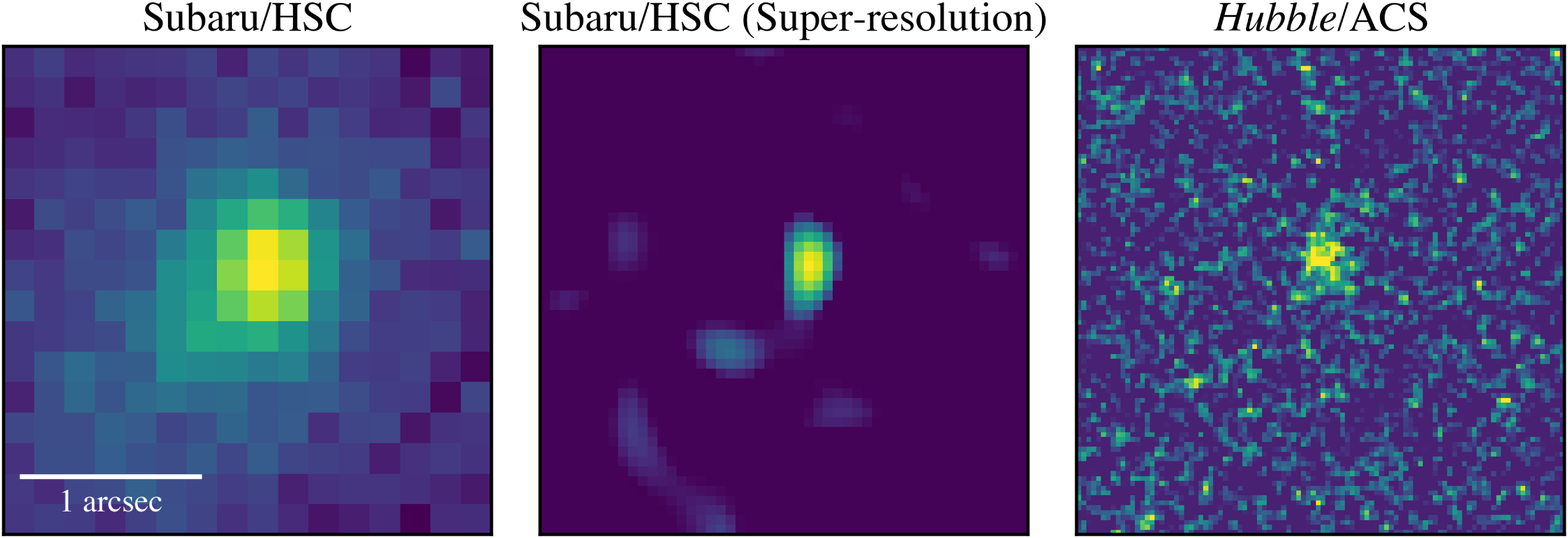}
  \end{center}
   \caption{Same as Figure \ref{fig_big_image}, but for examples of mis-classified galaxies. The top (bottom) three galaxies are galaxy mergers (isolated galaxies) that are incorrectly classified as isolated galaxies (galaxy mergers) in our analysis.}\label{fig_image_false}
\end{figure}

Figure \ref{fig_image_false} shows examples of mis-classified galaxies in the selection of galaxy major mergers.

\end{document}